\documentclass[fleqn,usenatbib]{mnras}

\usepackage[T1]{fontenc}

\DeclareRobustCommand{\VAN}[3]{#2}
\let\VANthebibliography\thebibliography
\def\thebibliography{\DeclareRobustCommand{\VAN}[3]{##3}\VANthebibliography}

\usepackage{graphicx}	% Including figure files
\usepackage{multicol}	% Advanced maths commands
\usepackage{wrapfig}	% Extra maths symbols
\usepackage{longtable}
\newcommand{\kms}{\mbox{km\,s$^{-1}$}}
\newcommand{\etal}{\mbox{\rm{et al.}~~}}
\title[WALLABY Pilot Survey Phase 2]{WALLABY Pilot Survey: the Tully-Fisher relation in the NGC 4808, Vela and NGC 5044 fields}

\author[Mould et al.]{
Jeremy Mould,$^{1,2}$\thanks{E-mail: jmould@swin.edu.au}
T. H. Jarrett,$^{3}$
H\'el\`ene Courtois$^{4}$
Albert Bosma$^{5}$
Nathan Deg$^{14}$
Alexandra Dupuy$^{8}$
\newauthor 
Lister Staveley-Smith$^{9,10}$
E.N. Taylor$^1$
Jayanne English$^{20}$
S. H. A. Rajohnson$^{21}$
Ren\'ee Kraan-Korteweg$^{21}$
\newauthor 
Duncan Forbes$^1$
Helga D\'enes$^{7}$ 
Karen Lee-Waddell$^{6,9,10}$
Austin Shen$^{18,19}$
O. I. Wong$^{6,9,16}$ 
\newauthor 
Benne Holwerda$^{13}$ 
B\"arbel Koribalski$^{11,12}$ 
Denis Leahy$^{22}$
Pavel Mancera Pi\~{n}a$^{15}$
Niankun Yu$^{17,23}$
\\
\\
$^{1}$Centre for Astrophysics \& Supercomputing, Swinburne University, Hawthorn, VIC 3122, Australia.\\
$^{2}$ARC Centre of Excellence for Dark Matter Particle Physics \\
$^{3}$Astronomy Department, University of Cape Town, Private Bag X3, Rondebosch 7701, South Africa\\
$^4$UCB Lyon 1, IUF IP21,  Lyon, 69622, Villeurbanne, France.\\
$^{5}$Aix Marseille University, CNRS, CNES, LAM Marseille, France.\\
$^6$CSIRO Space \& Astronomy, PO Box 1130, Bentley, WA 6102, Australia.\\
$^7$School of Physical Science \& Nanotechnology, Yachay Tech University, Hacienda San Jos\'e S/N 100119, Urcuqu\'i, Ecuador.\\
$^8$Korea Institute for Advanced Study, 85, Hoegi-ro, Dongdaemun-gu, Seoul 02455, Republic of Korea.\\
$^9$ICRAR-M468, University of Western Australia, 35 Stirling Highway, Crawley, WA 6009, Australia.\\
$^{10}$ICRAR - Curtin University, Bentley. WA 6102, Australia.\\
$^{11}$Australia Telescope National Facility, CSIRO Astronomy \& Space Science, PO Box 76, Epping, NSW 1710, Australia.\\
$^{12}$School of Science, Western Sydney University, Locked Bag 1797, Penrith, NSW 2751, Australia.\\
$^{13}$University of Louisville, Department of Physics \& Astronomy, Louisville KY 40292, USA.\\
$^{14}$Department of Physics, Engineering Physics \& Astronomy, Queen's University, Kingston, ON, K7L 3N6, Canada.\\
$^{15}$Leiden Observatory, Leiden University, P.O. Box 9513, 2300 RA Leiden,
The Netherlands\\
$^{16}$ARC Centre of Excellence for All Sky Astrophysics in 3 Dimensions (ASTRO 3D).\\
$^{17}$National Astronomical Observatories, Chinese Academy of Sciences, Beijing, 100101, P.R. China\\
$^{18}$CSIRO Space \& Astronomy, PO Box 1130, Bentley, WA 6102, Australia\\
$^{19}$Australian SKA Regional Centre (AusSRC)\\
$^{20}$Department of Physics and Astronomy, University of Manitoba, Winnipeg, Manitoba, Canada, R3T 2N2, Canada\\
$^{21}$Department of Astronomy, University of Cape Town, Private Bag X3, Rondebosch 7701, South Africa\\
$^{22}$Department of Physics and Astronomy, University of Calgary, Calgary, AB T2N 1N4, Canada\\
$^{23}$Key Laboratory of Radio Astronomy and Technology, Chinese Academy of Sciences, Beijing, 100101, P.R. China
}

\date{Accepted 2024 June 14. Received 2024 June 13; in original form 2024 May 8}

\pubyear{2024}

\begin{document}
\label{firstpage}
\pagerange{\pageref{firstpage}--\pageref{lastpage}}
\maketitle

\begin{abstract}
The Tully-Fisher Relation (TFR) is a well-known empirical relationship between the luminosity of a spiral
galaxy and its circular velocity, allowing us to estimate redshift independent distances. Here we use high
signal-to-noise HI 21-cm integrated spectra from the second pilot data release (PDR2, 180 deg$^2$) of the
Widefield ASKAP L-band Legacy All-sky Blind surveY (WALLABY). 
In order to prepare for the full WALLABY %hemisphere 
survey, we have investigated the TFR in phase 2 of the pilot survey with a further three fields.
	The data were obtained with wide-field
Phased Array Feeds on the Australian Square Kilometre Array Pathfinder (ASKAP) and have an angular
resolution of 30 arcsec and a velocity resolution of $\sim$4 \kms. Galaxy luminosities have been measured from the
Wide-field Infrared Survey Explorer (WISE), and optical galaxy inclinations from the Dark Energy Camera Legacy Survey.
% We present the TFR for 1000? spiral galaxies and find … 
%ratios.
	We present TFRs for wavelengths from 0.8--3.4$\mu$m. We examine sources of galaxy inclination data and investigate magnitudes from the DECam Local
	Volume Exploration Survey
	(DELVE) and DENIS catalogues and the 4HS target catalogue based on the VISTA Hemisphere Survey (VHS). We consider the baryonic TFR. 
	These are all of interest for TFR using the  full WALLABY survey of 200,000 galaxies.  %A few WALLABY detections have no optical counterpart, and only lower limits can currently be placed on their M$_{\rm{HI}}$/L ratios; these sources need further follow-up. 
	We demonstrate that WALLABY TFR distances can take their place among state of the art studies of the local velocity field.
\end{abstract}


%%%%%%%%%%%%%%%%%%%%%%%%%%%%%%%%%%%%%%%%%%%%%%%%%%

%%%%%%%%%%%%%%%%% BODY OF PAPER %%%%%%%%%%%%%%%%%%

\section{Introduction}
%{\Large DRAFT}

In the local Universe large-scale structure can be investigated using redshift-
independent distance indicators, such as the Tully-Fisher Relation (TFR),
which allow galaxies' peculiar velocities to be measured.
Peculiar velocities come about because mass density inhomogeneities in the Universe act on galaxies
and perturb the velocities given to them by the expansion of the Universe. To be exact,
the peculiar velocity of a galaxy arises from the volume integral of the overdensities divided
by the square of the distances to them, multiplied by the growth rate. Here the
over-density is $\delta$ = $\delta \rho / \bar\rho - 1$ where $\rho$ and $\bar\rho$ are
the density and mean density of the Universe, and the growth rate, $f$, is approximately $\Omega_m^{4/7}$
in the $\Lambda$CDM standard model, where $\Omega_m$ is the mass density parameter. Peculiar velocities have therefore been used
to constrain the cosmological parameter $f\sigma_8$ ($\sigma_8$ is the amplitude of the power spectrum on scales of 8$h^{-1}$ Mpc; Adams \& Blake 2017; Dupuy \etal
2019), thus providing a test of gravity on very large scales.  To measure peculiar velocities
we need redshift-independent distance indicators, such as the Fundamental Plane (Djorgovski \& Davis 1987),%Apj31359
the supernova standard candle (Kowal 1968) %AJ731021
and the Tully-Fisher relation (Tully \& Fisher 1977).
\begin{figure*}%[b]
\includegraphics[width=1.2\textwidth]{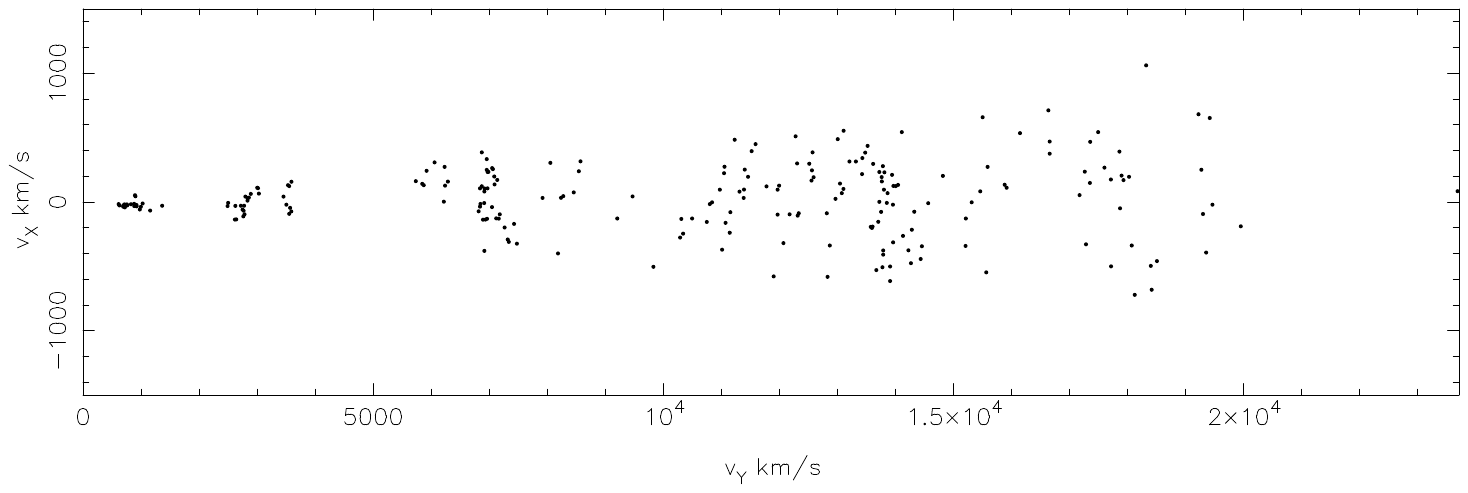}
\vskip -3 truein
\caption{WALLABY galaxies in the NGC4808 field in the supergalactic plane. The long axis
	is the heliocentric recession velocity in the SGY direction, and the vertical axis is the the heliocentric recession velocity in the SGX direction.% perpendicular to it. 
	  The structure is evident in galaxy groups and filaments.}
\end{figure*}

The Widefield ASKAP L-band Legacy All-sky Blind surveY (or WALLABY;
Koribalski \etal 2020) is being conducted on the Australian SKA Pathfinder (ASKAP, Hotan \etal 2021), an innovative imaging radio telescope located at the Murchison Radio-astronomy Observatory in Western Australia. The aim of WALLABY is to use the powerful widefield phased-array technology of ASKAP to observe initially 14,000 square degrees of the southern hemisphere in the 21-cm line of neutral hydrogen at 30-arcsec resolution (Westmeier \etal 2022)\footnote{Sky coverage is described at https://wallaby-survey.org/overview/}. The survey began in January 2023, but a number of fields were observed in 2022 as a pilot program, among them the phase 2 target fields centred on NGC 4808, Vela and NGC 5044.
 WALLABY pilot survey phase 2 data were released in DR2 by Murugeshan \etal (2024).

The goals of this paper are first to learn as much as possible about the Tully-Fisher
Relation (TFR) early on, since the full WALLABY Survey will be a hundred times larger than the pilot survey, and, second, to measure peculiar velocities to be compared  with expectations from the CosmicFlows program (Courtois \& Tully 2015).

In the paper we extract and measure HI spectra of galaxies in these fields and combine them ($\S\S$2,3 \& 4) 
with customized WISE total magnitudes (Jarrett \etal 2023). %The Tully Fisher relation (TFR) is presented in $\S$3. 
Infrared photometry has the advantage of being less affected by extinction in our Galaxy and the target galaxies (Aaronson, Huchra \& Mould 1979).
We also investigate available  I band photometry and optical diameters.
The utility of WALLABY for peculiar velocity measurement is considered in
$\S$5, and our conclusions are summarized in $\S$6.

\section{The NGC 4808 field}

Source finding using SoFiA (Westmeier \etal 2021; Serra \etal 2015) and HI measurements for the NGC 4808 field follow Courtois \etal (2022; Paper 1). 
SoFiA2 uses the Smooth + Clip algorithm for source finding, which operates by spatially and spectrally smoothing the data on multiple scales and applying a user-defined flux threshold relative to the noise level in each iteration. A wide range of useful preconditioning and post-processing filters is available, including noise normalization, flagging of artifacts and reliability filtering. 
Values of W$_{50}$ were measured directly from the flux density versus frequency spectra produced by the WALLABY pipeline.
All sources were inspected by Tobias Westmeier, and objects deemed to be questionable were deleted. An example would
be an optical galaxy split in two by SoFiA. Over the 3 fields between 7 and 14\% of the sources were eliminated in this way.
Half of these involve pairs or splits. It is customary to exclude such objects from the TFR.
The effects of radio frequency interference are also trapped at this stage.

Table~1 reports the TFR data for the NGC 4808 field. Column (1) is the name of the WALLABY survey HI detection. Columns (2) \& (3) are the SoFiA coordinates of the WALLABY HI detections. Column (4) is the mean
velocity of the HI profile integrated over the spectrum. Column (5)
is the axial ratio of the DECaLS\footnote{https://datalab.noirlab.edu/ls/decals.php} $g$ band galaxy image, which yields the inclination $i$ of the disk. The axial ratio was measured
at approximately 5\% of the sky background level with the ellipse task of IRAF\footnote{https://iraf-community.github.io}.
A minimum axial ratio of 0.2 was adopted. Zeros denote galaxies for which no axial ratio could be measured. 

Coordinates from the HyperLeda catalogue differ from WALLABY coordinates by 7.5$^{''}~rms$ after 2.5$\sigma$ deviates were removed, with equal contributions from RA and Dec. The WALLABY beam  diameter is 30$^{''}$ and this is a contributor\footnote{
Typical position uncertainties for HI point sources in WALLABY are estimated as the beam diameter (30$^"$)
divided by the source signal-to-noise, resulting in 7.5$"$ uncertainty only for 4$\sigma$ detections. Westmeier \etal (2022) found an $rms$ of 5$^{''}$ in the
phase~1 pilot survey}. In some cases there may also be a difference between the HI centroid and the centroid of the stellar light. Columns (6) \& (7) are the WISE\footnote{Wright \etal (2010)} W1 total galaxy magnitude after removal of contaminating stars and its uncertainty. Columns (8) \& (9)
are the width at half peak (or double peak) of the HI profile and its uncertainty. Column (10) is the signal to noise
ratio of the HI profile. Column (11) provides other names for the source from the HyperLeda catalogue. Double identifications were eliminated by choosing the closer of the two galaxies.
In the TFR we use Wmx, the width after correction for resolution and turbulence following Tully \& Fouqu\'e (1985).
The velocity axis of the TFR is $\Delta{\rm V(0)~ =~ Wmx~ /~ sin}~ i~/\rm{(1+z)}$, where $i$ is obtained from column (5)'s optical axial ratio.
Nineteen galaxies have Wmx both measured from WALLABY and compiled by Tully \etal (2009, EDD, the Extragalactic Distance Database) and Courtois \etal (2009).
The mean velocity width ratio, EDD/WALLABY, is 1.00 $\pm$ 0.01, and the $\chi^2$ per degree
of freedom for the identity relation is 0.27, indicating that for these bright galaxies the
velocity width uncertainties have been overestimated.
 The velocity structure of the NGC4808 field is shown in Figure~1. Some groups and filaments of galaxies are visible.

\subsection{Comparison with Arecibo observations}
Only a small portion of the WALLABY footprint protrudes into the northern
hemisphere. The NGC 4808 field therefore provides a rare opportunity to
compare observations with those obtained by the Arecibo Observatory
in the ALFALFA Survey (Haynes \etal 2018) and catalogued in EDD and labelled Wmx1. Figure 2 shows this comparison.

\begin{figure}%[h]
\includegraphics[width=1.2\columnwidth]{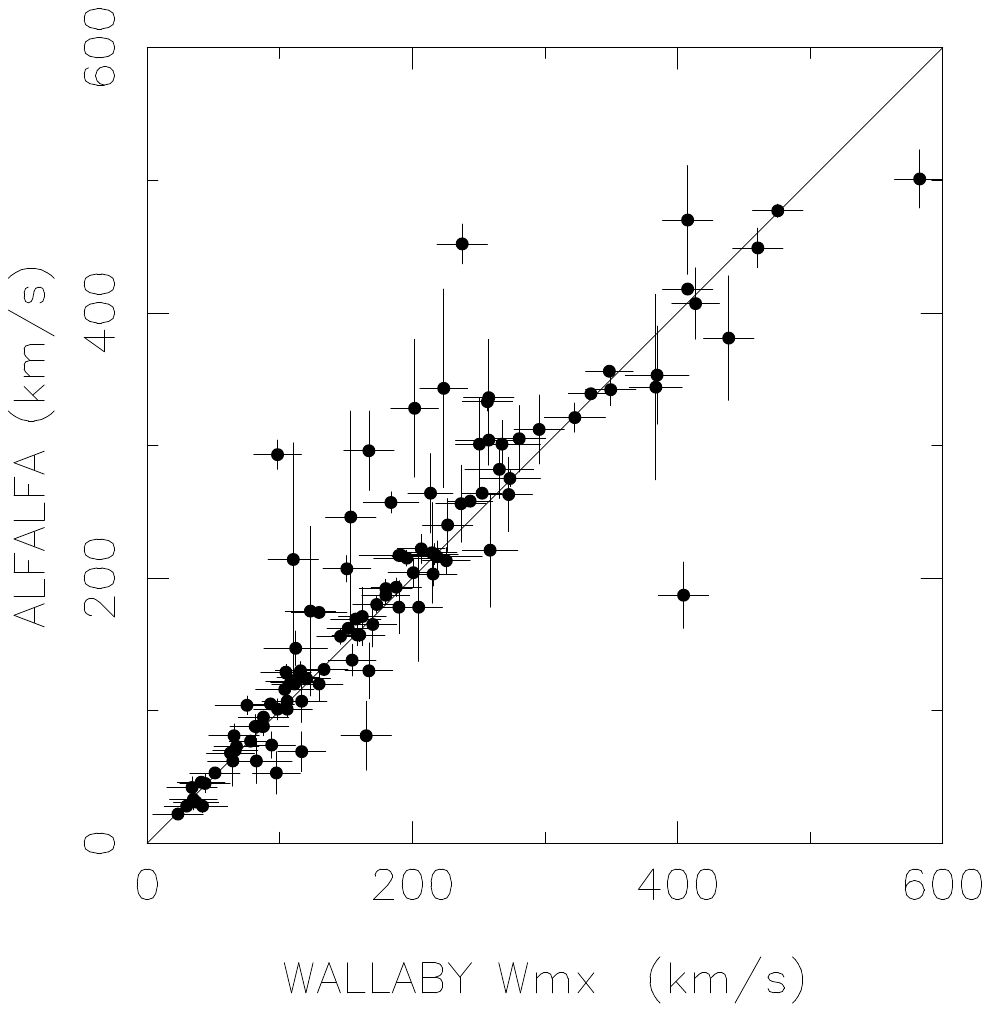}
	\caption{(a) Comparisons of velocity widths from Table 1 and those from ALFALFA.}
%{\bf Below:} A = ALFALFA, W = WALLABY. The vertical axis essentially shows the difference between WALLABY and ALFALFA velocity widths.}
\includegraphics[width=1.2\columnwidth]{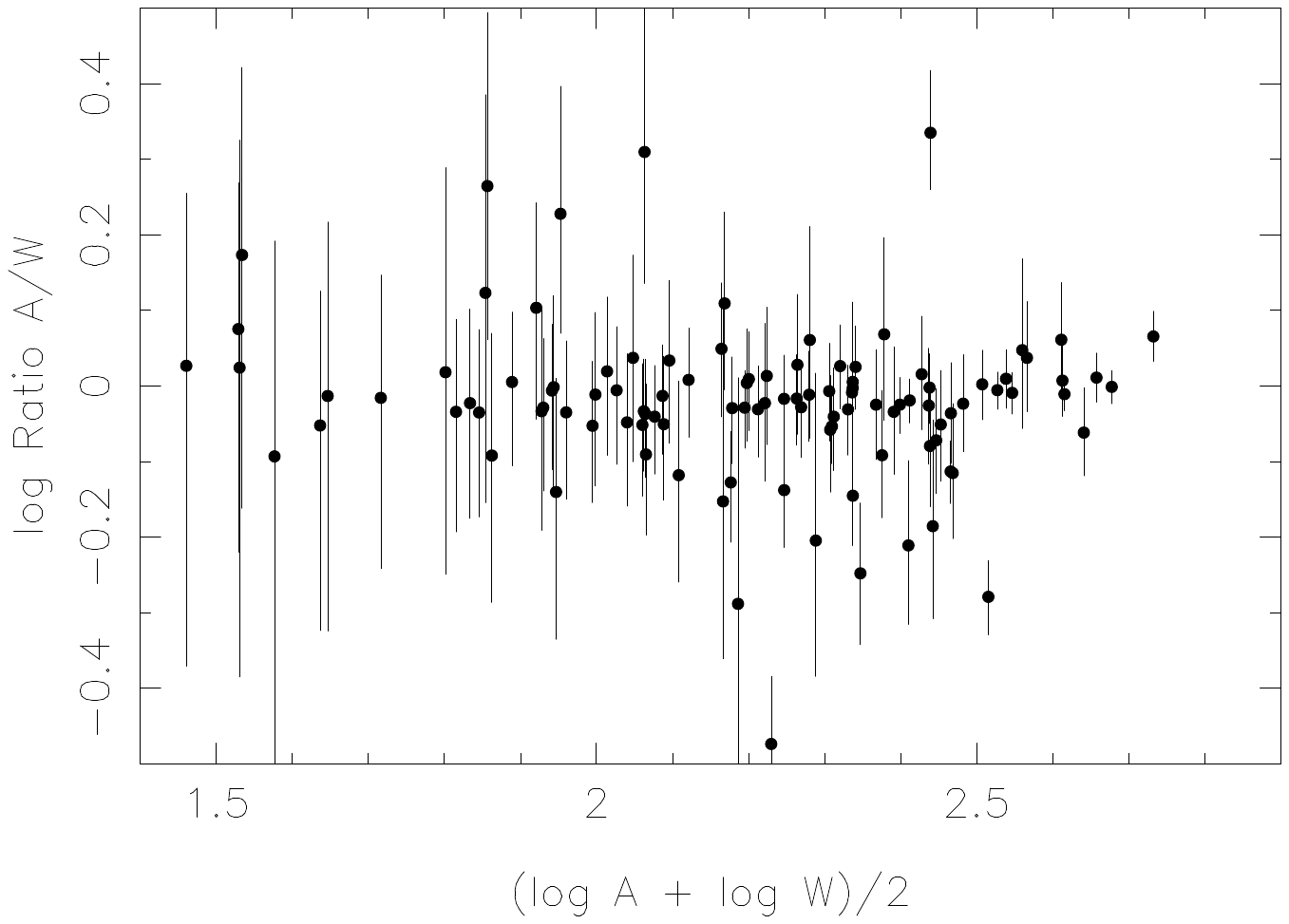}
\setcounter{figure}{1}
\caption{(b) A = ALFALFA, W = WALLABY. The vertical axis essentially shows the difference between WALLABY and ALFALFA velocity widths.}
\end{figure}

The measured redshifts of WALLABY and Arecibo are in good agreement. Although 6\% 
of the velocity widths are in disagreement by more than twice the joint uncertainty, %2$\sigma$, but
this is no more than expected statistically.
We note the more extreme cases, WALLABY J125258+073025, WALLABY J130412+43057, WALLABY J130012+054417 and WALLABY J130950+062601.
Examination of the HI profiles of the discrepant galaxies reveals that
both their WALLABY and ALFALFA profiles are of low signal to noise (S/N)and
they are all distant, cz $>$ 10,000 \kms. %We shall return to this issue in
%the next section. 
Another way of looking at this comparison is in the lower section of Figure~2.
Here the $hyperfit$ slope is significant, --0.072 $\pm$ 0.006.
\subsection{TFR for the NGC 4808 field}
The galaxy inclinations are calculated in the normal way $$\cos^2 i = \frac{(b/a)^2-0.2^2}{1 - 0.2^2}\eqno(1).$$ where $b/a$ is the axial ratio.
 The minimum axial ratio of 0.2 has been adopted since the original work of Tully \& Fisher (1977). Paturel \etal (2003) listed
no galaxies with $b/a~<$ 0.25 in their Principal Galaxies Catalogue. In $\S$A4 we note that we see exceptions with $b/a~<$ 0.2, but these are
recorded in Table 1 as $b/a$ = 0.2. This is not an issue, as they are clearly edge-on.
We use the data of Table 1 to construct the TFR. Excluded from Figure 3 are galaxies
with inclinations less than 45$^\circ$ for which sin $i$ is too uncertain, those with
velocity width errors greater than 20 \kms, and those with %signal to noise ratio less 
 S/N $<$ 3.7. Absolute magnitudes were calculated with a Hubble Constant of 73 \kms Mpc$^{-1}$
(Riess \etal 2022) after correction of heliocentric velocities to the cosmic microwave background frame, following Lineweaver \etal (1996). A simple least squares linear fit to Figure 3, excluding the outlier WALLABY J130436+045341 and 3 other 2.5$\sigma$ deviates, gives
a $\chi^2$ per degree of freedom of 5.1, its difference from unity indicating that the velocity width uncertainties
are partially responsible for the scatter in the TFR, but there is intrinsic scatter as well. In $\S$4.2 we quantify the uncertainties in $i$ which contribute 0.53 mag to the vertical errors, compared with 0.46 mag for the width uncertainties here. By comparison errors in W1 are negligible, growing slowly from 0.01 to 0.1 mag going from 14 to 17 mag in W1. We explore possible ways to reduce the uncertainties in $i$ in the Appendix. %A similar $\chi^2$ is obtained with the hyperfit (Robotham \& Obreschkow 2015). Pe
A similar $\chi^2$ is obtained from a fit with the {\em hyperfit} Bayesian regression package (Robotham \& Obreschkow 2015), including an assumed Gaussian scatter of 200 km/s from peculiar velocities. {\em Hyperfit} maximizes the likelihood of a linear fit in the presence of uncertainties in both coordinates. This is an important distinction from a least squares fit, when there is a range of errors in the data. Unbiased estimators for the population model of the vertical scatter  and its variance can be obtained. The website interface to {\em Hyperfit} was accessed at hyperfit.icrar.org. The intrinsic scatter in the TFR measured by {\em hyperfit} is 0.85 mag. Selection biases may also affect the fit, but these are not investigated here

\begin{figure}%[h]
\includegraphics[width=1.2\columnwidth]{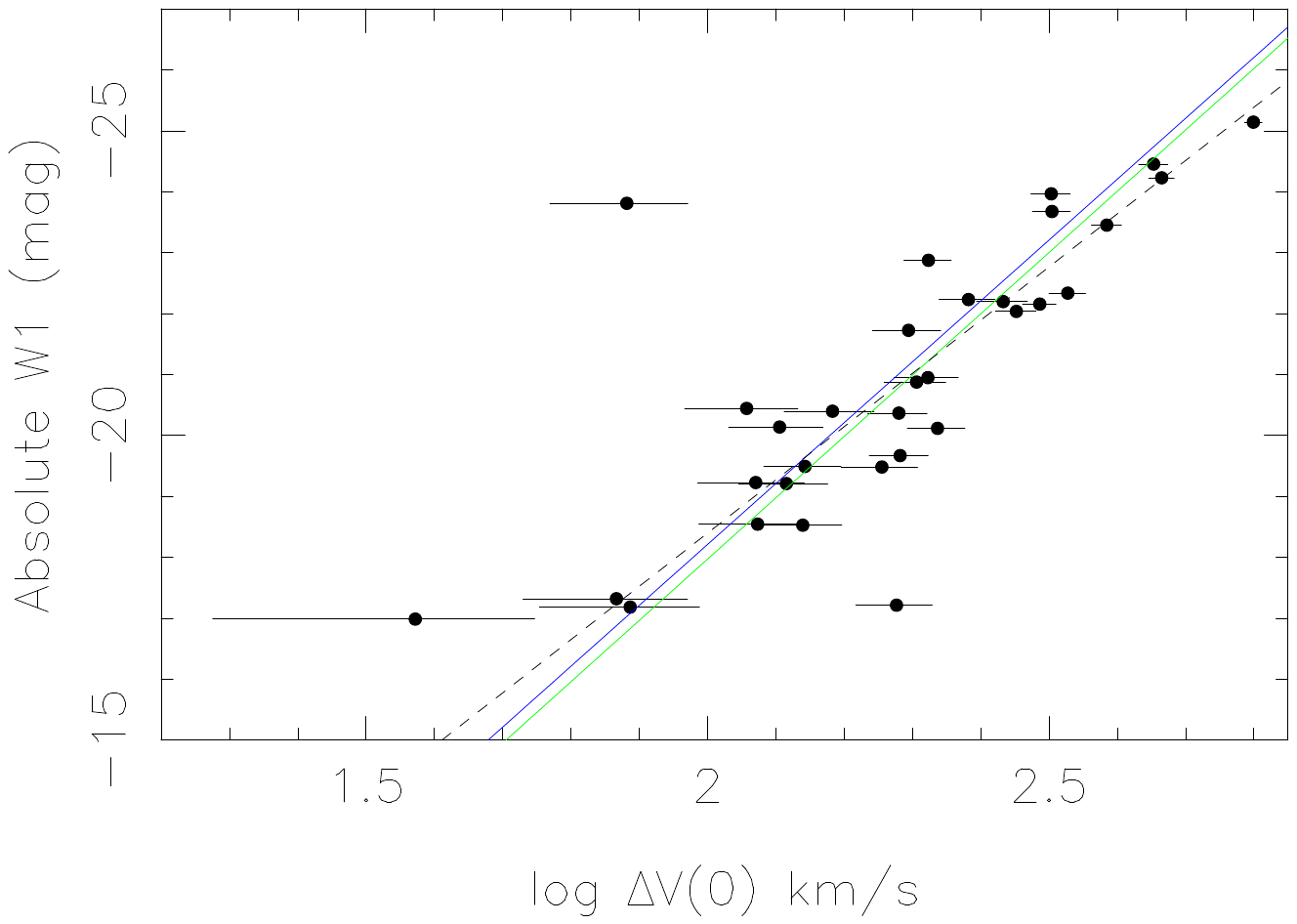}%FIG 3
\caption{TFR for the NGC4808 field. The dashed line is the best unweighted linear fit; the green line is the result from {\em hyperfit}, which weights the fitted points by their uncertainties and allows for intrinsic scatter. %The dashed line is the best linear fit.
	%The green line is the hyperfit which weights the fitted points by their uncertainties. 
	 The blue line is the TFR of Bell \etal (2023).}
\end{figure}
Photometry for galaxies with names in column (11) of Table 1  is also available from the Siena Galaxy Atlas (Moustakas \etal 2023). This is presented in the Appendix.

\subsection{Star formation}
The W1 WISE bandpass has some sensitivity to PAH emission (Cluver \etal 2017) and so we have
examined the W1--W2 colour in this sample. Figure 4 shows that the vast majority of these HI detections
has W1--W2 $<$ 0.2 mag, making it unlikely that there is a significant number of starburst galaxies
in the sample. We have located galaxies with W1--W2~$>$~0.25 mag in the TFR of Figure 3 and only one
of them, WALLABY J130903+065753 is significantly discrepant in the TFR (on the high luminosity side.)
 It may be a starburst galaxy.
\begin{figure}
\includegraphics[width=1.2\columnwidth]{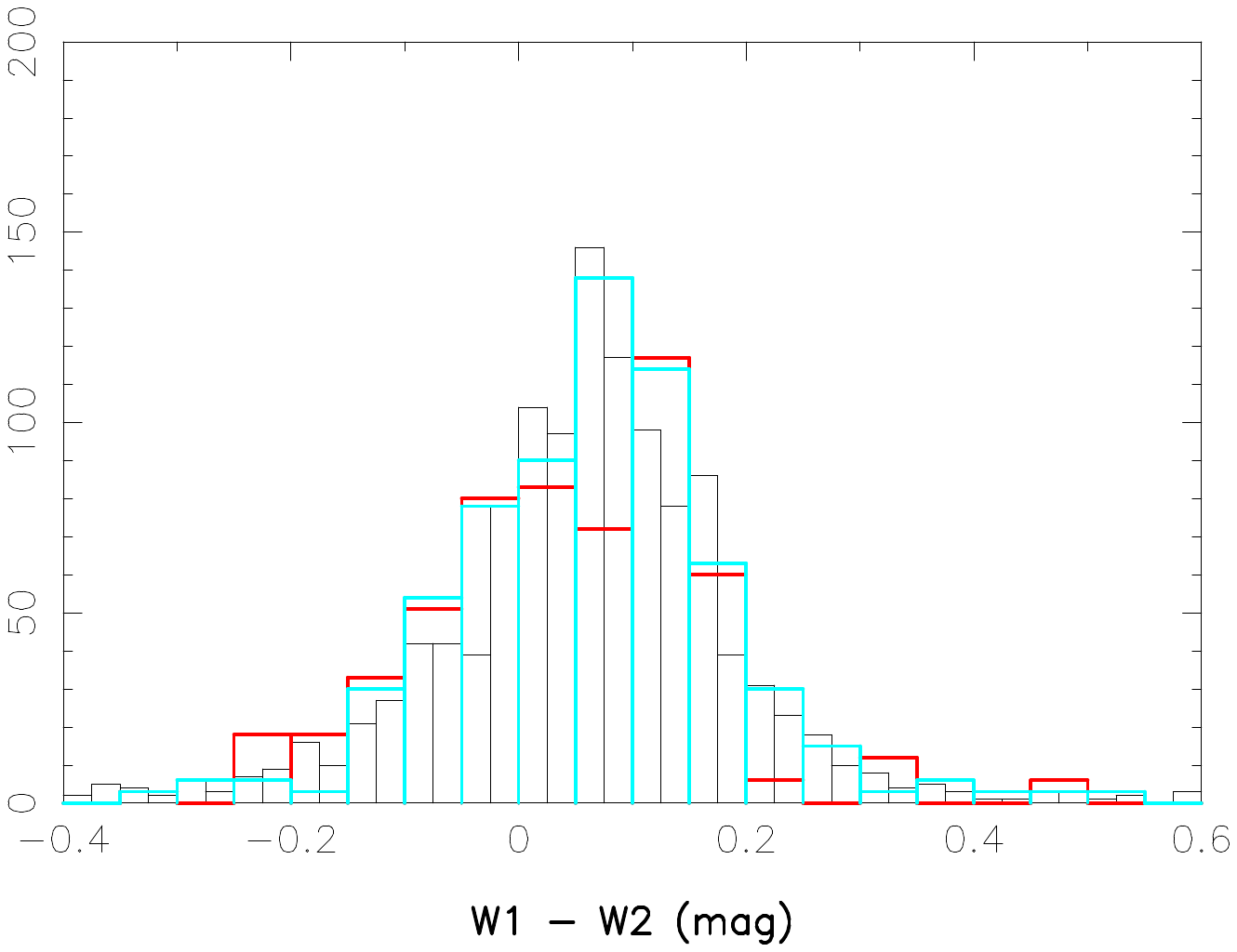} %FIG 4
	\caption{Distribution of W1--W2 colours for the galaxies in Table~1 (blue). The Vela and NGC 5044 fields are shown in red and black respectively. Vela and NGC 4808 have been upscaled by a factor of 3 for ease of comparison}
\end{figure}

\subsection{Baryonic TFR}
For galaxies with a large HI mass  relative to their stellar mass it stands to reason that adding it to the stellar mass will make a good alternative TFR 
(McGaugh \etal 2000). For our sample the HI mass is a product of the template science code\footnote{carnaby.aussrc.org:/mnt/shared/wallaby/notebooks/WALLABY\_
notebooks/user\_science.ipynb} run after the SoFiA pipeline and correction for the WALLABY flux deficit, and the stellar mass in solar masses can be
calculated from $$\log M_* = -0.04 + 1.12 ~\log \rm{L}_{\rm{W1}} \eqno(2)$$ (Wen \etal 2013), where L$_{W1}$ is the 3.4$\mu$ luminosity in solar units. We also considered the stellar mass formula of Cluver \etal (2014).That formula gives a stellar 
mass larger by a factor 1.42$\pm$0.02 than Wen's, and we have not used it because the uncertainties in WISE W2 are 50\% larger than in W1\footnote{The procedure described in the Appendix makes this choice moot in the use of the BTFR for CosmicFlows4.}. The HI mass is multiplied by 1.3 to account approximately for other species (McGaugh \etal 2021), and corrected to the H$_0$ value used for M$_*$.
Figure 5 is the baryonic TFR for galaxies with cz $<$ 10,000 \kms. With four 2.5$\sigma$ deviates removed, the scatter about the linear least squares fit and about the $hyperfit$ is 0.25 dex
or 0.63 $\pm$ 0.14 mag, compared with 0.65 $\pm$ 0.13 mag for the W1 TFR. Although the baryonic TFR is an alternative to the W1 TFR, it seems to be equal, rather than superior, for the purposes of measuring peculiar velocities.

\begin{figure}%[h]
\includegraphics[width=1.2\columnwidth]{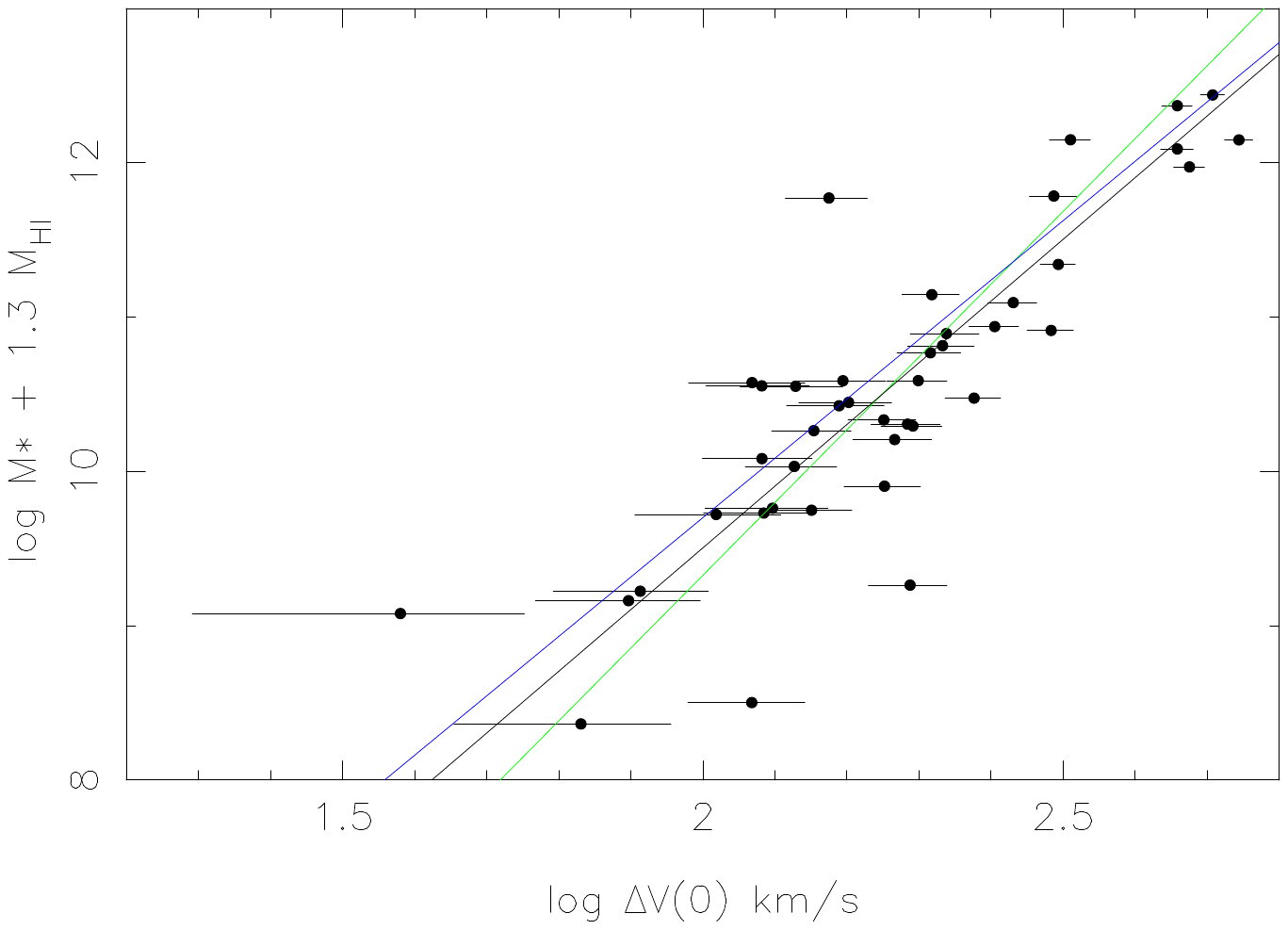}%FIG 5
\caption{Baryonic TFR for galaxies in the NGC 4808 field with cz $<$ 10,000 \kms. The vertical axis is the sum of stellar and gas masses. The black line has slope 4. The green line is the $hyperfit$ with slope 4.45 $\pm$ 0.29. The blue line is from Lelli \etal (2019).}
\end{figure}

\section{TFR for the Vela field}
Table 2 contains the optical, infrared and radio data for the Vela field, centred on 10$^h$ 3$^m$ and --45$^d$ 36$^m$, and has the same format as Table 1. The structure of this field is shown in Figure 6, strengthened by redshifts from the overlapping region of the MeerKAT Vela supercluster survey (Rajohnson %Steyn 
\etal (2024), and the TFR for the field with S/N $>$ 3 is shown in Fig 7. We dropped the S/N threshold from that used in the previous field because of the smaller number of galaxies available.
The vertical scatter   %$\chi^2$
for a linear fit excluding the lowest velocity width galaxy WALLABY J095710-485624, which is an outlier, is 0.86 mag. Although the Vela field is at low Galactic latitude (--8$^\circ$), it proved possible to measure accurate W1 total magnitudes by removing superposed stars. The field is close to the centre of the Laniakea Supercluster (Tully \etal 2014), and will be of importance in
the study of the velocity field in this region. We defer this to a future study when adjacent fields can be included.

\begin{figure*}%[h]
\includegraphics[width=1.15\textwidth]{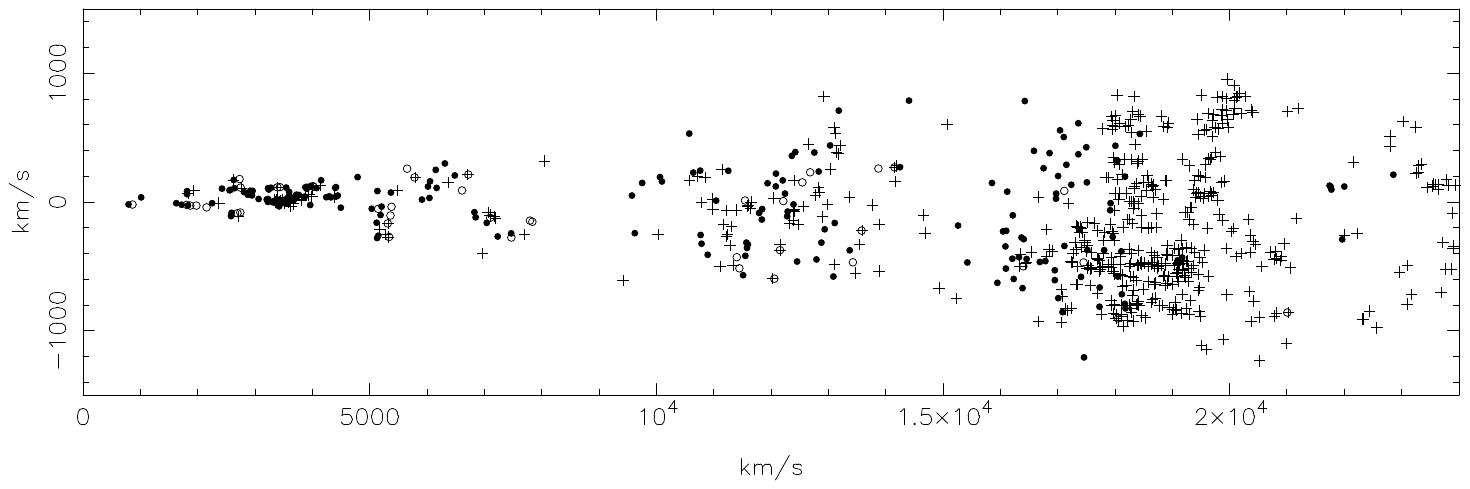}
\vskip -2.6 truein
\caption{WALLABY Vela galaxies in the supergalactic plane. The long axis
	is the heliocentric recession velocity in the SGY direction, the vertical axis, the SGX direction. The dynamically significant Vela Supercluster identified by Kraan-Korteweg \etal (2017) and Courtois \etal (2019) is visible at 18,000 \kms within WALLABY's range. Redshifts from Table 2 are solid symbols, MeerKAT redshifts (Rajohnson \etal 2024) are open symbols and published  and forthcoming optical Vela redshifts are plus signs.}
\end{figure*}

\begin{figure}%[h]
\hspace*{-1 cm}\includegraphics[width=1.2\columnwidth]{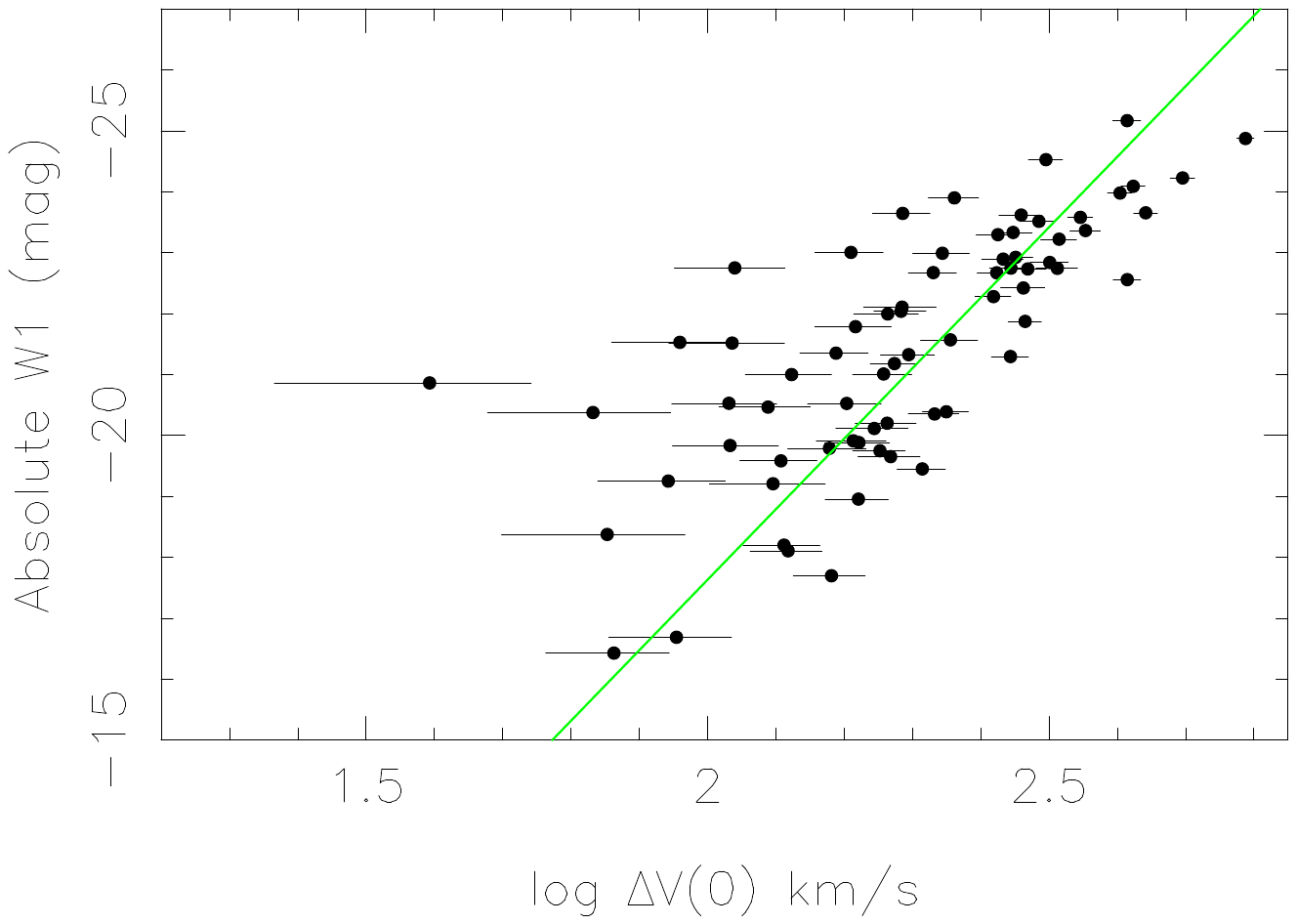}%FIG 7
\caption{TFR for the Vela field. %The dashed line is the least squares fit.
	The green line is the $hyperfit$.}
\end{figure}
\section{The NGC 5044 field}
\begin{figure*}%[h]
\includegraphics[width=1.15\textwidth]{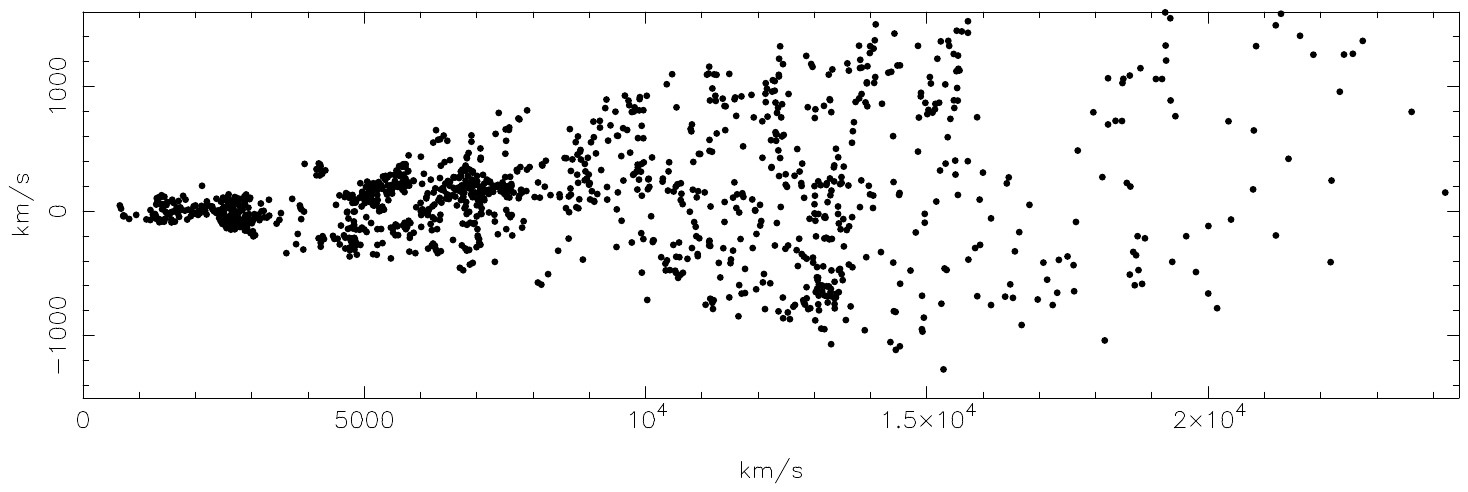}%FIG 8
\vskip -2.6 truein
\caption{WALLABY galaxies from the NGC 5044 field in the supergalactic plane. The long axis
is the heliocentric recession velocity in the SGY direction. The vertical axis is the SGX coordinate.}
\end{figure*}

This field is the largest of the pilot survey at 120 square degrees. Its structure is shown in Figure 8.
The WALLABY and WISE data are in Table~3, which has the same format as Table~1, and the TFR is Figure 9. 
Where the galaxy seemed too round to warrant ellipse fitting, a
 zero appears in the axial ratio column of Table 3. Zeros were also recorded in the fairly rare case
 that the source was just outside the boundaries of the closest DECaLS image.
Comparison with published data from Nancay, Greenbank and Parkes  
(Tully \etal 2009, EDD, Springob \etal 2005, Theureau \etal 2006, and Koribalski \etal 2004) is in Tables 4 and 5 and Figures 10 and 11, which compare WALLABY velocities with published velocities and WALLABY velocity widths with published velocity widths, respectively. Where there are two published HI profiles, both are plotted in the figures. The last two columns of these tables are the WALLABY values
from Table 3.% With the exception of WALLABY  J133314-160715, which is confused with AGC530316, t
The comparison is good. With the 3 most deviant points in Figure 11 removed, there is  an offset of 2 $\pm$ 2 \kms from the dashed line.
The velocity widths in Tables 4 \& 5 and Figures 10 \& 11 have not been
corrected for resolution and turbulence.
\begin{figure}
\includegraphics[width=1.2\columnwidth]{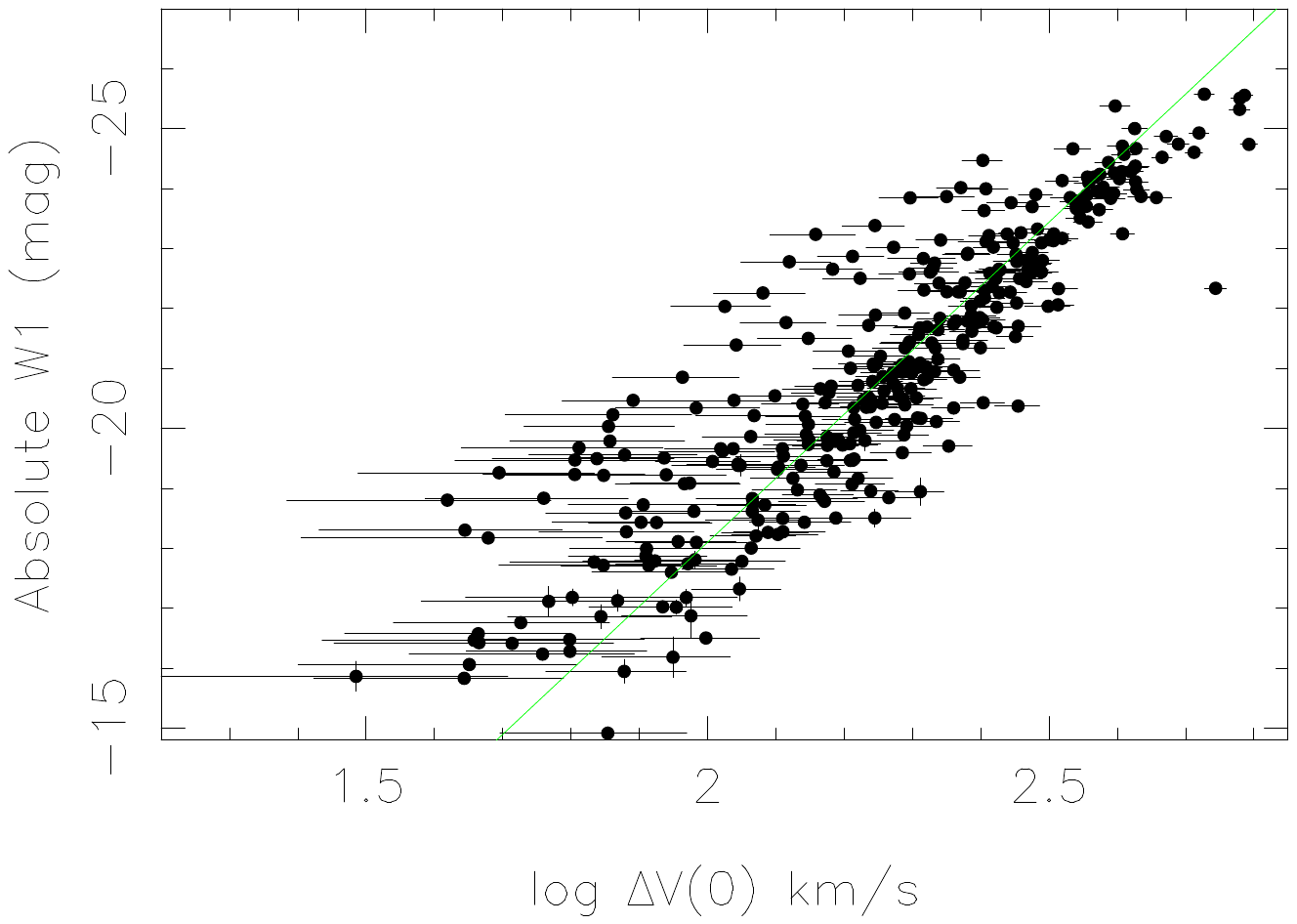}%FIG 9
\caption{TFR for the NGC5044 field. Fourteen 3$\sigma$ deviates from the regression line have been removed. Only two of these are starburst candidate galaxies with W1 -- W2 $>$ 0.25 mag. Magnitude error bars are mostly vanishingly small. The green line is the $hyperfit$. There is a skew distribution around the fit,
	but this does not seem to be related to unusual redshifts or inclinations.}
\end{figure}
\begin{figure}%[h]
\includegraphics[width=1.2\columnwidth]{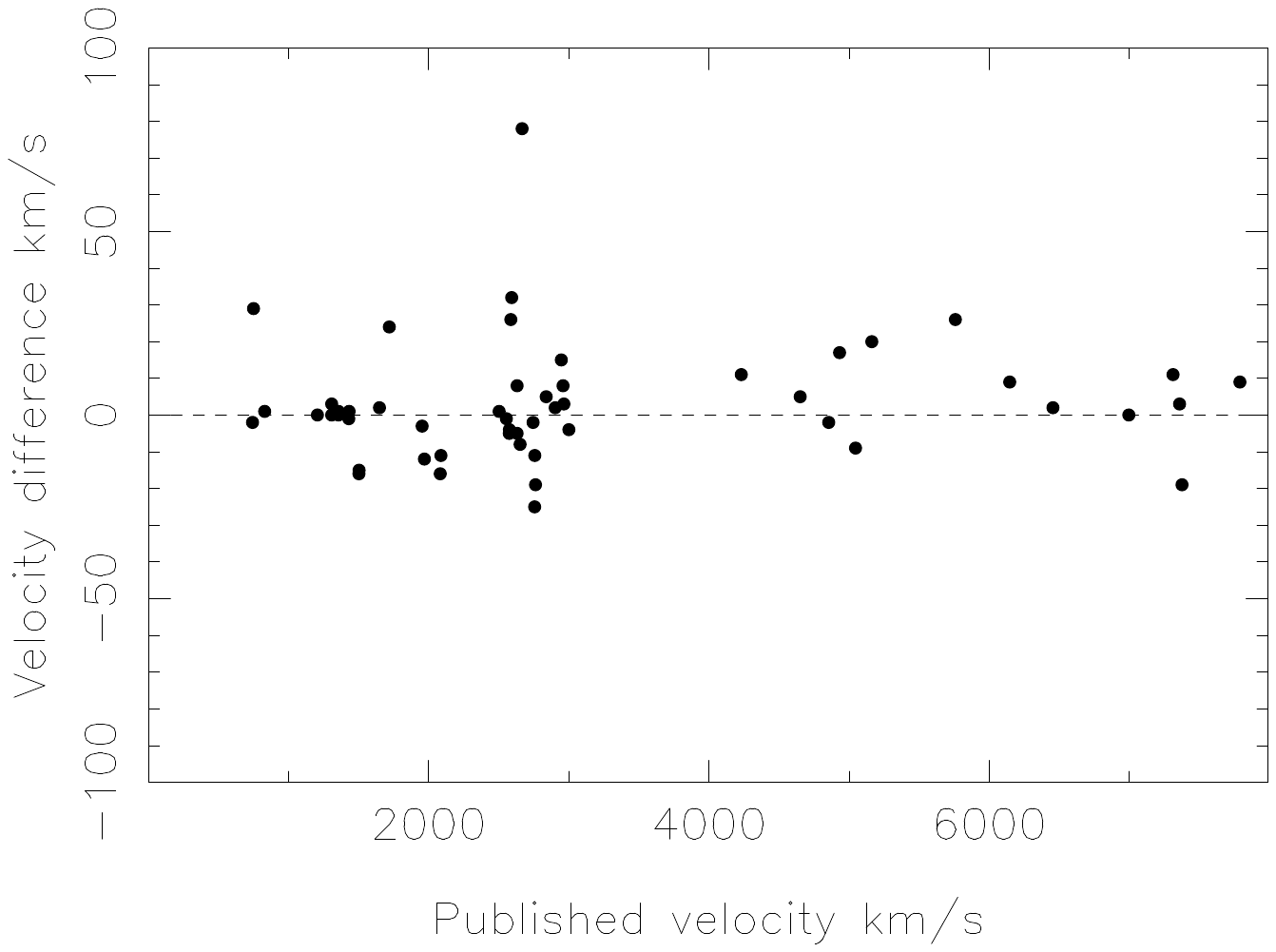}%FIG 10
\caption{Comparison of WALLABY velocities from Tables 4 and 5 in the NGC 5044 field with published heliocentric velocities.}
\end{figure}
\begin{figure}%[h]
\includegraphics[width=1.2\columnwidth]{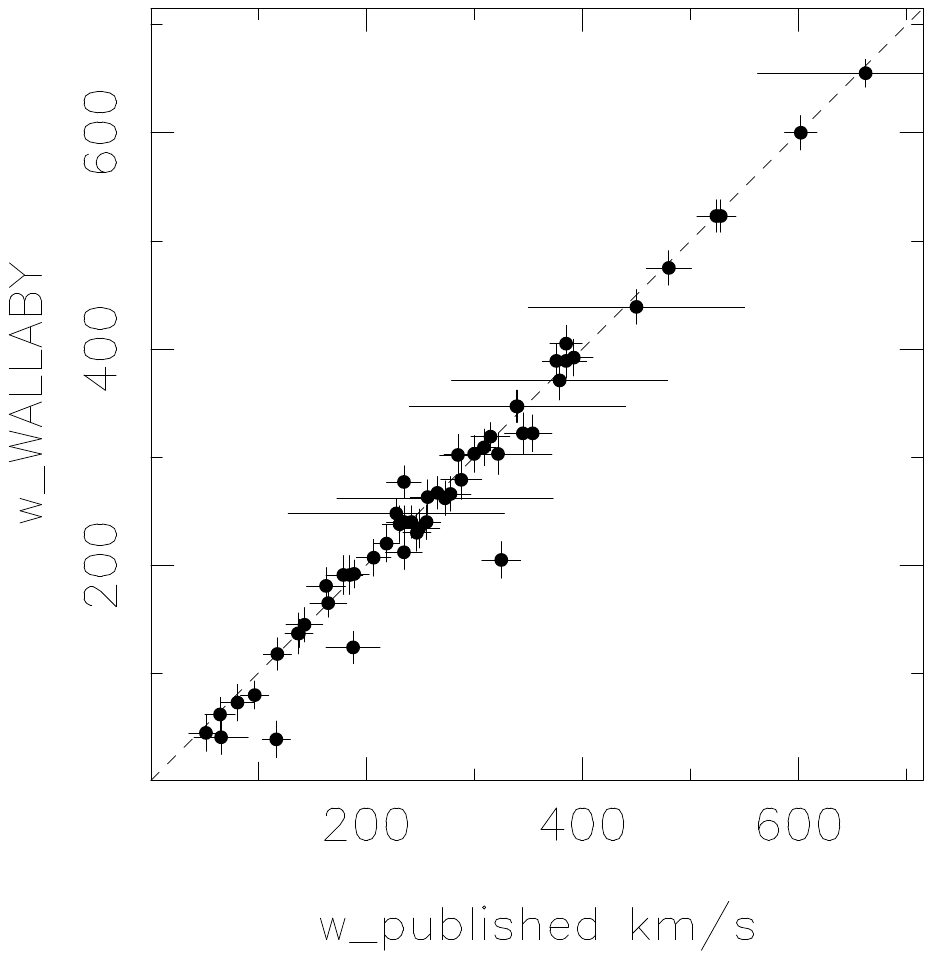}%FIG 11
\caption{Comparison of WALLABY velocity widths from Tables 4 and 5 in the NGC 5044 field with  velocity widths compiled in EDD (Tully \etal 2009). The dashed line is the 1:1 line. {\bf Below:} The difference plot with the 3 most deviant points removed.}
\includegraphics[width=1.2\columnwidth]{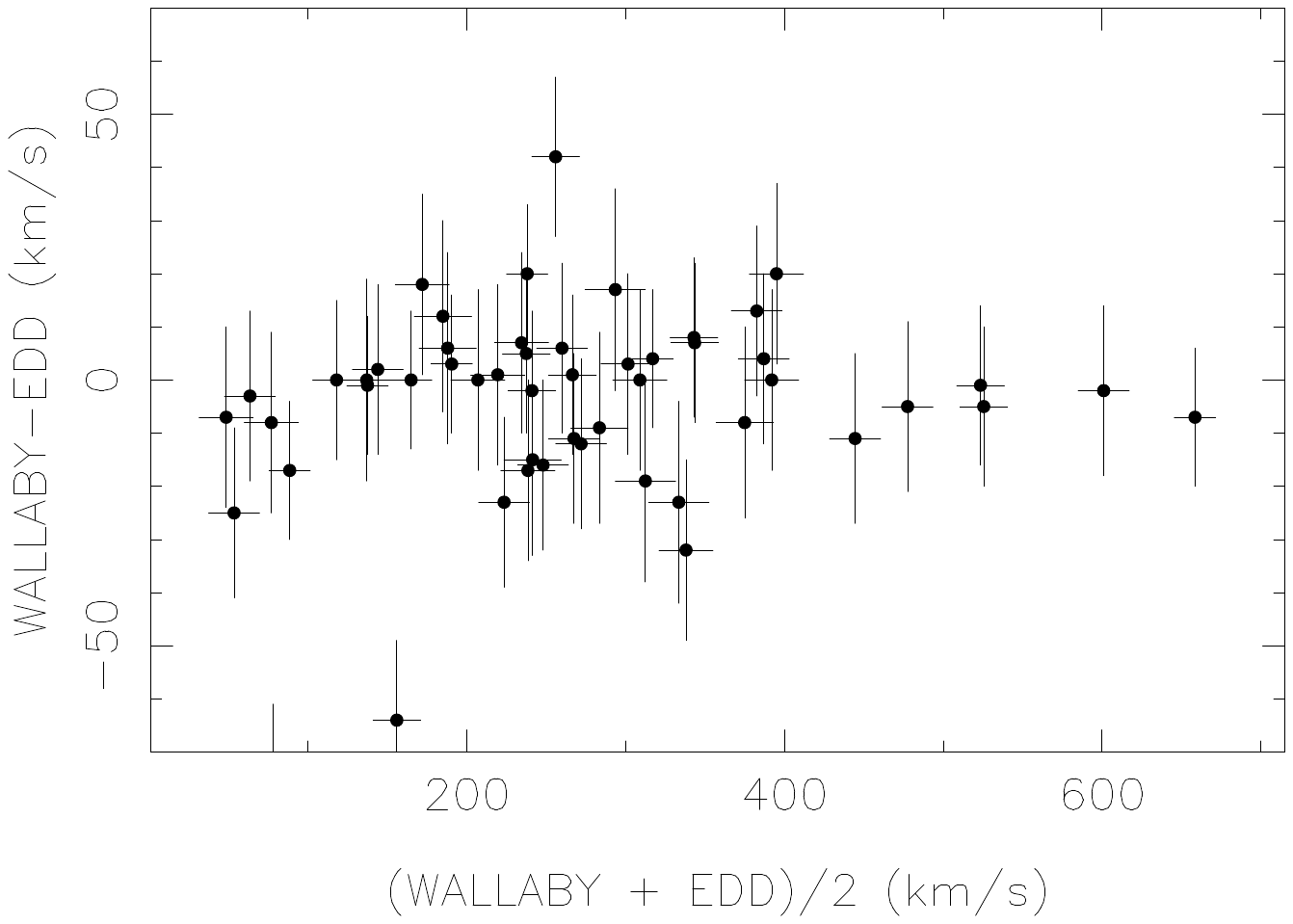}%FIG 11
\end{figure}
The fitted TFRs in our three WALLABY fields are similar to one another within the errors. 

Fu (2024) has developed
a Lucy rectification algorithm for cases where axial ratios are highly uncertain, which produces a TFR based on the statistical distribution of galaxy inclinations. We applied this algorithm to the NGC 5044 field, as it offered by far the largest sample. One starts with an initial un-inclination-corrected (M$_{W1}$, W$_{50}$) plane, and the algorithm then converges 
and reduces the scatter. % in the initial un-inclination-corrected (M$_{W1}$, W$_{50}$) plane. 
However, the scatter in the final statistical TFR obtained was 1.79 mag rms,
compared with 1.08 mag in Figure 9 with the $\sin~i$ term included.  Although it would save a lot of painstaking measurements with uncertainties that are not well characterized,
we elected not to use this algorithm. Axial ratio information is valuable for WALLABY TFR data.

\subsection{HyperLeda diameters}
The Principal Galaxies Catalogue (Paturel \etal 1992, PGC) is a very good match to WALLABY source lists with 1173 out of 1672 sources with PGC or other bright galaxy names in Tables 1--3 (70\%). 
Paturel \etal (2003) have measured diameters and axis ratios on the RC2 system (de Vaucouleurs \etal 1976) at a limiting surface brightness of 25 B magnitudes per arcsec$^2$.
A related 21 cm galaxy scaling relation to the TFR is that between diameter and velocity width. PGC galaxies
have D$_{25}$ and R$_{25}$ in HyperLeda (Makarov \etal 2014), and so the diameters can be corrected to uniform face-on surface brightness by a prescription in the RC2. For a Hubble Constant of 73 \kms Mpc$^{-1}$ the relation
is shown in Figure 12. The scatter in the relation is 0.18 dex, which is equivalent to 0.9 mag in the standard candle version of the TFR, since $\delta$d/d
= (ln 10)$\delta\log$D, where d is distance and D is diameter, whereas 2$\delta$d/d = 0.4(ln 10)$\delta$m, where m is magnitude. Nevertheless, the diameter relation is competitive in this context, and
 opens the possibility of using both luminosity and diameter information along with velocity width. That would be analogous to the Fundamental Plane for early-type galaxies. $Hyperfit$ reports a significant correlation between diameter and axial ratio uncertainty, in the sense that the largest physical diameter galaxies
 have 0.14 larger R$_{25}$ uncertainties than those ten times smaller, but this  would not\footnote{A change of 0.14 in R$_{25}$ of 0.14 only makes  a difference of 2 degrees in inclination on average.}
 seem to act to suppress the scatter in Figure 12. If the luminosity TFR and the diameter TFR are used jointly as they stand, however, errors in velocity width and inclination will produce errors in distance that correlate.
\begin{figure}
\includegraphics[width=1.2\columnwidth]{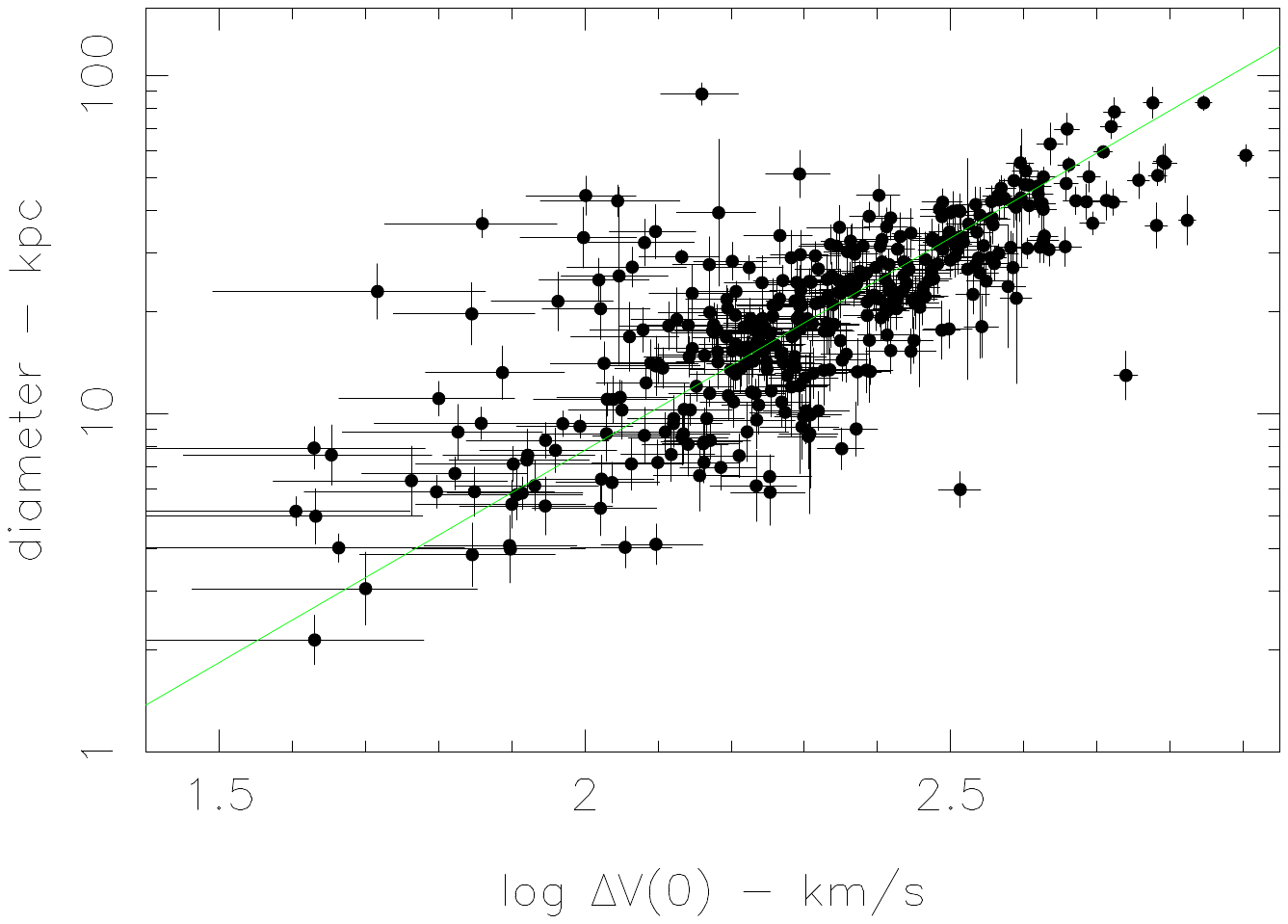}%FIG 12
\caption{The relation between PGC diameter and inclination corrected velocity width for the NGC5044 field.
	A Hubble constant of 73 \kms Mpc$^{-1}$ was adopted.
	The green line is the $Hyperfit$.}
\end{figure}
\subsection{Kinematic inclinations}
Deg \etal (2022) and Murugeshan \etal (2024) have made kinematic maps, including rotation curves, of many WALLABY galaxies, 108 of which are in the present work.  In this work the inclinations of two flat disk models (FAT and 3DBAROLO)  are averaged together and the uncertainty  is set to half the difference between the two fits. In the flat disk algorithm geometric parameters are averaged across all radii.
We have matched those with the galaxies in Tables 1--3 and the comparison of inclinations is shown in Figure 13. The solid correlation gives us confidence in our ellipse fitted axial ratios. There is also a good correlation between
the axial ratios in Table 3 and HyperLeda axial ratios (Figure 14). If we assume the inclinations derived
from Table 3 have an uncertainty of 12.5 degrees, we obtain $\chi^2$ = 1 in a 1:1 fit. There is no significant offset between kinematic and axial ratio inclinations down to the 1 degree level. Further discussion of axial ratios is in the Appendix.
\begin{figure}%[h]
\includegraphics[width=1.3\columnwidth]{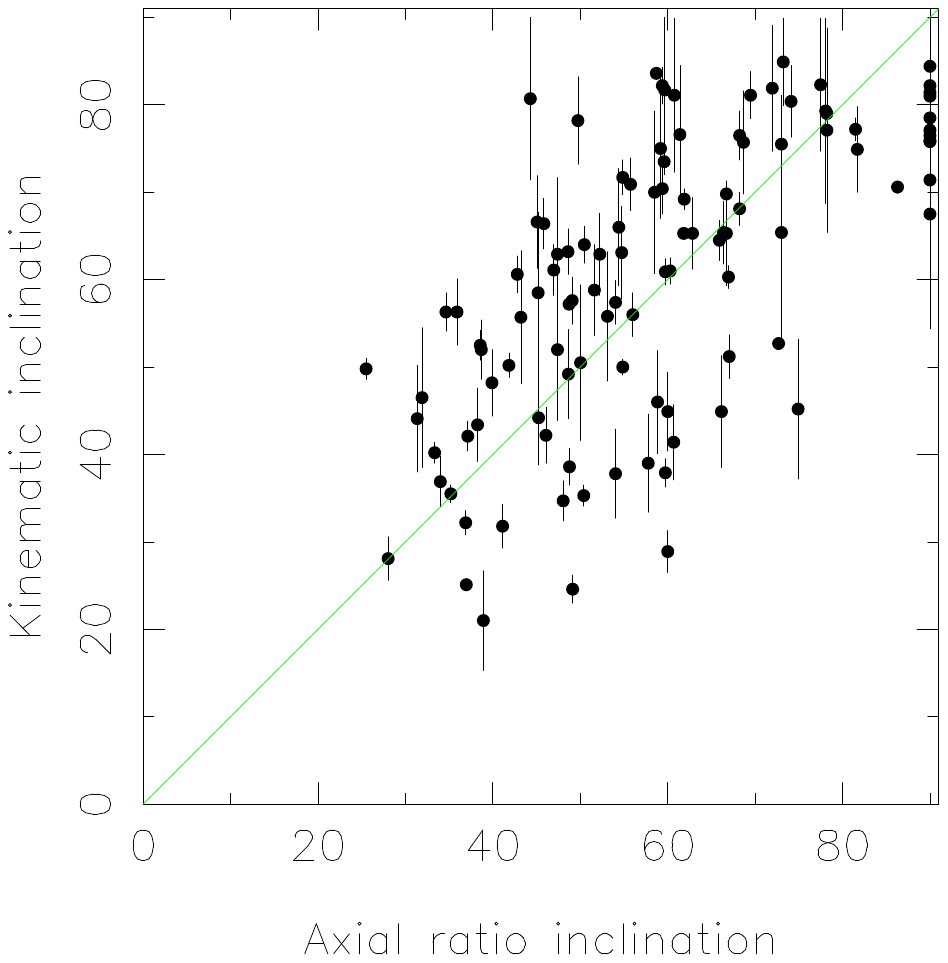}%FIG 13
\caption{Inclinations from WALLABY kinematic modelling (Deg \etal 2022) and those used here from ellipse fitting. The latter have not had uncertainties measured. The green line is equality.}
\end{figure}
\begin{figure}%[h]
\includegraphics[width=1.2\columnwidth]{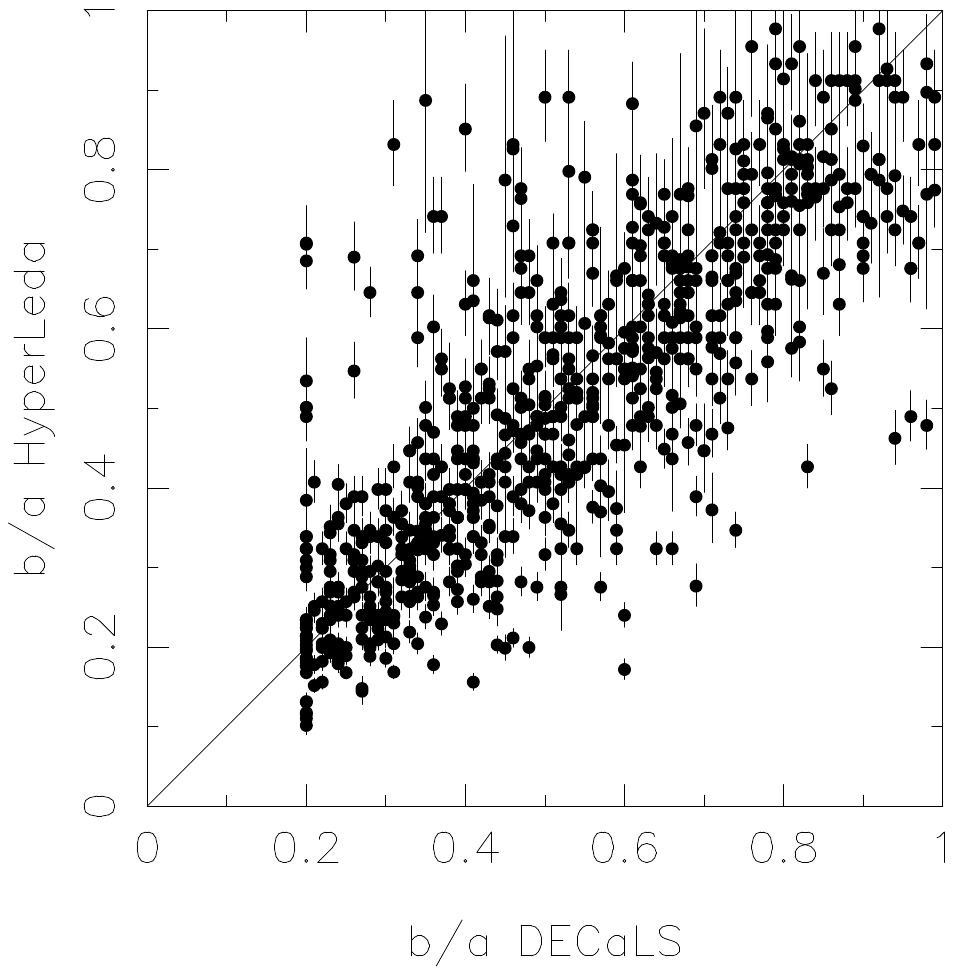}%FIG 14
\caption{Axial ratios from HyperLeda for PGC galaxies and those used here from ellipse fitting.
	The line is the 1:1 line.}
\end{figure}
\subsection{I and J magnitudes}
An additional resource %to HyperLeda
is the DECam Local Volume Exploration Survey (DELVE DR2 catalog,
Drlica-Wagner \etal 2022). We explore this because W1 total magnitudes may
not be available for the volume of data generated by the full WALLABY survey. To reduce multiple matching, PGC positions were used, and cuts were made to SExtractor parameters (Bertin \& Arnouts 1999)
semi-major axis ($>$ 5 arcsec), $g$ magnitude ($<$ 19) and positional error $<$ 15 arcsec.
Where multiple matching occurred, the closest galaxy to the PGC position was taken to be the match.
Inspection of the DECaLS images confirmed that this is an effective procedure in finding the galaxy that is most likely to be the HI source.
A TFR is shown in Figure~15, where the $i$ magnitude is mag\_auto\_i and the axial ratio is b\_image\_g/
a\_image\_g. 
The correlation is strong, as seen in Table~6, where the hyperfit $\chi^2$ is shown.
The $rms$ difference in inclinations between DELVE and Table 3 is 15 degrees,
similar to what was found in the previous section, with DELVE delivering a slightly better $\chi^2$. With 2.5$\sigma$ deviates rejected, the vertical scatter is 1.27 mag.

\begin{figure}
\includegraphics[width=1.2\columnwidth]{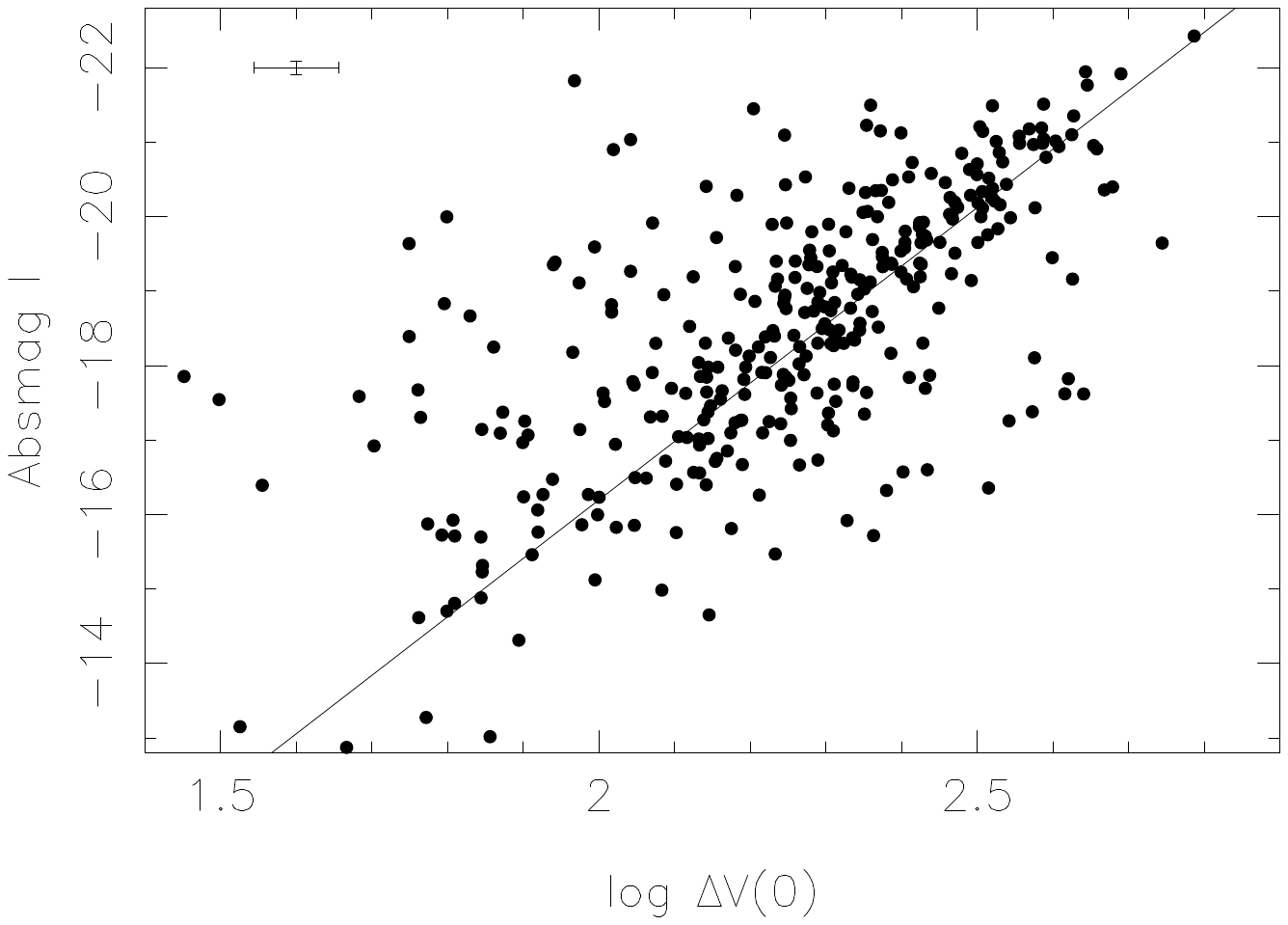} %FIG 15
	\caption{TFR from DELVE DR2 $i$ magnitudes for galaxies with PGC positions in the NGC 5044 field. The error bar is an average velocity width uncertainty and a peculiar velocity uncertainty of 200 \kms. The line is the TFR of Springob \etal (2018).}
\end{figure}

Armed with accurate positions from HyperLeda and DELVE, we can perform a match with a catalogue of galaxy magnitudes
prepared for the 4MOST collaboration (Taylor \etal 2023) and based on the VISTA Hemisphere Survey. We
present the J-band TFR in Figure 16. The vertical scatter is 1.45 mag. Also available from HyperLeda are I magnitudes from the DENIS survey
(Paturel \etal 2005). Figure~17 shows this TFR, and its vertical scatter is 0.81 mag. This catalogue covers the southern hemisphere with cutouts for
the Magellanic Clouds and low Galactic latitudes. The $\chi^2$ per degree of freedom for these four TFRs
is given in Table~6.
The TFR for the DENIS data is well fit by equation (14) of the SFI++ sample of Springob \etal (2018). The magnitude extinction correction (1.05 $\log (a/b)$ ) of Giovanelli \etal (1994) does not change $\chi^2$ significantly. We  provide slopes and zeropoints of the $hyperfits$ in Table~8, but note that these quantities will be determined with larger samples in the full WALLABY survey.
\begin{figure}
\includegraphics[width=1.2\columnwidth]{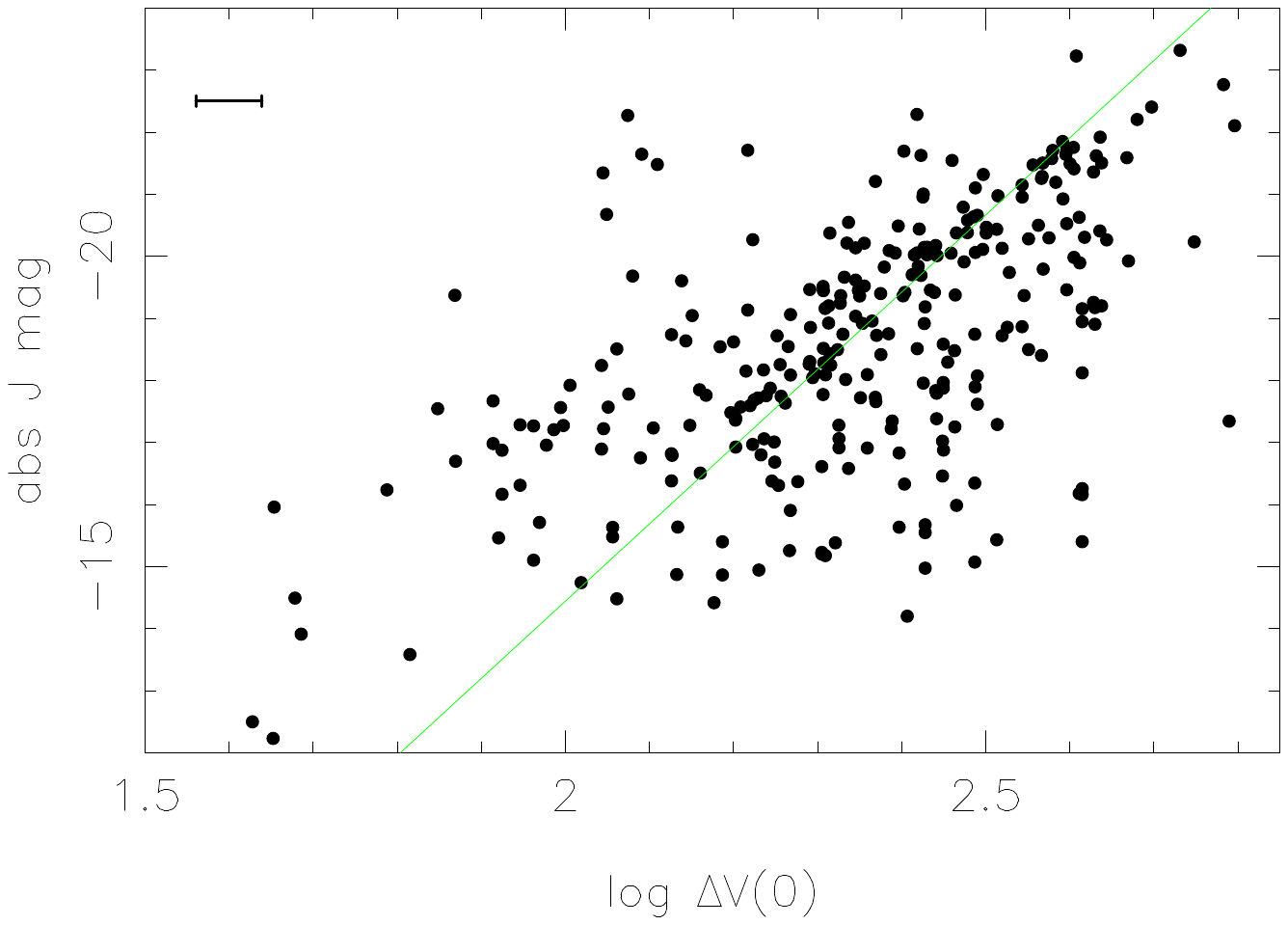} %FIG 16
\caption{TFR in J magnitudes from 4HS (the 4MOST Hemisphere Survey) for the NGC 5044 field. The error bar shows an average velocity width uncertainty.}
\end{figure}
\begin{figure}
\includegraphics[width=1.2\columnwidth]{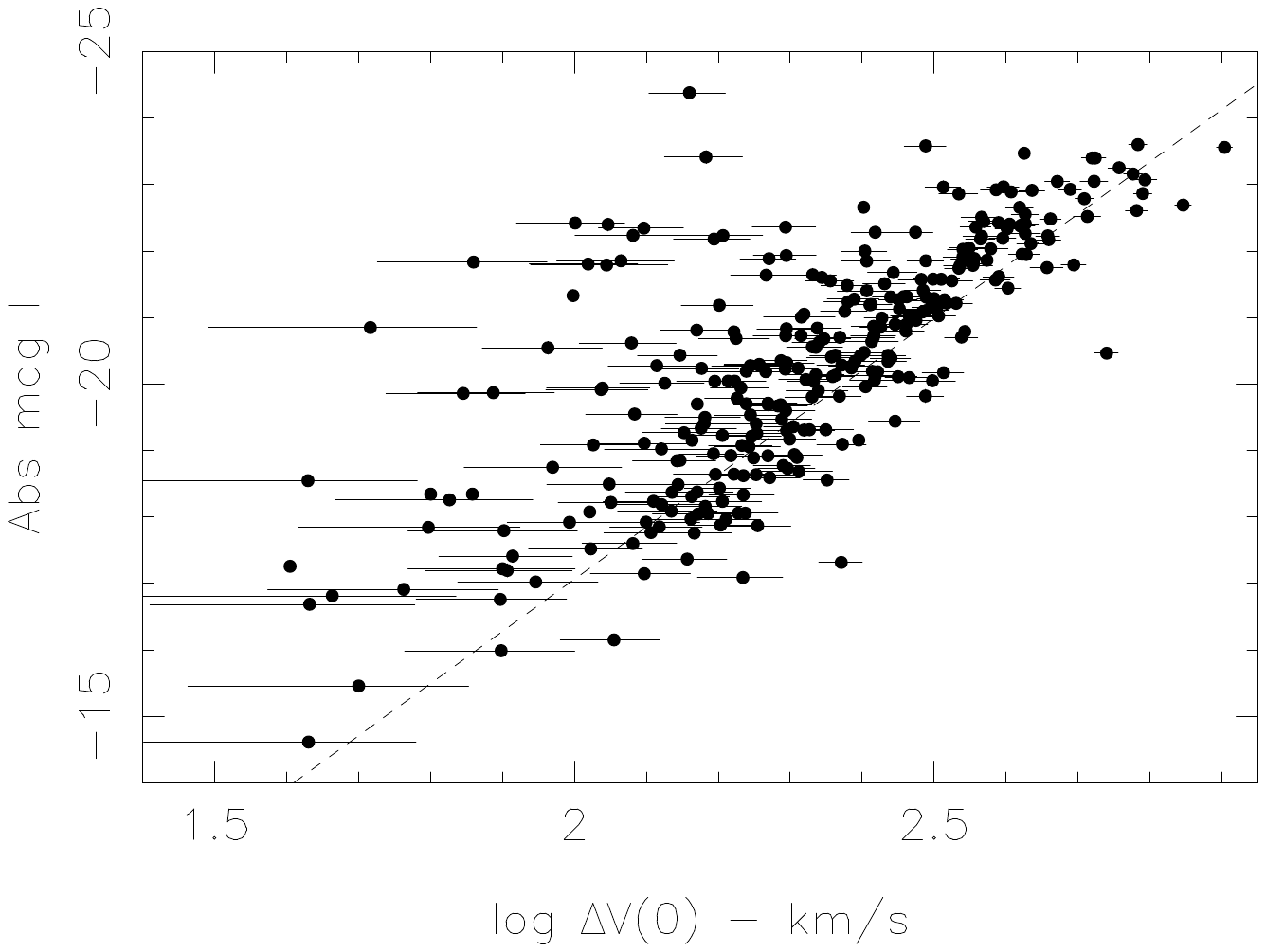} %FIG 17
	\caption{TFR in I magnitudes from the DENIS survey for the NGC 5044 field. The dashed line is from Springob \etal (2018).}
\end{figure}
\begin{figure}
%\vspace{-1 in}
 	\setcounter{figure}{18}
\begin{wrapfigure}{l}[\dimexpr1\width+\columnsep]{\columnwidth}
\includegraphics[width=\columnwidth]{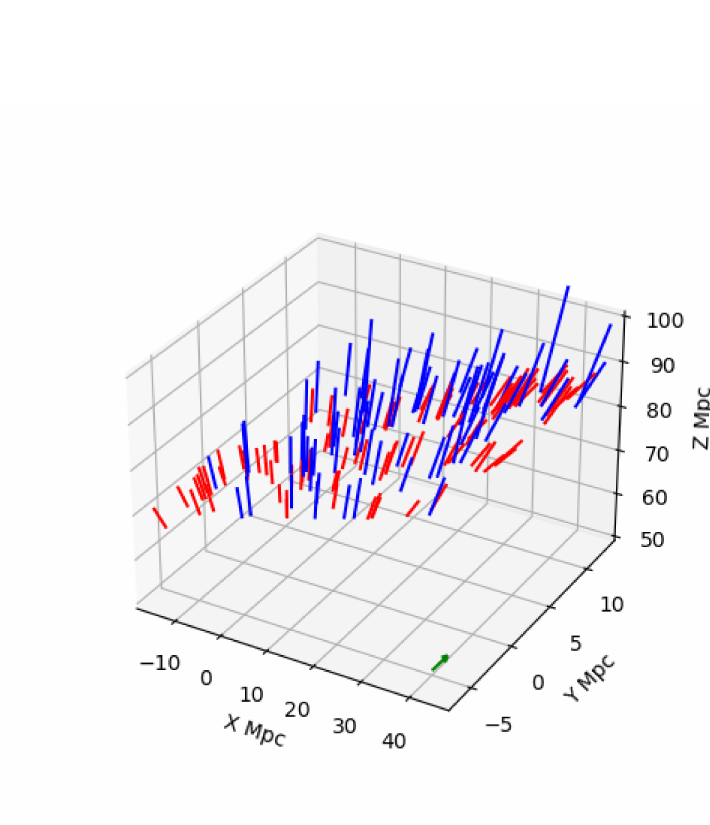}%}%FIG 20
	\vspace{-1cm}
\caption{A portion of the NGC 5044 field in real space, where the Z
	coordinate is in the radial direction, and the tangential coordinates point N (Y) and E (X). The difference between the TFR distance times H$_0$ and the redshift is the peculiar velocity, shown by the length of the arrows. Red is redshift and blue is approaching.
These have been scaled down by a factor of ten for clarity. A bulk flow of 500 \kms depicted by the barely visible green arrow in the foreground, added to these data can be retrieved, as described in $\S$5.2.}
\end{wrapfigure}%$figure}
 \end{figure}
\begin{table}%[ht]
\setcounter{table}{5}
\caption{\bf  TFR Quality of linear fit, NGC 5044 field}
\begin {tabular}{|c|c|c|c|c|c|}
\hline
TFR&Source&Figure&S/N cutoff&$\chi^2$&n\\
$i$ mag&DELVE DR2& 15&none&1.48&325\\
J mag&4MOST&16&3.7&5.34&189\\
I mag&DENIS& 17&3.0&3.30&303\\
W1 mag&WISE& 9&3.7&3.20&344\\
\hline
\multicolumn{6}{l}{Linear fits were used, not $hyperfits$.}\\

\end{tabular}
\end{table}
The $\chi^2$ was calculated with an intrinsic TFR scatter of 0.40 mag (Masters \etal 2014) and an assumed inclination error of 12.5 degrees (see below).
The photometric uncertainties in DELVE and 4MOST are included but are negligible.

\subsection{WALLABY detections with no optical counterpart}
In Tables 1-3 WALLABY pilot survey galaxies not detectable in  WISE or DECaLS and free from close companions are  a fraction of a percent of the total. %shown in Figure 18. 
In a sample of 1700 galaxies statistical juxtaposition of noise in the datacubes will occur, and, although all SoFiA detections have been examined, confirmation is needed. Deeper optical imaging and higher resolution HI maps are desirable for these objects.
%In general, the amount of rotation seen in the first moment
%maps of Figure 18 is small. 
These sources extend over multiple WALLABY beams, so the velocity range is not smeared out. They are being studied intensively in another WALLABY program (O'Beirne \etal 2024), and thus we shall not go into detail here. %\footnote{
 Their relationship to those Ultra Diffuse Galaxies (UDGs; Mancera Pina \etal 2019, 2020), which are not HI rich (For \etal 2023) remains to be explored. 
%They are discussed in the Appendix.%listed in Table~7.
%Figure 18 gives the zeroth and first moment maps of HI for these optical non-detections.% in WALLABY.
%\begin{figure*}%[h]
%\includegraphics[width=\textwidth,natwidth=2200,natheight=1407]{BlankGalaxiesPlot.png}
%\caption{Zeroth and first moment maps of optical non-detections in the NGC 5044 field. %The number in \kms in the lower right hand corner of first moment maps
%is the range of the colour code from cyan to maroon.%purple to red.  
%Three  seem to be rotators and one is not.} %The
%	\vspace{-6in}
%\includegraphics[width=\textwidth]{fig18.pdf}%FIG 18
%\end{figure*}
%\pagebreak
\section{Discussion}
\subsection{Combined fields}
We combine the TFRs for the three fields in Figure 18.
The intercepts of the three TFRs are not different within the uncertainties.
According to the IRSA dust map viewer the Schlafly \& Finkbeiner (2011)
extinction in Vela varies from E(B-V) = 0.17 mag in the NE region of our field
to 0.475 mag in the SW. In the W1 bandpass this range is (0.037, 0.103) mag in A$_{W1}$. This is not responsible for the scatter in Figure 7 or Figure 19.
The other two fields are high latitude fields with E(B-V) = 0.03 and 0.06 mag.
\subsection{A simulation}
The full WALLABY survey will address the growth of structure in the Universe, measuring $f\sigma_8$ with high
statistical accuracy (Said \etal 2020) and discovering large scale flows in the accessible volume. As a simple test we inserted
a 500 \kms bulk flow in a 10$^4$ Mpc$^3$ volume of the NGC 5044 field at 5500 \kms redshift, as depicted in Figure 20. Scatter of 1.04 mag
rms from the (W1, $\Delta V(0)$) TFR was added to the Hubble flow and the input flow velocity was recovered with 2.5$\sigma$
significance from 156 galaxies. This bodes well for the full survey and suggests that expectations based on the simulations of Boubel \etal (2024)
with 7341 ALFALFA galaxies will be realised. The average TFR scatter assumed in their error model for these 156 galaxies
was 0.93 mag.

The full WALLABY survey will cover regions of high and low extinction, and therefore the W1 magnitudes have a significant advantage for measuring peculiar velocities on large scales. If these measurements can be automated for $\sim$10$^5$ galaxies, this approach would be preferred. If not, Table 6 suggests that DELVE DR2 would provide a good alternative, better still if that can be combined with  an extinction map to correct the I magnitudes and a higher throughput filter 
on the DELVE survey %an optical survey just a little deeper than the PGC catalogue, 
than the PGC catalogue to ensure unique matching. % to close to 100\% of galaxies with accurate positions.
\begin{figure}%[b]
\setcounter{figure}{18}
\includegraphics[width=1.2\columnwidth]{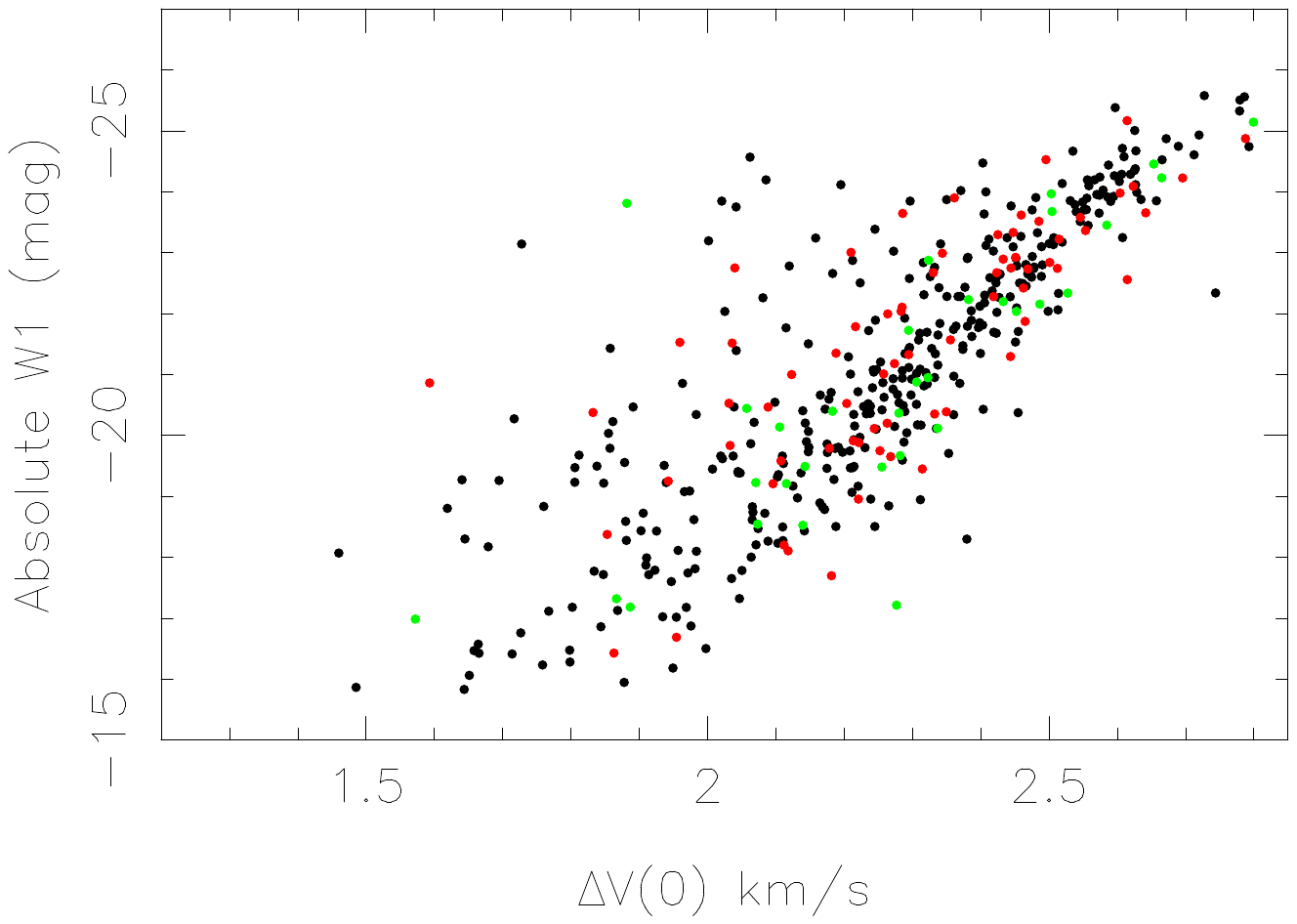}%}%FIG 19
\caption{The three WALLABY phase 2 pilot survey TFR relations combined. NGC5044 is in black, Vela is in red and NGC4808 in green.}
\end{figure}
%\vfill\break
%\clearpage

\subsection%{Addition of WALLABY pilot survey TFR data to CosmicFlows4}
{New positions of the Great Attractor and Vela from the WALLABY addition to CosmicFlows4}
Figure~20 shows comparatively a map of the Norma region as seen from the computation of the linear velocity field and corresponding density field, thanks to peculiar velocities (Courtois \etal 2023).
The left panel shows the reconstruction with CosmicFlows4 (CF4) galaxies only (black dots) while the right panel shows the reconstruction with the WALLABY pilot data added (coloured dots) to CF4.
	Before the addition of WALLABY data the Great Attractor was located a bit closer to us at SGX-SGY=(-30,0) Mpc/h. The new location is at SGX-SGY=(-40,+10) Mpc/h (see Tables 7a\&b). The signal to noise of the Linear density field reconstruction at the Great Attractor location is a strong 15$\sigma$. 
We also note that the velocity flow around Coma cluster is modified. Now Coma appears as a clearly detached cluster.

This figure corresponds to the Great Attractor position in the Supergalactic plane SGX-SGY. It  is a slice of $\pm$5 Mpc/h thickness, centred at supergalactic SGZ= 0 Mpc/h.
The blue and red background colors show the linear density field recovered from the CosmicFlows4 compendium of peculiar velocities (Courtois \etal 2023). The black dots are galaxies from the CosmicFlows4 catalogue.  
The coloured dots are the WALLABY pilot data.
%	Before the addition of WALLABY data the Great Attractor was located a bit closer to us at SGX-SGY=(-30,0) Mpc/h. The new location is  at SGX-SGY=(-40,+10) Mpc/h (see Table 7).
Also the velocity flow around Coma is modified. Now Coma appears as a clearly detached cluster. 
	This figure corresponds to the Vela supercluster position in the Supergalactic plane SGY-SGZ. It is a slice of $\pm$5 $i.e.$ 10 %+/- 5
 Mpc/h thickness, centered at supergalactic SGX= -130 Mpc/h.
The blue and red background colours show the linear density field recovered from the CosmicFlows4 compendium of peculiar velocities (Courtois \etal 2023). The green square is the position of Vela as found by Kraan-Korteweg \etal 2017 from redshift surveys. 
The red triangle was the position of Vela as found by Courtois \etal 2019 using the V-web of CF3. The black dots are galaxies from the CosmicFlows4 catalogue.  The coloured dots are the WALLABY pilot data.

The region scrutinized in Figure 21 is a slice at supergalactic SGX=- 130 Mpc/h.
A clear over-density of galaxies is seen on both sides of the zone obscured by the Milky Way disc (vertical on this figure) at SGZ=-130 Mpc/h. 
This is what was named the "Vela wall" of the "Vela supercluster": a filament running across the ZOA and extending in total to more than 6,000 \kms from the redshift surveys' point of view.
The black arrows on this figure correspond to the peculiar (gravitational) velocity field.
	A striking result is that, quite independently, the new WALLABY data confirms the Vela position published by Kraan-Korteweg \etal (2017). The Vela supercluster extends from side to side across the ZOA. 
Since it is a three-dimensional velocity field, on the figure these vectors are projected onto the plane SGY-SGZ.
The red triangle on this Figure 21, was published in 2019 by Courtois \etal as a point of convergence of the V-web located at SGX-SGY-SGZ (-13,000; 7,000; -9,000) \kms. This location is the central region of a gravitational basin of attraction located on the sky in the Vela constellation direction. 
This Vela large-scale structure is not part of the larger Shapley attractor as one can see that there is a clear delimitation of the streamlines splitting between these two basins.
	Consequently, the cross-analysis of redshift surveys data (Kraan-Korteweg 2017) and the new addition of WALLABY data in the southern hemisphere to the peculiar velocity surveys data confirms the position of the Vela supercluster at  SGY-SGZ=(40; -140) Mpc/h.
The signal to noise of the linear density field reconstruction at the Vela location has strong 4$\sigma$ significance.
We also note that before the WALLABY data the cluster NGC5044 was not well reconstructed by Cosmic-Flows-4 catalogue. Even worse there was an under-density (a cosmic void) reconstructed at its location. The addition of the WALLABY pilot data, causes NGC5044 now to appear on the map at SGY-SGZ=(100;0) Mpc/h.

\begin{table}
	\caption{(a) Large scale structure parameters Positions of the Great Attractor (GA) and Vela supercluster %}\\
	in CF4 alone in black and with the %}\\{ 
	addition of WALLABY in bold. %}\\ 
	The radius of the corresponding %}\\{ 
	spherical overdensity is given %xxxx}\\
	in the last column.}%}\\
 \begin{tabular}{|l|c|c|c|c|c|}
 \hline
 Stru-&SGX&SGY&SGZ&overdensity&distance\\
 cture&&&&at the peak&to $\delta$=0 \\
	 & Mpc& Mpc& Mpc&&Mpc\\
	 & /h& /h& /h&&/h\\
 \hline
GA&-49/{\bf -40}&4/{\bf 5}&8/{\bf 0}&0.7/{\bf 0.9}&30/{\bf 40}\\ 
 Vela&-130/&-15/&-150/&-0.4/&60/60\\
 &~~~/{\bf -130}&~~~/{\bf 40}&~~~/{\bf -140}&~~~/{\bf 0.5}&\\
%GA&-40/{\bf   5}&0/{\bf  0}&8/{\bf 0}&0.65/{\bf 0.87}&30/{\bf 40}\\ 
%Vela&-130/{\bf 40}&-140/{\bf -50}&-150&-0.2/{\bf 0.53}&60/60\\
%\hline
%\multicolumn{6}{l}{
\end{tabular}
\setcounter{table}{6}
\begin{tabular}{lcccc}
 \hline
 Structure             &  vx                       &   vy
&      vz                    & Vbulk          \\
                  &  \kms                   &\kms                  &\kms                      & \kms            \\
 Great Attractor   &   -153/\bf{-102}    &   42/\bf{78}       &      52/\bf{37}
       & 206/\bf{188} \\
Vela.                  &    220/\bf{-349}    &   82/\bf{-39}       &     252/\bf
{-83}     & 353/\bf{394} \\
\hline
\end{tabular}
\caption{(b) Bulk flow velocities computed  within sphere of radius 20 Mpc/h centere
d around GA [-40,5,0] Mpc/h and Vela [-130,40,-140] Mpc/h. The addition of pilot
 phase 2 WALLABY data modifies the amplitude and direction of the velocity field
. The CF4 alone values (in black) and strongly modified when adding WALLABY data
 (in bold) in the region of Vela supercluster where the bulk flow changes direct
ion towards negative supergalactic X and Z axis and is enlarged by an amount of
50km/s reaching 394 $\kms$ This shows that Vela is a main gravitational actor in
 the southern sky large-scale structures.}
\end{table}

%\begin{table}
%\end{table}
CosmicFlows4 is the most advanced representation of the flow field of galaxies within 20,000 \kms, using redshift independent distances. A multiwavelength baryonic TFR has been adopted for this purpose, and in the Appendix we describe the calibration $pro-tem$ of our single wavelength (W1) baryonic TFR, so that pilot survey distances can be introduced into the CF4 database, and a new reconstruction run. Reconstruction is the process of calculating the gravitational field of galaxies with known distances, and thus, in the linear approximation,  the peculiar velocity field.
	At this stage we are not able to include WALLABY selection effects in the reconstruction, as we are still learning what these are. The S/N ratio cutoff of the WALLABY survey is now known (Westmeier \etal 2022; Murugeshan \etal 2024), which will allow selection effects to be simulated.
\pagebreak
\begin{figure*}
	\begin{wrapfigure}{l}[\dimexpr1\width+\columnsep]{\textwidth}%columnwidth}
	\setcounter{figure}{19}
	\includegraphics[width=\textwidth]{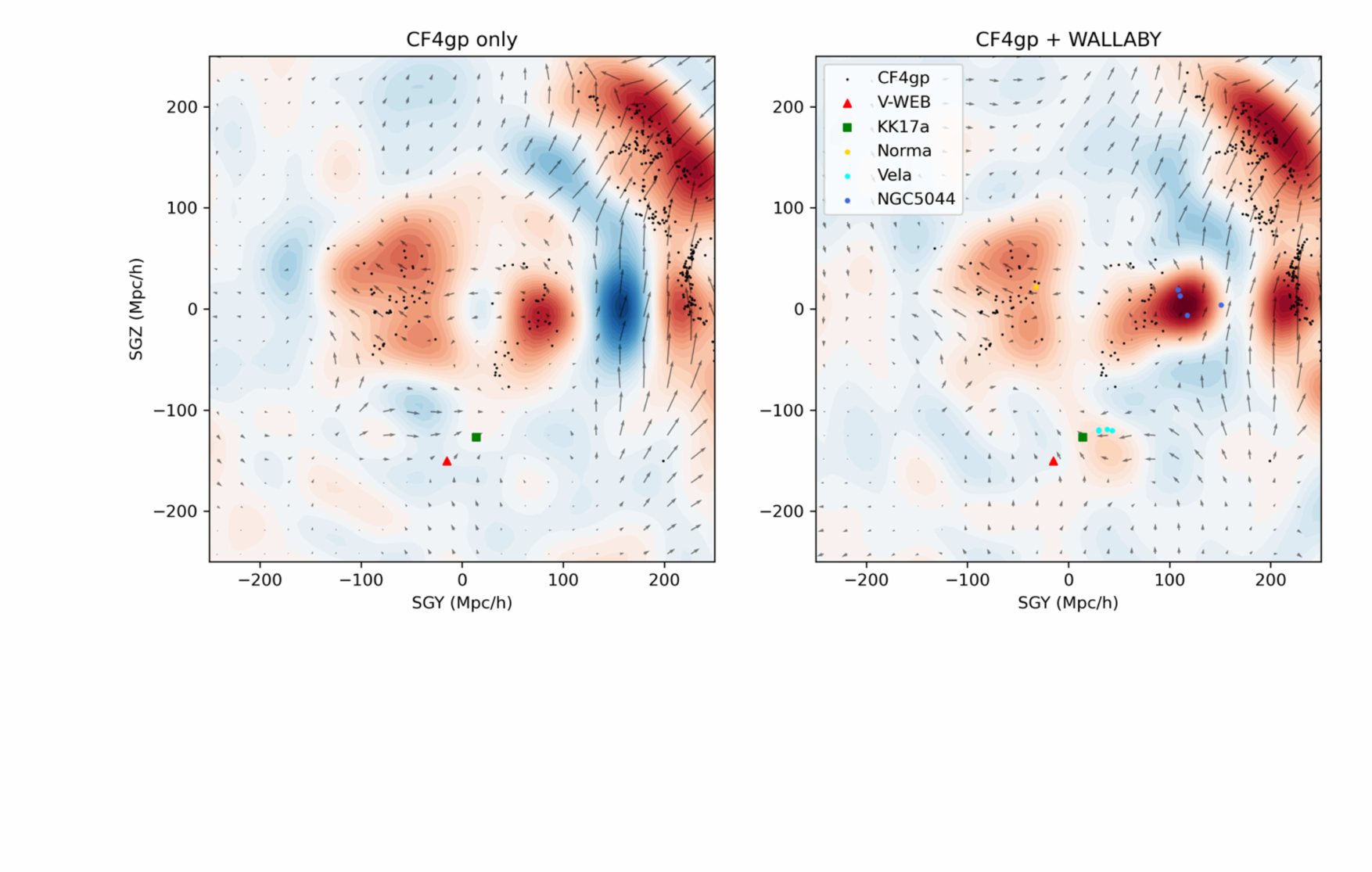}%CF4gp_WALLABY_norma_240607.pdf}%FIG 21
%	\vspace{-1.3in}
\caption{In the left panel is the flow field solution computed with only CF4 data. In the right panel is the solution using distances from  CF4 + WALLABY.
WALLABY data points are in colour. The background red shading corresponds to
	mass overdensity and blue to relative underdensity. The arrows
	show the direction of the consequent flow field.}
%\end{figure*}
%\begin{figure*}
\caption{A SGY-SGZ slice at SGX=-130 Mpc/h.  Left and right panels are as in the previous figure. The Vela supercluster is near the ZOA (which is at SGY =0 vertical). %It is worth to wait for the full convergence of the velocity field solution, because we might see something new at the VELA position.
	The position of Vela from Kraan-Korteweg \etal (2017) is a green square.
	And the position of Vela as a knot of the V-web (Courtois \etal 2019) is a red triangle.}
%\vspace{-1in}
\includegraphics[width=1.1\textwidth]{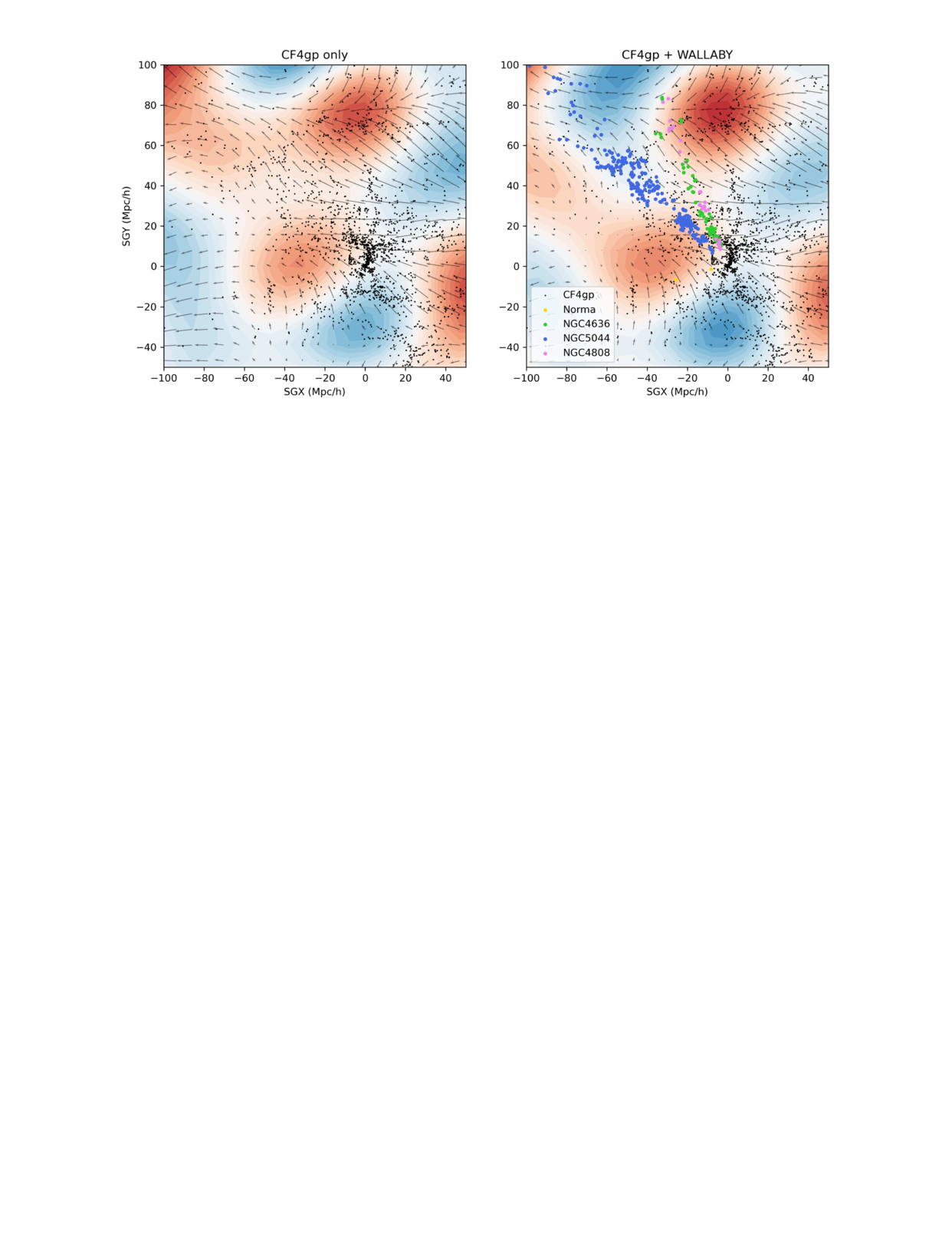}%thumbnail_CF4gp_WALLABY_vela_Z-130_240607.pdf}%FIG 22
\end{wrapfigure}
\end{figure*}
The first plot (Figure 20) is a SGX-SGY slice at SGZ=0. The red blob across the Zone Of Avoidance (ZOA)  (which is horizontal SGY=0)  and SGX=-40 Mpc/h is the Great Attractor. Figure 21 is perpendicular to it.
The effect of adding the WALLABY pilot survey galaxies is to greatly increase the spatial resolution of the flow field.
\begin{table}
\caption{\bf Summary of data and cuts}
\begin{tabular}{|l|r|r|r|r|r|r|r|}
	\hline
Target&Fie- & Area& PDR2& Table& SNR& incl& used\\
group &lds& deg$^2$&& tally&cut&cut\textdagger&TFR\\
N4808 &1& 30 &231& 201 T1&3.7&96&29\\
Vela &1 &30 &203&143 T2&3&34&71\\
N5044& 4& 120& 1326&1238 T3&3.7*&26&348\\
	Total&6& 180& 1760&1582&&398&448\\
\hline
\multicolumn{6}{l}{*also a 3$\sigma$ TFR deviates cut}\\
\multicolumn{6}{l}{\textdagger Numbers cut for $i~>~45^\circ$}\\
\end{tabular}
%\end{table}
%\begin{table}
\caption{$Hyperfits$*}
\begin{tabular}{|l|r|r|r|r|r|r|}
\hline
Field&Slope&$\pm$&$\beta$&$\pm$&$\sigma$&$\pm$\\
&&&mag&mag&mag&mag\\
NGC 4808 Fig 3&-10.20&0.76&2.54&1.82&0.65&0.13\\
Vela Figure 7&-11.57&0.84&5.51&1.99&0.90&0.12\\
NGC 5044 Fig 9&-10.68&0.30&3.26&0.69&0.94&0.06\\
\hline
\multicolumn{7}{l}{*$M(W1) = \alpha \log \Delta V(0)~ +~ \beta$. Slope = $\alpha$; $\sigma$ = scatter.}\\
\multicolumn{7}{l}{Columns headed $\pm$ are uncertainties in columns to their left.}\\
\end{tabular}
\end{table}
\pagebreak
\section{Conclusions}
The WALLABY Pilot Survey has taught us a number of things that can be built on in the full survey, which is now under way.\\
(1) The TFR is clearly observed in the three fields of the WALLABY Pilot Survey Phase 2 and can be used for redshift independent distances. A summary is given in Table 9. % and a plot of the TFR for the 3 fields combined is in the Appendix.
There are ten times the number of galaxies in these fields as there are for these parts of the sky in CosmicFlows4.\\
(2) Axial ratios of galaxies from isophotal ellipse fitting in the optical correlate with disk inclinations from WALLABY kinematic modelling and with HyperLeda axial ratios. The scatter in this relation allows our axial ratio uncertainties to be estimated, and this comes close to explaining the TFR scatter we see. Measuring and controlling errors in inclination are crucial to understanding the scatter in the TFR and estimating distances.\\
(3) A diameter/velocity-width scaling relation with scatter comparable to the TFR can be readily constructed from HyperLeda diameters. Further work is warranted on investigating the origin of this relation and potentially combining it with the TFR to obtain better distance estimates.\\
(4) For WALLABY sources matched to PGC galaxies, 4MOST magnitudes based on the VISTA Hemisphere Survey can be used to make a J magnitude TFR, and DELVE (DENIS)  magnitudes can be used to make an $i$ (I)  magnitude TFR. These would require calibrations before they could be used to investigate cosmic flows. By
calibrations we mean the slope of the TFR and any nonlinearity. The zeropoint
affects Hubble constant determinations, but not peculiar velocities.\\
(5) A number of optical non-detections have HI properties similar in hydrogen mass to luminosity ratios seen in one or two extreme objects
in the Local Volume HI Survey of Koribalski (2018). These number $\sim$0.3\% of the HI detections in the NGC 5044 field. Confirmation of these objects is required.\\
(6) Forecasts for constraints from the full WALLABY survey on the growth rate of structure at the current epoch and for the measurement of large scale
flows seem sound based on the pilot survey.\\
(7) Addition of WALLABY galaxies to CosmicFlows4 increases the spatial resolution of the flow field because the higher sampling density affords more detail in the reconstruction.
%\vfill\break
%\pagebreak
%\clearpage
\onecolumn
\setcounter{table}{0}
\begin{table}
\caption{WALLABY data and WISE photometry NGC 4808 field}
\begin{tabular}{|l|l|l|r|l|l|l|r|r|l|r|}
\hline
{Name}& {RA}&{Dec}&{v$_{helio}$}&{b/a}&{W1}&{$\delta$W1}& {W$_{50}$}& {$\delta$W$_{50}$}&S/N&
{alias}\\
& J2000&&km s$^{-1}$&&mag&mag&\kms&\kms&&\\
(1)&(2)&(3)&(4)&(5)&(6)&(7)&(8)&(9)&(10)&(11)\\
\hline
 WALLABY\_J124747+042017 & 191.94598&   4.33818&   987&   0.878&  10.28&   0.03&   46&   18&  4.2& NGC 4688\\
 WALLABY\_J124805+065910 & 192.02299&   6.98568& 13424&   0.760&  12.83&   0.02&  130&   19&  2.1& PGC043228\\
 WALLABY\_J124822+082925 & 192.09509&   8.49036&  1011&   0.666&   7.35&   0.01&  424&   19&  2.9& NGC 4698\\
 WALLABY\_J124911+032310 & 192.29604&   3.38610&   722&   0.739&   9.50&   0.01&  167&   16&  6.0& NGC 4701\\
 WALLABY\_J124915+043926 & 192.31544&   4.65741&  2668&   0.860&  12.92&   0.03&   75&   17&  4.9& UGC 7976\\
 WALLABY\_J124923+050820 & 192.34853&   5.13909& 12410&   0.725&  13.83&   0.01&  212&   25&  1.8& PGC3395349\\
 WALLABY\_J124925+042328 & 192.35762&   4.39118&  2643&   0.550&  13.82&   0.04&  104&   18&  3.9& PGC1267592\\
 WALLABY\_J124935+033620 & 192.39824&   3.60573&  7085&   0.897&  12.46&   0.01&  166&   18&  4.3& PGC043361\\
 WALLABY\_J124937+074725 & 192.40498&   7.79054& 16420&   0.739&  14.94&   0.02&  273&   19&  2.0& \\
 WALLABY\_J124944+044608 & 192.43346&   4.76898&  8441&   0.370&  12.28&   0.01&  146&   19&  2.6& PGC043389\\
 WALLABY\_J124947+035042 & 192.44933&   3.84521&   698&   0.440&  13.75&   0.03&   35&   17&  5.4& UGC 7983\\
 WALLABY\_J124950+025100 & 192.45959&   2.85022&  1159&   0.220&   9.82&   0.01&  208&   19&  2.2& UGC 7982\\
 WALLABY\_J124957+040438 & 192.49109&   4.07742& 10202&   0.680&  13.60&   0.02&  218&   18&  4.1& PGC1263188\\
 WALLABY\_J124959+054921 & 192.49675&   5.82272&   630&   0.720&  13.80&   0.02&  109&   18&  3.3& PGC1289538\\
 WALLABY\_J125012+073442 & 192.55397&   7.57855& 11451&   0.760&  10.73&   0.01&  274&   19&  3.1& AGC225050\\
 WALLABY\_J125103+072753 & 192.76622&   7.46492& 15171&   0.640&  12.37&   0.01&  297&   20&  2.0& PGC1322513\\
 WALLABY\_J125112+045132 & 192.80307&   4.85892&  7532&   0.884&  10.18&   0.01&  233&   18&  3.5& NGC 4734\\
 WALLABY\_J125114+052031 & 192.81215&   5.34211&  7514&   0.977&  13.45&   0.02&   74&   17&  4.6& AGC225201\\
 WALLABY\_J125121+063829 & 192.84126&   6.64150& 14998&   0.740&  14.21&   0.02&  124&   17&  5.4& PGC1304685\\
 WALLABY\_J125127+054748 & 192.86482&   5.79682& 14439&   0.710&  12.50&   0.01&  455&   19&  2.6& PGC1289326\\
 WALLABY\_J125133+080241 & 192.88995&   8.04494&  3597&   0.888&  13.49&   0.08&  130&   18&  4.0& PGC043556\\
 WALLABY\_J125134+055147 & 192.89421&   5.86332& 14592&   0.674&  12.03&   0.02&  401&   24&  1.8& PGC1290429\\
 WALLABY\_J125156+035954 & 192.98697&   3.99838&  7684&   0.574&  15.41&   0.02&  206&   28&  1.7& PGC1261946\\
 WALLABY\_J125211+043045 & 193.04642&   4.51261& 19307&   0.790&  13.03&   0.01&  271&   19&  2.6& PGC4345126\\
 WALLABY\_J125215+042728 & 193.06422&   4.45782&   696&   0.570&  15.42&   0.11&   57&   19&  2.5& AGC226122\\
 WALLABY\_J125224+071046 & 193.10010&   7.17953& 14948&   0.325&  13.73&   0.03&  230&   17&  6.6& PGC1315615\\
 WALLABY\_J125233+031517 & 193.13998&   3.25474& 14620&   0.666&  12.38&   0.01&  295&   22&  1.9& PGC1248106\\
 WALLABY\_J125243+075643 & 193.17969&   7.94529& 11588&   0.740&  13.27&   0.02&  201&   22&  1.9& PGC1335032\\
 WALLABY\_J125258+073025 & 193.24283&   7.50694& 16023&   0.790&  12.56&   0.02&  421&   19&  3.0& PGC043729\\
 WALLABY\_J125303+070944 & 193.26555&   7.16228& 15196&   0.600&  16.16&   0.06&  111&   19&  2.3& PGC5509705\\
 WALLABY\_J125308+081035 & 193.28596&   8.17653& 14834&   0.490&  16.03&   0.05&  123&   41&  1.1& PGC434576\\
 WALLABY\_J125311+032639 & 193.29700&   3.44418&  2789&   0.900&  16.54&   0.20&   29&   19&  2.6& PGC166147\\
 WALLABY\_J125313+042746 & 193.30641&   4.46279&   725&   0.688&  10.35&   0.01&   87&   16&  6.2& NGC 4765\\
 WALLABY\_J125334+060947 & 193.39287&   6.16324& 13460&   0.410&  12.69&   0.02&  160&   19&  2.5& \\
 WALLABY\_J125339+040434 & 193.41631&   4.07620&   887&   0.710&  14.69&   0.04&   40&   18&  3.7& AGC224229\\
 WALLABY\_J125340+064728 & 193.42010&   6.79115&  7630&   0.808&  11.88&   0.01&  206&   19&  2.2& PGC043817\\
 WALLABY\_J125341+034603 & 193.42392&   3.76761& 19644&   0.851&  13.03&   0.02&  145&   19&  3.0& PGC1258221\\
 WALLABY\_J125343+040920 & 193.43007&   4.15566&   772&   0.825&  14.74&   0.05&   39&   19&  2.6& PGC1264260\\
 WALLABY\_J125413+052149 & 193.55656&   5.36375& 14499&   0.544&  12.01&   0.01&  312&   19&  2.6& PGC4676508\\
 WALLABY\_J125419+064115 & 193.57948&   6.68763&  3633&   0.785&  13.64&   0.01&   92&   27&  1.7& PGC3091848\\
 WALLABY\_J125431+050959 & 193.63210&   5.16657& 19563&   0.604&  12.72&   0.01&  484&   45&  1.0& PGC1279102\\
 WALLABY\_J125443+050110 & 193.68056&   5.01958& 10661&   0.533&  16.33&   0.08&  120&   18&  3.4& \\
 WALLABY\_J125443+080310 & 193.68300&   8.05288&  2520&   0.715&  11.23&   0.01&  243&   19&  3.1& NGC 4791\\
 WALLABY\_J125445+045357 & 193.69009&   4.89937&  7453&   0.650&  15.17&   0.06&  123&   19&  2.8& PGC1274939\\
 WALLABY\_J125450+072542 & 193.71011&   7.42845& 15011&   0.617&  13.43&   0.01&  274&   25&  1.8& PGC1321584\\
 WALLABY\_J125500+051140 & 193.75122&   5.19446& 19741&   0.790&  15.73&   0.03&   79&   18&  4.5& PGC3480219\\
 WALLABY\_J125507+040229 & 193.78165&   4.04148& 14467&   0.972&  12.04&   0.01&   91&   19&  2.7& PGC1262572\\
 WALLABY\_J125509+075453 & 193.79150&   7.91487&  2655&   0.551&  12.18&   0.02&  178&   18&  3.7& UGC 8042\\
 WALLABY\_J125515+081543 & 193.81377&   8.26203& 21390&   0.850&  12.30&   0.01&  169&   17&  4.9& PGC1343076\\
 WALLABY\_J125516+025347 & 193.81685&   2.89662&  2803&   0.664&   9.59&   0.01&  365&   18&  3.3& NGC 4799\\
 WALLABY\_J125520+050620 & 193.83734&   5.10573& 10706&   0.548&  13.85&   0.02&  200&   21&  1.9& PGC3480237\\
 WALLABY\_J125548+041805 & 193.95232&   4.30158&   760&   0.537&   8.71&   0.02&  269&   15&  7.7& NGC 4808\\
 WALLABY\_J125548+045901 & 193.95166&   4.98387& 14632&   0.525&  12.33&   0.01&  430&   18&  4.3& PGC1276334\\
 WALLABY\_J125549+040049 & 193.95634&   4.01383&   712&   0.518&  12.24&   0.03&  124&   17&  5.5& UGC 8053\\
 WALLABY\_J125556+075542 & 193.98351&   7.92859& 12464&   0.495&  12.68&   0.01&  299&   34&  1.4& PGC4538018\\
 WALLABY\_J125603+045202 & 194.01451&   4.86747&  1375&   0.800&  14.06&   0.05&   42&   17&  4.5& PGC5057592\\
\hline
\multicolumn{11}{r}{continued in full in MNRAS}\\
\end{tabular}
\end{table}
%\vfill\break
\pagebreak
\onecolumn
\begin{table}
\caption{WALLABY data and WISE photometry Vela field}
\begin{tabular}{|l|l|l|r|l|l|l|r|r|l|r|}
\hline
{Name}& {RA}&{Dec}&{v$_{helio}$}&{b/a}&{W1}&{$\delta$W1}& {W$_{50}$}& {$\delta$W$_{50}$}&{S/N}&
{~~alias~~}\\
& J2000&&km s$^{-1}$&&mag&mag&\kms&\kms&&\\
(1)&(2)&(3)&(4)&(5)&(6)&(7)&(8)&(9)&(10)&(11)\\
\hline
 WALLABY\_J094317-452105 & 145.82089& -45.35157&  5340& 0.313&   14.04&    0.13&  229& 17& 5.7&          \\
 WALLABY\_J094325-453019 & 145.85829& -45.50539& 12065& 0.616&   13.72&    0.05&  251& 18& 3.7&          \\
 WALLABY\_J094333-452200 & 145.89011& -45.36670& 18231& 0.511&   11.49&    0.02&  143& 19& 2.6& PGC527583\\
 WALLABY\_J094429-464515 & 146.12170& -46.75434& 17251& 0.499&   13.55&    0.05&  277& 19& 2.3&          \\
 WALLABY\_J094524-480828 & 146.35260& -48.14116&   880& 0.200&   14.56&    0.09&   75& 15& 8.0&          \\
 WALLABY\_J094533-462130 & 146.39050& -46.35859&  7502& 0.430&   15.30&    0.07&  114& 18& 3.4& PGC515934\\
 WALLABY\_J094613-471709 & 146.55690& -47.28599& 14082& 0.500&   12.57&    0.05&  217& 18& 4.3&          \\
 WALLABY\_J094622-463856 & 146.59392& -46.64899&  2704& 0.300&   11.18&    0.02&  291& 16& 6.7&          \\
 WALLABY\_J094723-431306 & 146.84944& -43.21851&  5318& 0.479&   12.93&    0.03&  154& 32& 1.7&          \\
 WALLABY\_J094814-464424 & 147.05835& -46.74026& 17507& 0.422&   14.26&    0.02&  218& 17& 5.7&          \\
 WALLABY\_J094819-442224 & 147.08121& -44.37339& 18139& 0.281&   12.27&    0.01&  272& 35& 2.0& PGC538024\\
 WALLABY\_J094828-472216 & 147.11841& -47.37133& 17106& 0.470&   12.35&    0.01&  297& 19& 2.5& PGC502430\\
 WALLABY\_J094830-464320 & 147.12833& -46.72243& 17352& 0.548&   14.02&    0.04&  249& 17& 4.5& PGC511516\\
 WALLABY\_J094840-452959 & 147.16727& -45.49980& 13467& 0.226&   13.01&    0.02&  376& 19& 3.1&          \\
 WALLABY\_J094854-462741 & 147.22504& -46.46154& 17222& 0.536&   13.28&    0.02&  279& 19& 3.2&          \\
 WALLABY\_J094855-435404 & 147.23100& -43.90133& 19445& 0.873&   13.13&    0.02&  350& 17& 4.6& PGC543149\\
 WALLABY\_J094924-454230 & 147.35231& -45.70839&  5284& 0.378&   12.88&    0.03&   98& 18& 4.5&          \\
 WALLABY\_J094926-435730 & 147.36195& -43.95849& 19448& 0.859&   13.61&    0.02&  203& 18& 4.0&          \\
 WALLABY\_J094940-454322 & 147.41983& -45.72292& 16372& 0.290&   11.68&    0.01&  458& 19& 2.3& PGC096455\\
 WALLABY\_J094959-481705 & 147.49632& -48.28491&  5212& 0.220&   12.10&    0.03&  270& 16& 6.1& PGC489114\\
 WALLABY\_J095019-452605 & 147.57959& -45.43482& 12136& 0.506&   11.70&    0.01&  192& 19& 2.2& PGC309881\\
 WALLABY\_J095019-441017 & 147.58247& -44.17142& 19387& 0.864&   12.99&    0.02&  120& 19& 2.5& PGC309880\\
 WALLABY\_J095031-461322 & 147.63036& -46.22299& 13532& 0.385&   16.03&    0.06&  220& 18& 4.5& PGC517483\\
 WALLABY\_J095053-432902 & 147.72421& -43.48414&  2699& 0.717&   14.04&    0.05&  151& 31& 3.8& PGC548188\\
 WALLABY\_J095101-451536 & 147.75775& -45.26022& 11295& 0.274&   14.65&    0.05&  161& 18& 4.5& PGC528651\\
 WALLABY\_J095126-452831 & 147.85852& -45.47532& 12124& 0.341&   13.48&    0.02&  269& 17& 5.0& PGC526385\\
 WALLABY\_J095131-440121 & 147.88051& -44.02256& 13106& 0.415&   12.22&    0.01&  477& 19& 2.5& PGC028395\\
 WALLABY\_J095159-443357 & 147.99901& -44.56604& 19297& 1.000&   15.02&    0.03&  101& 19& 2.2&          \\
 WALLABY\_J095205-443809 & 148.02348& -44.63591& 18041& 0.536&   12.68&    0.01&  254& 32& 1.5& PGC535167\\
 WALLABY\_J095211-464251 & 148.04860& -46.71441& 13589& 0.321&   13.47&    0.02&  294& 17& 4.8& PGC511580\\
 WALLABY\_J095234-475517 & 148.14436& -47.92139& 17230& 0.738&   13.98&    0.03&  150& 15& 7.3&          \\
 WALLABY\_J095239-470647 & 148.16661& -47.11310&  1820& 0.900&   13.03&    0.07&   73& 19& 2.0& PGC506395\\
 WALLABY\_J095244-474722 & 148.18340& -47.78953&  5390& 0.390&   13.92&    0.10&  115& 18& 3.6& PGC495718\\
 WALLABY\_J095311-451624 & 148.29988& -45.27333& 11263& 0.762&   13.69&    0.04&  206& 19& 3.1& PGC528542\\
 WALLABY\_J095316-431054 & 148.31738& -43.18174& 18933& 0.600&   14.28&    0.02&   89& 22& 1.9&          \\
 WALLABY\_J095347-455759 & 148.44792& -45.96661&  7089& 0.200&   11.14&    0.01&  367& 19& 2.4& PGC028528\\
 WALLABY\_J095354-433845 & 148.47746& -43.64592&  7727& 0.432&   14.19&    0.03&  175& 17& 4.8&          \\
 WALLABY\_J095423-443318 & 148.59874& -44.55511& 10029& 0.853&   13.65&    0.04&  150& 18& 3.5& PGC536076\\
 WALLABY\_J095433-440456 & 148.63873& -44.08222& 17770& 0.948&   13.77&    0.02&  270& 19& 2.8&          \\
 WALLABY\_J095441-440640 & 148.67125& -44.11118& 17882& 0.483&   12.71&    0.02&  252& 19& 2.3& PGC540776\\
 WALLABY\_J095448-462159 & 148.70078& -46.36654&  7074& 0.824&   12.91&    0.02&  186& 17& 5.0& PGC515876\\
 WALLABY\_J095449-461903 & 148.70421& -46.31764& 16160& 0.721&   14.67&    0.04&  275& 19& 2.4& PGC516415\\
 WALLABY\_J095457-480416 & 148.73886& -48.07116& 11541& 0.561&   11.82&    0.02&  440& 18& 3.5& PGC308536\\
 WALLABY\_J095532-443931 & 148.88661& -44.65867& 18673& 0.799&   11.88&    0.01&  336& 18& 3.5& PGC534960\\
 WALLABY\_J095552-455731 & 148.96976& -45.95863& 12425& 0.781&   14.85&    0.04&   93& 18& 4.0&          \\
 WALLABY\_J095558-472929 & 148.99477& -47.49157& 19141& 0.933&   14.40&    0.02&  120& 15& 7.9&          \\
 WALLABY\_J095635-453219 & 149.14613& -45.53882& 13794& 0.531&   14.43&    0.03&  174& 18& 3.8&          \\
 WALLABY\_J095648-450253 & 149.20219& -45.04817& 19190& 0.438&   15.53&    0.04&  228& 19& 2.3&          \\
 WALLABY\_J095710-485624 & 149.29468& -48.94017&  3727& 0.250&   12.83&    0.04&   40& 16& 6.5& PGC580713\\
 WALLABY\_J095719-465024 & 149.32941& -46.84004& 13024& 0.617&   13.56&    0.02&  277& 19& 3.1&          \\
 WALLABY\_J095802-444730 & 149.51060& -44.79173& 23768& 0.729&   14.67&    0.02&  244& 36& 3.7& PGC533495\\
 WALLABY\_J095809-472057 & 149.54031& -47.34937&  3644& 0.896&   14.38&    0.05&  124& 17& 5.6&          \\
 WALLABY\_J095813-430817 & 149.55484& -43.13825&  7307& 0.396&   14.88&    0.06&  184& 18& 3.9&          \\
 WALLABY\_J095824-433328 & 149.60017& -43.55786& 18236& 0.608&   12.85&    0.01&  130& 19& 2.1& PGC547363\\
 WALLABY\_J095836-470454 & 149.65350& -47.08191&  3628& 0.412&   10.89&    0.01&  109& 19& 3.2&          \\
 WALLABY\_J095848-455156 & 149.70287& -45.86573&  3664& 0.598&   12.09&    0.01&  196& 18& 4.0&          \\
\hline
\multicolumn{11}{r}{Continued in full in MNRAS}
\end{tabular}
\end{table}
\pagebreak
\onecolumn
\begin{table}
\caption{WALLABY data and WISE photometry NGC 5044 field}
\begin{tabular}{|l|l|l|r|l|l|l|r|r|l|r|}
\hline
{Name}& {RA}&{Dec}&{v$_{helio}$}&{b/a}&{W1}&{$\delta$W1}& {W$_{50}$}& {$\delta$W$_{50}$}&S/N&
{alias}\\
& J2000&&km s$^{-1}$&&mag&mag&\kms&\kms&&\\
(1)&(2)&(3)&(4)&(5)&(6)&(7)&(8)&(9)&(10)&(11)\\
\hline
%WALLABY\_J125656-202606 & 194.23398& -20.43518&  3676& 0.000&   --&   --&   17& 18& 3.6&          \\% J1256      
WALLABY\_J125700-171910 & 194.25180& -17.31950&  3972& 0.620&   11.65&    0.01&  220& 19& 2.6&PGC044234 \\%J1257 0-1719
WALLABY\_J125706-182159 & 194.27628& -18.36646& 16185& 0.706&   11.29&    0.01&   88& 29& 1.6&ESO575-036\\%J1257 6-1821
%WALLABY\_J125721-171102 & 194.33859& -17.18393&  4836& 0.700&   16.06&    0.05&   64& 24& 1.8&          \\% J1257      
%WALLABY\_J125724-153912 & 194.35216& -15.65339&  5593& 0.200&   13.35&    0.01&   72& 19& 2.6&PGC908309 \\%J125724-1539
WALLABY\_J125757-130339 & 194.48877& -13.06087&  5022& 0.556&    9.53&    0.01&  215& 19& 2.8&NGC4838   \\%J125757-13 3
%WALLABY\_J125831-181843 & 194.62988& -18.31210& 15236& 0.854&   11.54&    0.01&  272& 34& 2.2&ESO575-040\\%J125831-1818
%WALLABY\_J125832-161038 & 194.63490& -16.17740&  5331& 0.836&   14.43&    0.01&  157& 19& 2.3&PGC901483 \\%J125832-1610
WALLABY\_J125834-164819 & 194.64464& -16.80551&  3843& 0.386&   13.61&    0.01&  147& 19& 3.1&PGC044478 \\%J125834-1648
WALLABY\_J125836-161333 & 194.65154& -16.22606&  5266& 0.000&   12.41&    0.01&  131& 19& 2.7&          \\% J1258      
WALLABY\_J125843-170913 & 194.67995& -17.15386&  8361& 0.485&   12.88&    0.01&  318& 19& 2.6&PGC044499 \\%J125843-17 9
WALLABY\_J125848-114647 & 194.70290& -11.77985&  4849& 0.200&   13.61&    0.02&  154& 19& 2.2&PGC104987 \\%J125848-1146
WALLABY\_J125850-145324 & 194.71039& -14.89014&  4874& 0.330&   14.38&    0.02&  144& 18& 4.1&PGC918593 \\%J125850-1453
WALLABY\_J125855-142319 & 194.73131& -14.38869& 19381& 0.000&   19.04&    0.46&  148& 18& 3.4&          \\% t:J12      
WALLABY\_J125859-135141 & 194.74915& -13.86140&  4883& 0.496&   15.01&    0.02&   55& 19& 2.1&PGC932206 \\%J125859-1351
WALLABY\_J125907-121329 & 194.77928& -12.22482&  1312& 0.430&   13.37&    0.03&   80& 16& 6.0&PGC044549 \\%J1259 7-1213
WALLABY\_J125910-161749 & 194.79179& -16.29707& 12931& 0.686&   13.15&    0.02&  180& 19& 3.0&PGC899955 \\%J125910-1617
WALLABY\_J125912-182326 & 194.80096& -18.39075& 15297& 0.000&   14.13&    0.01&   89& 19& 2.1&PGC868681 \\%J125912-1823
WALLABY\_J125912-124101 & 194.80396& -12.68385&  3875& 0.389&   14.65&    0.03&  134& 17& 4.5&PGC949183 \\%J125912-1241
%WALLABY\_J125915-150108 & 194.81252& -15.01891&  1210& 0.697&   --&   --&  162& 17& 5.2&          \\% t:J12      
WALLABY\_J125923-120455 & 194.84590& -12.08215&  6349& 0.460&   12.24&    0.01&  191& 18& 4.5&PGC3082105\\%J125923-12 4
%WALLABY\_J125924-150253 & 194.85040& -15.04830&  1466& 0.522&   --&   --&  223& 21& 1.9&          \\% J1259      
WALLABY\_J125930-135159 & 194.87573& -13.86649& 13996& 0.000&   13.58&    0.01&   97& 18& 3.4&PGC932150 \\%J125930-1351
WALLABY\_J125930-180055 & 194.87677& -18.01554&  4959& 0.430&   15.59&    0.06&  120& 17& 4.6&PGC873976 \\%J125930-18 0
WALLABY\_J125930-120652 & 194.87744& -12.11468&  6387& 0.960&   12.91&    0.01&  129& 17& 4.7&PGC957468 \\%J125930-12 6
WALLABY\_J125931-140755 & 194.87987& -14.13200&  6451& 0.759&   11.50&    0.01&  145& 17& 4.9&NGC4862   \\%J125931-14 7
WALLABY\_J125932-182203 & 194.88594& -18.36763& 15110& 0.603&   13.85&    0.02&  119& 17& 4.9&          \\% t:J12      
WALLABY\_J125932-151419 & 194.88669& -15.23881&  1418& 0.571&   12.78&    0.01&   95& 19& 2.1&PGC3082108\\%J125932-1514
WALLABY\_J125939-145813 & 194.91435& -14.97029&  4953& 0.221&   10.80&    0.01&  360& 17& 5.2&PGC044645 \\%J125939-1458
WALLABY\_J125940-151600 & 194.91905& -15.26692&  4860& 0.914&   13.54&    0.01&   97& 19& 2.7&PGC913440 \\%J125940-1516
WALLABY\_J125942-140140 & 194.92767& -14.02790&  4368& 0.466&   10.64&    0.01&  462& 18& 3.6&NGC4863   \\%J125942-14 1
WALLABY\_J125942-164952 & 194.92847& -16.83133&  1368& 0.450&   15.81&    0.07&   43& 18& 4.2&          \\% t:J12      
WALLABY\_J125943-150804 & 194.93118& -15.13447& 13409& 0.945&   15.41&    0.05&   86& 19& 2.5&PGC915253 \\%J125943-15 8
WALLABY\_J125945-145159 & 194.93814& -14.86639&  2659& 0.398&   13.20&    0.01&  147& 19& 2.7&PGC918920 \\%J125945-1451
WALLABY\_J125949-200826 & 194.95747& -20.14056& 13972& 0.000&   14.17&    0.02&   52& 29& 1.6&          \\% t:J12      
WALLABY\_J125951-150446 & 194.96494& -15.07954& 12208& 0.676&   13.05&    0.02&  271& 18& 3.4&PGC916026 \\%J125951-15 4
WALLABY\_J125956-192430 & 194.98334& -19.40843&   826& 0.360&   14.48&    0.03&   45& 16& 6.7&PGC044681 \\%J125956-1924
WALLABY\_J125959-143124 & 194.99905& -14.52352&  2576& 0.247&   13.79&    0.01&   67& 17& 4.7&PGC923490 \\%J125959-1431
%WALLABY\_J130002-183046 & 195.00870& -18.51289& 13819& 0.564&   16.11&    0.12&  345& 19& 3.1&          \\% J1300      
%WALLABY\_J130003-183028 & 195.01302& -18.50791& 13770& 0.579&   13.80&    0.01&  249& 18& 3.8&PGC867354 \\%J1300 3-1830
WALLABY\_J130004-154055 & 195.01862& -15.68214&  1135& 0.200&   14.47&    0.05&  121& 18& 3.3&PGC907834 \\%J1300 4-1540
WALLABY\_J130004-152151 & 195.02060& -15.36439&  1589& 0.200&   10.74&    0.01&  117& 18& 4.3&PGC044701 \\%J1300 4-1521
WALLABY\_J130010-130042 & 195.04393& -13.01183&  4735& 0.201&   11.64&    0.01&  166& 19& 2.6&PGC044718 \\%J130010-13 0
WALLABY\_J130016-170140 & 195.06892& -17.02799&  4908& 0.416&   11.14&    0.01&  240& 19& 3.0&PGC170439 \\%J130016-17 1
WALLABY\_J130017-122041 & 195.07176& -12.34494&  1581& 0.596&   11.20&    0.01&  168& 16& 6.1&PGC044735 \\%J130017-1220
WALLABY\_J130018-115758 & 195.07573& -11.96620& 15261& 0.978&   11.69&    0.01&  181& 18& 3.3&PGC959514 \\%J130018-1157
WALLABY\_J130026-151708 & 195.10945& -15.28555&  4920& 0.345&    8.74&    0.01&  600& 17& 4.9&NGC4877   \\%J130026-1517
WALLABY\_J130029-181400 & 195.12294& -18.23359& 14628& 0.810&   15.99&    0.05&   98& 19& 2.1&          \\% t:J13      
WALLABY\_J130029-133942 & 195.12305& -13.66192&  6392& 0.310&   16.35&    0.07&   76& 40& 1.2&          \\% J1300      
WALLABY\_J130029-133839 & 195.12349& -13.64420&  4331& 0.265&   12.62&    0.01&  158& 19& 3.0&PGC044776 \\%J130029-1338
WALLABY\_J130032-123039 & 195.13338& -12.51109&  6359& 0.680&   14.34&    0.01&  136& 17& 5.3&PGC951730 \\%J130032-1230
WALLABY\_J130033-122307 & 195.14067& -12.38545&  2500& 0.750&   16.43&    0.13&   45& 16& 5.8&PGC953585 \\%J130033-1223
WALLABY\_J130037-143948 & 195.15587& -14.66346&  2701& 0.562&   10.61&    0.01&  150& 17& 5.5&NGC4887   \\%J130037-1439
WALLABY\_J130043-185429 & 195.18138& -18.90821& 10404& 0.498&   14.43&    0.03&  225& 17& 5.1&PGC862238 \\%J130043-1854
WALLABY\_J130043-154253 & 195.18234& -15.71495&  1383& 0.000&   11.74&    0.02&   85& 17& 4.9&PGC044812 \\%J130043-1542
WALLABY\_J130044-142906 & 195.18745& -14.48505&  6422& 0.200&   14.48&    0.03&  181& 17& 5.1&PGC924035 \\%J130044-1429
WALLABY\_J130047-130944 & 195.19746& -13.16238& 18291& 0.631&   13.71&    0.01&  264& 18& 4.1&PGC941853 \\%J130047-13 9
WALLABY\_J130047-141534 & 195.19971& -14.25961&  4596& 0.880&   15.48&    0.05&  117& 17& 4.8&          \\% t:J13      
WALLABY\_J130049-143608 & 195.20488& -14.60221&  6500& 0.550&   11.60&    0.01&  259& 18& 4.3&PGC3098292\\%J130049-1436
WALLABY\_J130049-170522 & 195.20738& -17.08964&  4826& 0.485&   16.45&    0.03&  101& 19& 2.3&          \\% J1300      
WALLABY\_J130052-164149 & 195.21719& -16.69698&  4286& 0.919&   15.21&    0.04&   79& 19& 2.9&          \\% t:J13      
WALLABY\_J130052-145506 & 195.21899& -14.91849& 13859& 0.200&   14.99&    0.04&   78& 19& 3.1&          \\% t:J13      
WALLABY\_J130053-132655 & 195.22302& -13.44861&  2551& 0.531&    9.37&    0.01&  230& 16& 6.1&NGC4897   \\%J130053-1326
WALLABY\_J130056-135640 & 195.23347& -13.94450&  2654& 0.677&    9.25&    0.01&  277& 16& 6.2&NGC4899   \\%J130056-1356
WALLABY\_J130057-172247 & 195.23891& -17.37990&  4889& 0.731&   14.74&    0.03&   87& 18& 3.3&PGC884267 \\%J130057-1722
\multicolumn{11}{r}{Continued in full in MNRAS}
\end{tabular}
\end{table}
\vfill\break
\pagebreak
\begin{table*}
\caption{ N5044-DR3 galaxies with 2 published spectra}
\begin{tabular}{l c l l c c r r r r r } 
 \hline
WALLABY   &       PGC no&      Name&       AGC/HIPASS no&     Vsys&  Width&  err&     S/N&  flux&      Vwal&  Wwal\\  
	  &             &          &                    & \kms&\kms&\kms&&Jy\kms&\kms&\kms \\
 J130943-163617&  45650  &PGC045650  &      HIPASSJ1309-16&     2575 &   340&    &  5.4 &27.18    &2574    &347   \\ % -5       12
J130943-163617 & 45650  &PGC045650  &    HIPASSJ1309-16   &  2576   & 339  &   14   &  8.9  &7.62    &2574    &347  \\  % -4       11
J131234-173225  &45877  &PGC045877  &     AGC530053 shg05       &    2756 &   385 &    19 &    2.2 &12.68  &  2757   & 389\\  %  -25       40
J131234-173225&  45877  &PGC045877  &    HIPASSJ1312-17   &  2762   & 376  &   13   & 10.0  &2.48    &2757    &389  \\  %-19       31
J132019-123420 & 46535  &NGC5088    &    HIPASSJ1320-12 &    1430&    235   &  16   &  6.6 &39.67   & 1431   & 240 \\%    -1          3
J132019-123420  &46535  &NGC5088    &   AGC530112 shg05      &      1432&    242  &   13 &   14.1& 42.77 &   1431  &  240\\%        1       10
J132043-220256&  46574  &ESO576-040 &    AGC029511 shg05        &    2087  &  185 &    12  &  11.6 &16.20   & 2089    &191   \\% -11          2
J132043-220256 & 46574  &ESO576-040 &    HIPASSJ1320-22   &  2082    &179   &  16    & 5.8  &3.75     &2089    &191   \\% -16       -4
J132947-175747  &47394  &NGC5170    &      AGC029724      &      1503&    524  &   18&     3.9 &71.37  &  1503 &   523  \\%  -15       14
J132947-175747&  47394  &NGC5170    &  HIPASSJ1329-17 &    1502  &  528   &  13 &   11.8  &5.95   & 1503   & 523    \\%-16       18
J133541-240428 & 47948  &ESO509-074 &    AGC029872 shg05        &    2592  &  354  &   18&     4.1  &2.45   & 2572   & 322    \\% 32       42
J133541-240428  &47948  &ESO509-074 &    HIPASSJ1335-24   &  2586    &345   &  17&     4.9    &8.66   & 2572   & 322     \\%26       33
J133802-175254&  48171  &NGC5247    & AGC029924 shg05      &      1356&    137 &    8 &    53.5& 57.73 &   1354 &   137   \\ % -1          3
J133802-175254&  48171  &NGC5247    &  HIPASSJ1338-17 &    1355  &  138 &   13   &  32.9 &56.62   & 1354   & 137   \\    % 0          4
\hline
\end{tabular}
\end{table*}
\begin{table*}
\caption{ N5044-DR3 galaxies with 1 published spectrum}
\begin{tabular}{l l l l r r r r r r r }
\hline
    WALLABY        &   PGC no       &     Name        &     AGC/HIPASS no&       Vsys&  Width&  err&   S/N&  flux&    Vwal&  Wwal\\%  Vs-Vw  Ws-Ww
	  &             &          &                    & \kms&\kms&\kms&&Jy\kms&\kms&\kms \\

 J125907-121329    &  44549         & PGC044549       & HIPASSJ1259-12   &   1308&    97&   13&   11.0&  14.13&   1312&80\\%      0     46
 J125931-140755    &  44610         & NGC4862         & AGC520377 shg05         &  6456 &  143 &  17 &   4.7 &  3.84&   6451 &  145\\%      2     14
 J125956-192430    &  44681         & PGC044681       & HIPASSJ1259-19     &  829  &  52  & 16  &  6.3  & 5.03 &   826  &  45 \\%     1     14
 J130026-151708    &  44761         & NGC4877         & HIPASSJ1300-15      &4933   &602   &15   & 7.4&  16.39  & 4920   &600  \\%   17     32
 J130053-132655    &  44829         & NGC4897         & HIPASSJ1300-13&      2555&   247&   13&   12.7 & 29.33   &2551&   230   \\%  -1     31
 J130056-135640    &  44841         & NGC4899         & HIPASSJ1301-13B&     2653 &  273 & 100 &  34.7  &21.09&   2654 &  277    \\% -8     11
 J130059-143042    &  44847         & NGC4902         & HIPASSJ1300-14  &    2631  & 256  & 13  & 12.5&  27.93 &  2624  & 240     \\% 8     43
 J130107-133053    &  936912        & PGC936912       & HIPASSJ1300-13B  &   1308   & 65   &14   & 8.7 &  7.15  & 1306   & 62      \\%3     17
 J130213-145817    &  44977         & NGC4924         & HIPASSJ1302-14A   &  4856&   266&   14    &9.3  & 8.36   &4864&   267\\%     -2     29
 J130213-171416    &  44982         & PGC044982       & HIPASSJ1302-17     &  750 &  117 &  13&   11.5&  27.24&    726 &   39 \\%    29     93
 J130225-174046    &  45006         & PGC045006       & AGC520412 shg05           &4652  & 480  & 21 &   5.7 & 11.18 &  4649  & 475  \\%    5     37
 J130311-172230    &  45073         & PGC045073       & HIPASSJ1303-17A&     2965   & 66   &25  &  7.7  &10.69  & 2963   & 41   \\%   3     37
 J130314-172514    &  45084         & PGC045084       & HIPASSJ1303-17B &     742&   118&   13   &63.1&  93.27   & 741&   118    \\% -2      9
 J130815-210002    &  45524         & ESO575-061      & AGC029274        &   1649 &  165 &  17&    5.0 &  2.94&   1649 &  168     \\% 2      2
 J130947-101916    &  45652         & PGC045652       & HIPASSJ1309-10    &  1206  & 207  & 16 &   5.8  &14.02 &  1210  & 196      \\%0     32
 J131035-214450    &  45721         & ESO576-003      & AGC029327 shg05          & 2959   &231   &16  &  6.6&  11.27  & 2962&   238\\%      8      6
 J131305-195830    &  45911         & ESO576-011      & AGC029361           &2757&   309&   11   &12.4 & 13.23   &2759 &  306 \\%   -11     12
 J131321-185558    &  45935         & ESO576-012      & AGC029366 shg05&           6148 &  235 &  17&    4.9  & 3.19&   6149  & 237  \\%    3     23
 J131330-193244    &  45952         & NGC5022         & AGC029371 &          3001  & 392   &18 &   4.3   &8.46 &  3005   &398   \\%  -4     36
 J131334-152553    &  45958         & PGC045958       & HIPASSJ1313-15&      2503   & 81  & 15  &  7.9&  12.19  & 2501&    73    \\%  1     11
 J131702-104612    &  46252         & IC4216          & HIPASSJ1316-10 &     2839&   257&   16   & 5.9 &  7.06   &2838 &  263     \\% 5      6
 J131706-161516    &  46261         & PGC046261       & AGC530085 shg05       &    2631 &  288 &  19    &2.9  & 8.27&   2636  & 279\\%     -5     19
 J131739-213708    &  46299         & ESO576-023      & AGC029435 shg05       &   2947  & 188  & 25&    2.7   &6.34 &  2933   &124 \\%    15     88
 J131835-211758    &  46373         & ESO576-026      & AGC029453         &  1953   &163   &18 &   3.9&   2.40  & 1955&   181  \\%   -3    -10
 J131920-145037    &  46441         & NGC5073         & AGC530097 shg05          & 2745&   385&   15  &  8.0 & 16.07   &2746 &  405   \\%  -2     15
 J131934-124159    &  46473         & NGC5079         & AGC530104           &2228 &  235 &  16   & 6.6  & 1.92&   2223  & 249 \\%
 J132018-214925    &  46525         & NGC5084         & HIPASSJ1320-21&      1719  & 662  &100    &8.7& 102.72 &  1719   &655  \\%   24     24
 J132148-131222    &  46664         & NGC5105         & AGC530120 shg05     &     2903   &219   &12&   11.1 & 14.30  & 2903&   220   \\%   2     10
 J132430-210729    &  46878         & IC4237          & AGC029585 shg05      &    2667&   325&   18 &   3.8  &14.03   &2607 &  205    \\% 78    237
 J132441-194214    &  46889         & ESO576-050      & AGC029590        &   1969&   189 &  14  &  8.7&  17.01&   1967  & 192\\%    -12    152
 J132500-240035    &  46920         & ESO508-071      & AGC029599 shg05        &  7378 &  315  & 18   & 4.5 &  3.14 &  7388   &319 \\%   -19     12
 J133017-214446    &  47450         & ESO576-072      & AGC029740          & 7314  & 228  &100    &5.6  & 2.45  & 7312&   248  \\%   11     16
 J133058-215044    &  47505         & ESO577-001      & AGC029757           &7361   &379&  100&    3.4   &4.35   &7366 &  371   \\%   3     23
 J133103-150604    &  47514         & PGC047514       & AGC530277 shg05&           4232   &300 &  18 &   4.2&   6.22&   4231  & 303    \\% 11     12
 J133208-225710    &  47599         & ESO509-045      & AGC029792 &          5047&   278  & 19  &  2.8 &  3.20 &  5056   &266     \\%-9     23
 J133209-245132    &  47600         & ESO509-044      & AGC029789  &         7791 &  450&  100   & 2.8  & 4.09  & 7790&   439\\%      9     44
	J133314-160715    &  47717         & PGC047717       & tmc06    &6342   &  206&  100   & 3.4  & 2.35  & 6120&   683\\%AGC530316 shg05  &        2169  & 276 &  35    &3.0   &5.39   &    &     \\%??
 J133933-222957    &  48300         & ESO577-020      & AGC029955 shg05   &       6998   &249  & 19&    2.7&   2.67   &7000  & 235\\%      0     24
 J133949-221815    &  48323         & ESO577-022      & AGC029961     &      5760&   285   &17 &   4.7 &  2.62&   5746   &302 \\%    26      7
 J134002-252831    &  48346         & IC4315          & AGC029965      &     5163 &  322   &50  &  3.8  & 3.10 &  5134&   303  \\%   20     50
\hline
\multicolumn{6}{l}{shg05 = Springob \etal (2005)}\\
\multicolumn{6}{l}{ctf09 = Courtois \etal (2009)}\\
\multicolumn{6}{l}{tmc06 = Theureau \etal (2006)}
\end{tabular}
\end{table*}

\begin{multicols}{2}
\section*{Acknowledgements}
The Australian SKA Pathfinder is part of the Australia Telescope
National Facility (https://ror.org/05qajvd42) which is
managed by CSIRO. Operation of ASKAP is funded by the
Australian Government with support from the National Collaborative
Research Infrastructure Strategy. ASKAP uses
the resources of the Pawsey Supercomputing Centre. Establishment
of ASKAP, the Murchison Radio-astronomy Observatory
and the Pawsey Supercomputing Centre are initiatives
of the Australian Government, with support from
the Government of Western Australia and the Science and
Industry Endowment Fund. We acknowledge the Wajarri
Yamatji as the traditional owners of the Observatory site.
WALLABY acknowledges technical support from the
Australian SKA Regional Centre (AusSRC) and Astronomy
Data And Computing Services (ADACS). AD is supported by a KIAS Individual Grant
PG 087201 at the Korea Institute for Advanced Studies. AB acknowledges support from the
Centre National d'Etudes Spatiales (CNES), France. %We thank Tobias
%Westmeier for his support. 
PEMP acknowledges support from the Dutch Research Council (NWO) through the Veni grant VI.Veni.222.364.
This research was supported partially by the Australian Government through the Australian Research Council Centre of Excellence for Dark Matter Particle Physics (CDM, CE200100008).

This publication makes use of data products from the Wide-field Infrared Survey Explorer, 
which is a joint project of the University of California Los Angeles, and the
Jet Propulsion Laboratory/California Institute of Technology, 
funded by the National Aeronautics and Space Administration.
This research uses services and data provided by the Astro Data Lab at NSF's
National Optical-Infrared Astronomy Research Laboratory. %NOIRLab is operated
%by the Association of Universities for Research in Astronomy (AURA) Inc
%under a cooperative agreement with the National Science Foundation.
The Legacy Surveys consist of three individual and complementary projects: the Dark Energy Camera Legacy Survey (DECaLS; Proposal ID \#2014B-0404; PIs: David Schlegel and Arjun Dey), the Beijing-Arizona Sky Survey (BASS; NOAO Prop. ID \#2015A-0801; PIs: Zhou Xu and Xiaohui Fan), and the Mayall z-band Legacy Survey (MzLS; Prop. ID \#2016A-0453; PI: Arjun Dey). DECaLS, BASS and MzLS together include data obtained, respectively, at the Blanco telescope, Cerro Tololo Inter-American Observatory, NSF's NOIRLab; the Bok telescope, Steward Observatory, University of Arizona; and the Mayall telescope, Kitt Peak National Observatory, NOIRLab. Pipeline processing and analyses of the data were supported by NOIRLab and the Lawrence Berkeley National Laboratory (LBNL). The Legacy Surveys project is honored to be permitted to conduct astronomical research on Iolkam Du'ag (Kitt Peak), a mountain with particular significance to the Tohono O'odham Nation.

NOIRLab is operated by the Association of Universities for Research in Astronomy (AURA) under a cooperative agreement with the National Science Foundation. LBNL is managed by the Regents of the University of California under contract to the U.S. Department of Energy.

This project used data obtained with the Dark Energy Camera (DECam), which was constructed by the Dark Energy Survey (DES) collaboration. Funding for the DES Projects has been provided by the U.S. Department of Energy, the U.S. National Science Foundation, the Ministry of Science and Education of Spain, the Science and Technology Facilities Council of the United Kingdom, the Higher Education Funding Council for England, the National Center for Supercomputing Applications at the University of Illinois at Urbana-Champaign, the Kavli Institute of Cosmological Physics at the University of Chicago, Center for Cosmology and Astro-Particle Physics at the Ohio State University, the Mitchell Institute for Fundamental Physics and Astronomy at Texas A\&M University, Financiadora de Estudos e Projetos, Fundacao Carlos Chagas Filho de Amparo, Financiadora de Estudos e Projetos, Fundacao Carlos Chagas Filho de Amparo a Pesquisa do Estado do Rio de Janeiro, Conselho Nacional de Desenvolvimento Cientifico e Tecnologico and the Ministerio da Ciencia, Tecnologia e Inovacao, the Deutsche Forschungsgemeinschaft and the Collaborating Institutions in the Dark Energy Survey. 
 The Collaborating Institutions are Argonne National Laboratory, the University of California at Santa Cruz, the University of Cambridge, Centro de Investigaciones Energeticas, Medioambientales y Tecnologicas-Madrid, the University of Chicago, University College London, the DES-Brazil Consortium, the University of Edinburgh, the Eidgenossische Technische Hochschule (ETH) Zurich, Fermi National Accelerator Laboratory, the University of Illinois at Urbana-Champaign, the Institut de Ciencies de l'Espai (IEEC/CSIC), the Institut de Fisica d'Altes Energies, Lawrence Berkeley National Laboratory, the Ludwig Maximilians Universitat Munchen and the associated Excellence Cluster Universe, the University of Michigan, NSF's NOIRLab, the University of Nottingham, the Ohio State University, the University of Pennsylvania, the University of Portsmouth, SLAC National Accelerator Laboratory, Stanford University, the University of Sussex, and Texas A\&M University.

BASS is a key project of the Telescope Access Program (TAP), which has been funded by the National Astronomical Observatories of China, the Chinese Academy of Sciences (the Strategic Priority Research Program The Emergence of Cosmological Structures Grant \# XDB09000000), and the Special Fund for Astronomy from the Ministry of Finance. The BASS is also supported by the External Cooperation Program of Chinese Academy of Sciences (Grant \# 114A11KYSB20160057), and Chinese National Natural Science Foundation (Grant \# 12120101003, \# 11433005).
We acknowledge the use of the HyperLeda database (http://leda.univ-lyon1.fr
 and the Siena Galaxy Atlas. That was made possible by funding support from the U.S. Department of Energy, Office of Science, Office of High Energy Physocs under award number DE-SC0020086 and from the National Science Foundation under grant AST-1616414.

We thank WALLABY team member Matthew Colless %, %Denis Leahy 
%	and Lister Staveley-Smith %and Ren\'ee Kraan-Korteweg 
	for reading drafts of this paper and Tobias Westmeier for his management of the WALLABY project. We thank the referee for pointing out some things that needed clarifying. 
%%%%%%%%%%%%%%%%%%%%%%%%%%%%%%%%%%%%%%%%%%%%%%%%%%
\section*{Data Availability}

When released, 21 cm profiles in this paper will be curated
by the CSIRO ASKAP Science Data Archive. These profiles can be viewed at https://github.com/jrmould/wallaby-HI-profiles.

%\columnbreak


\section*{References}
\noindent
Aaronson, M., Huchra, J. \& Mould, J. 1979, ApJ, 229, 1\\
Adams C. \& Blake, C. 2017, %{\it Improving constraints on the growth rate of
%structure by modelling the density-velocity cross-correlation in the 6dF
%galaxy survey} 
	MNRAS, 471, 839\\
Bell, R., Said, K., Davis, T. \& Jarrett, T. 2023, MNRAS, 519, 102\\
%Benitez-Llambay \& Frenk, C. 2020, %{\it The detailed structure and onset of galaxy formation in low mass gaseous dark matter haloes}, 
%	MNRAS, 498, 4887\\
Bertin, E. \& Arnouts, S. 1999, A\&AS, 117, 393\\
%Cardelli, J. \etal 1989, {\it The relationship between optical, IR and UV extinction} ApJ, 345, 245\\
Boubel, P., Colless, M., Said, K. \& Staveley-Smith, L. 2024,%{\it Cosmic growth rate measurements from Tully Fisher peculiar velocities}, 
	MNRAS, 531, 84\\
Cluver, M. \etal 2014, ApJ, 782, 90\\
Cluver, M., Jarrett, T., Dale, D., Smith, J., August, T. \& Brown, M. 2017, %{\it Calibrating star formation in WISE using total infrared luminosity} 
	ApJ, 850, 68\\
Courtois, H. \& Tully, R.B. 2015, %{\it Update on HI data collection from the Green Bank,
%Parkes \& Arecibo telescopes for the Cosmic Flows Project} 
	MNRAS, 447, 1531\\
Courtois, H., Tully, R.B., Fisher, J.R., Bonhomme, N., Zavodny, M. \& Barnes, A. 2009,%{\it The Extragalactic Distance Database: All digital HI profile catalog}  
  AJ, 138, 1938\\
Courtois, H. \etal 2023, %{\it WALLABY pre-pilot \& pilot survey: the Tully-Fisher
%Relation in Eridanus, Hydra, Norma \& NGC4636 fields} 
MNRAS, 519, 4589\\
Courtois, H., Dupuy, A., Guinet, D., Baulieu, G., Ruppin, F., \& Brenas, P. 2023, %{\it Gravity in the local universe: density \& velocity fields using CosmicFlows4} 
A\&A, 670, 15\\
Deg, N. \etal 2022, %{\it WALLABY pilot survey: Public release of HI kinematics models for more than 100 galaxies from phase 1 of pilot observations with ASKAP} 
PASA, 39, 59\\
de Vaucouleurs, G.,  de Vaucouleurs, A., \& Corwin, H. 1976, {Second Reference Catalogue of Bright Galaxies}, Univ. Texas Press, Austin\\%tal properties of elliptical galaxies} ApJ, 313, 59\\
Djorgovski, S. \& Davis, M. 1987, ApJ, 313, 59\\
Drlica-Wagner, A. \etal 2022, %{\it The DECam Local Volume Exploration Survey Data Release 2}, 
ApJS, 261, 38\\
Dupuy, A., Courtois, H. \& Kubik, B. 2019, %{\it An estimation of the local growth rate from Cosmicflows peculiar velocities} 
MNRAS, 486, 440\\
For, B.-Q. \etal 2023, %{\it WALLABY Pre-pilot survey: Ultra Diffuse Galaxies in the Eridanus subgroup}, arxiv 2309.11799\\
MNRAS,  526, 3130\\
Fu, H. 2024,%{\it Restoration of the Tully-Fisher Relation by statistical restoration}, 
arxiv 2401.13748\\
%Gault, L. \etal 2021, %{\it VLA imaging of HI bearing UDGs from the ALFALFA survey}, 
%ApJ, 909, 19\\
Giovanelli, R., Haynes, M., Salzer, J., Wegner, G., da Costa, L. \& Freudling, W. 1994, %{\it Extinction in Sc galaxies}, 
AJ, 107, 2036\\
Haynes, M. \etal 2018, %{\it The Arecibo Legacy Fast ALFA survey: the ALFALFA
%Extragalactic HI source catalog} 
ApJ, 861, 49\\
Hotan, A. \etal 2021, %{\it The Australian SKA pathfinder I. System description} 
PASA, 38, 9\\
Jarrett, T., Cluver, M., Taylor, E.N., Bellstedt, S., Robotham, A. \& Yao, H. 2023, 
%{\it A new WISE calibration of stellar mass} 
ApJ 946, 95\\
%Lagattuta, D. , Mould, J., Staveley-Smith, L. Hong Tao, Springob, C., Masters,
%Kim, S. \etal 2016 {\it Large scale filamentary structures around the Virgo cluster revisited} ApJ, 833, 207\\
Koribalski, B. \etal 2004, %{\it The 1000 brightest HIPASS galaxies and HI properties}, 
AJ, 128, 16\\
Koribalski, B. \etal 2020, %{\it WALLABY - the ASKAP HI All-Sky Survey} 
ApSS, 365, 118\\
Koribalski, B. \etal 2018 %{\it The Local Volume HI Survey (LVHIS}, 
MNRAS, 478, 1611\\
Kourkchi, E., Tully, R.B., Courtois, H., Dupuy, A. \&  Guinet, D. 2022, %{\it CosmicFlows4, the baryonic Tully Fisher relation providing 10,000 distances } 
MNRAS, 511, 6160\\
Kowal, C. 1968, AJ, 73, 1021\\
Kraan-Korteweg, R. \etal 2017, MNRAS, 466, L29\\
%Kourkchi, E. \etal 2020, {\it CosmicFlows3: two distance velocity calculators} AJ, 159, 67\\
Leisman, L. \etal 2017, %{\it (Almost) Dark Galaxies in the ALFALFA Survey}, 
ApJ, 842, 133\\
Lelli, F., McGaugh, S., Schombert, J., Desmond, H. \& Katz, H. 2019, MNRAS, 484, 3267\\
Lineweaver, C., Tenorio, L., Smoot, G., Keegstra, P., Banday, A. \& Lubin, P. 1996, ApJ, 470, 38\\
McGaugh, S., Schombert, J., Bothun, G. \& de Blok, W. 2000, %{\it The baryonic Tully Fisher relation} 
ApJ, 533, 99\\
McGaugh, S. \etal 2021, %{\it The baryonic Tully Fisher relation in the Local Group and the equivalent circular velocity} 
AJ, 162, 202\\
Makarov, D., Prugniel, P., Terekhova, N., Courtois, H. \& Vauglin, I. 2014, %{\it HyperLeda  III. The catalogue of extragalactic distances} 
A\&A, 530, 17\\
Mancera Pina, P. \etal 2023, ApJL, 833, L33\\
Mancera Pina, P. \etal 2020, MNRAS, 495, 3636\\
Masters, K., Springob, C. \& Huchra, J. 2014, %{\it Erratum: 2MTF I. The Tully Fisher Relations in the 2 micron all sky survey. 
%J, H \& K bands} 
AJ, 147, 124\\
Moustakas, J. \etal 2023, ApJS, 269, 3\\
Murugeshan, C. \etal 2024, submitted to PASA\\
O'Beirne, T. \etal 2024, in preparation\\
Paturel, G., Bottinelli, L. \& Gougenheim, L. 1993, %{\it The PGC-ROM 1992, a huge galaxy catalogue on a standard CDROM}, 
Bulletin d'Information du Centre de Donn\'ee Stellaires, 42, 33\\
Paturel, G. \etal 2003, %{\it HyperLeda I Identification and designation of galaxies}, 
A\&A, 412, 45\\
Paturel, G. \etal 2005, %{\it A catalog of LEDA galaxies with DENIS measurements}, 
A\&A, 430, 751\\
Rajohnson, S. \etal 2024, MNRAS, in press\\
%Rees, M. 1986 %{\it Gravitationally confined gas in dark minihaloes} 
%MNRAS, 218, 25\\
Riess, A. \etal 2022, %{\it A comprehensive measurement of the Hubble Constant with 1 \kms/Mpc uncertainty from the Hubble Space Telescope and SHOES team} 
ApJ, 934, 7\\
Robotham, A. \& Obreschkow, D. 2015, %{\it Fitting linear models to multidimensional data with multivariate gaussian uncertainties}, 
PASA, 32, 33\\
Said, K., Colless, M., Magoulas, C., Lucey, J., \& Hudson, M., 2022, MNRAS, 497, 1275\\
Schlafly, E. \& Finkbeiner, D. 2011, ApJ, 737, 103\\
Serra, P. \etal 2015, MNRAS, 448, 1922\\
%Spekkens, K. \& Karunakaran, A. 2018, %{\it Atomic Gas in blue UDGs around Hickson Compact Groupdf}, 
%ApJ, 855, 28\\
Springob, C. \etal 2009 %{\it SFI++ II: A new Tully-Fisher catalog, Derivation of peculiar velocities and dataset properties}, 
ApJS, 182, 474\\%Schlegel, D. \etal 1998, {\it Maps of dust infrared emission for use
%in estimating of reddening and cosmic microwave background radiation foregrounds}  ApJ, 500, 525\\
Springob, C. \etal 2005, %{\it A digital archive of HI 21 cm line spectra of optically targeted galaxies}, 
ApJS, 160, 149\\
%Steyn, N. \etal 2024, MNRAS, 529, L88\\
Taylor, E.N. \etal 2023, %{\it The 4MOST Survey of the Nearby Universe (4HS)}, 
The Messenger, 190, 46\\
Theureau, G., Hanski, M.~O., Coudreau, N., Hallet, N. and Martin, J. -M.,
2007 A\&A, 465, 71\\
%       Theureau, G. \etal 2006, %{\it The Nancay Radio Telescope archive}, 
%ASP Conf. Ser., 351, 429\\
Tully, R.B., Rizzi, L., Shaya, E., Courtois, H., Makarov, D. \& Jacobs, B. 2009 %{\it The Extragalactic Distance Database} 
AJ, 138, 323\\
Tully, R.B. \& Fisher, J.R. 1977, A\&A, 54, 661\\
Tully, R.B., Courtois, H., Hoffman, Y., \& Pomarede, D. 2014, %\etal {\it The Laniakea Supercluster of galaxies}, 
Nature, 513, 71\\
Tully, R.B. \& Fouqu\'e, P. 1985, %{\it The extragalactic distance scale. I - Corrections to fundamental observables} 
ApJS, 58, 67\\
Wen, X.-Q. \etal 2013, %{\it The stellar masses of galaxies from the 3.4$\mu$m band of the WISE all-sky survey} 
MNRAS, 433, 2946\\
Westmeier, T. \etal 2021, %{\it SOFIA2: an automated parallel HI source finding pipeline for the WALLABY survey} 
MNRAS, 506,~3962\\
Westmeier, T. \etal 2022, %{\it WALLABY Pilot Survey: Public release of almost 600 galaxies from phase 1 of ASKAP pilot observations} 
PASA, 39, 58\\
Wright, E. \etal 2010, %{\it The Wide-Field Infrared Survey Explorer (WISE): Mission description and initial on-orbit performance} 
AJ, 140, 1868\\
\end{multicols}
\clearpage
%%%%%%%%%%%%%%%%% APPENDICES %%%%%%%%%%%%%%%%%%%%%

\twocolumn
\appendix
\renewcommand{\thesection}{A\arabic{section}}
\section*{Appendix}
\subsection{SoFiA velocity widths}
In addition to the velocity widths we have measured and presented in Table 3, the SoFiA pipeline for WALLABY
provides Wm50\footnote{This is distinct from w50 in the WALLABY catalogue.}, and the relation between the two is shown in Figure A1. Both are HI profile widths at 50\%, but there is a systematic difference
which is mostly unrelated to signal-to-noise. Both velocity widths are from velocities on the optical
convention. A linear fit has a slope of 0.92 and a $\chi^2$ of 1.3. In other words, the scatter is close to that
expected on the grounds of measurement errors, but the non-unity slope is a quantity to be noted
for investigations that require a calibrated TFR.
The mean difference W50 -- Wm50 is --7.7 $\pm$ 0.7 \kms with no significant difference in the mean between the high and lower SNR data. We use our measured W50s in this paper, because
measurement errors are computed. Not so for Wm50.

The Wm50 measurements are made by interpolating the WALLABY spectrum at 50\% of the peak flux. This is automated. The w50 values on the other hand are measured by isolating the region of the spectrum where the galaxy flux is by setting a cursor. Based on the signal to noise a level of smoothing is chosen (2, 4, 8 pixels), and the smoothed spectrum is then interpolated similarly. The conversion of w50 to Wmx takes account of the broadening introduced by this smoothing. For the full WALLABY survey it may be necessary to adopt the Wm50 approach. But it will be important to include uncertainty calculation in the process.
\renewcommand{\thefigure}{A\arabic{figure}}
\begin{figure}
\includegraphics[width=1.2\columnwidth]{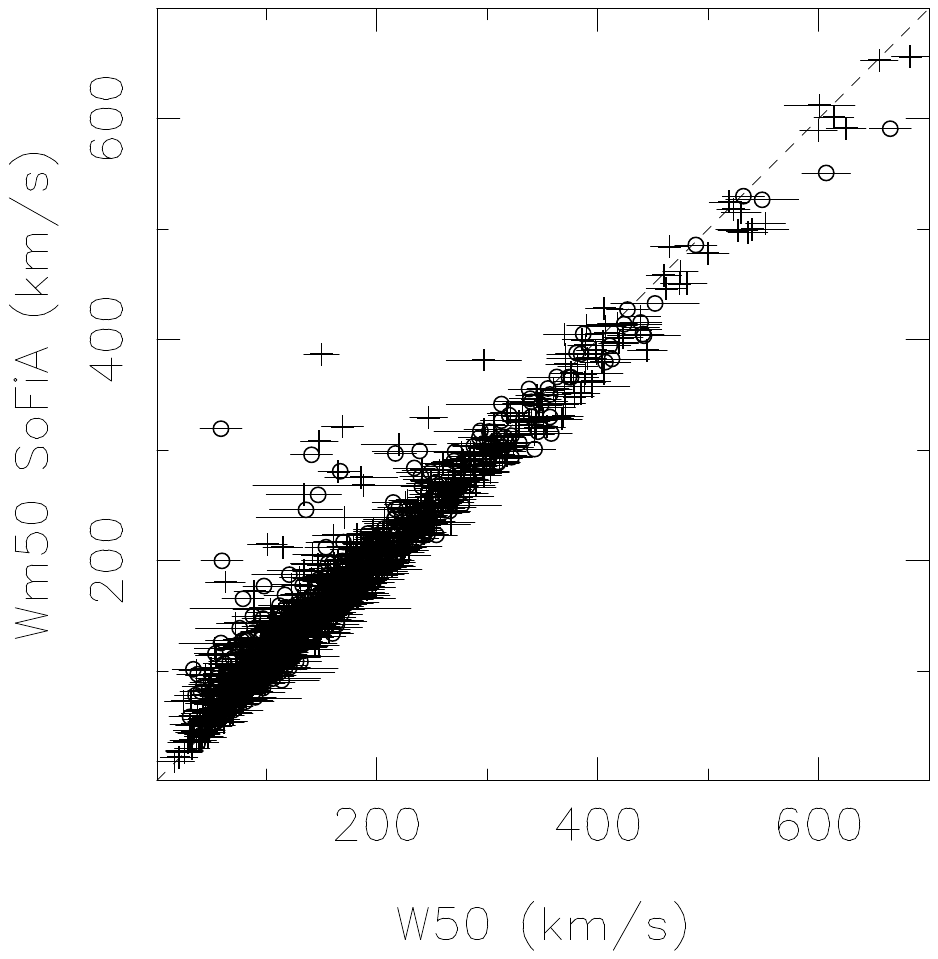} %FIG A1
\caption{Comparison of SoFiA Wm50s with those in Table 3 for the NGC 5044 field. Crosses are for S/N $>$ 3, open circles for S/N $<$ 3. The figure below shows the differences, with the higher S/N in green. The red line is a running mean.}
\includegraphics[width=1.2\columnwidth]{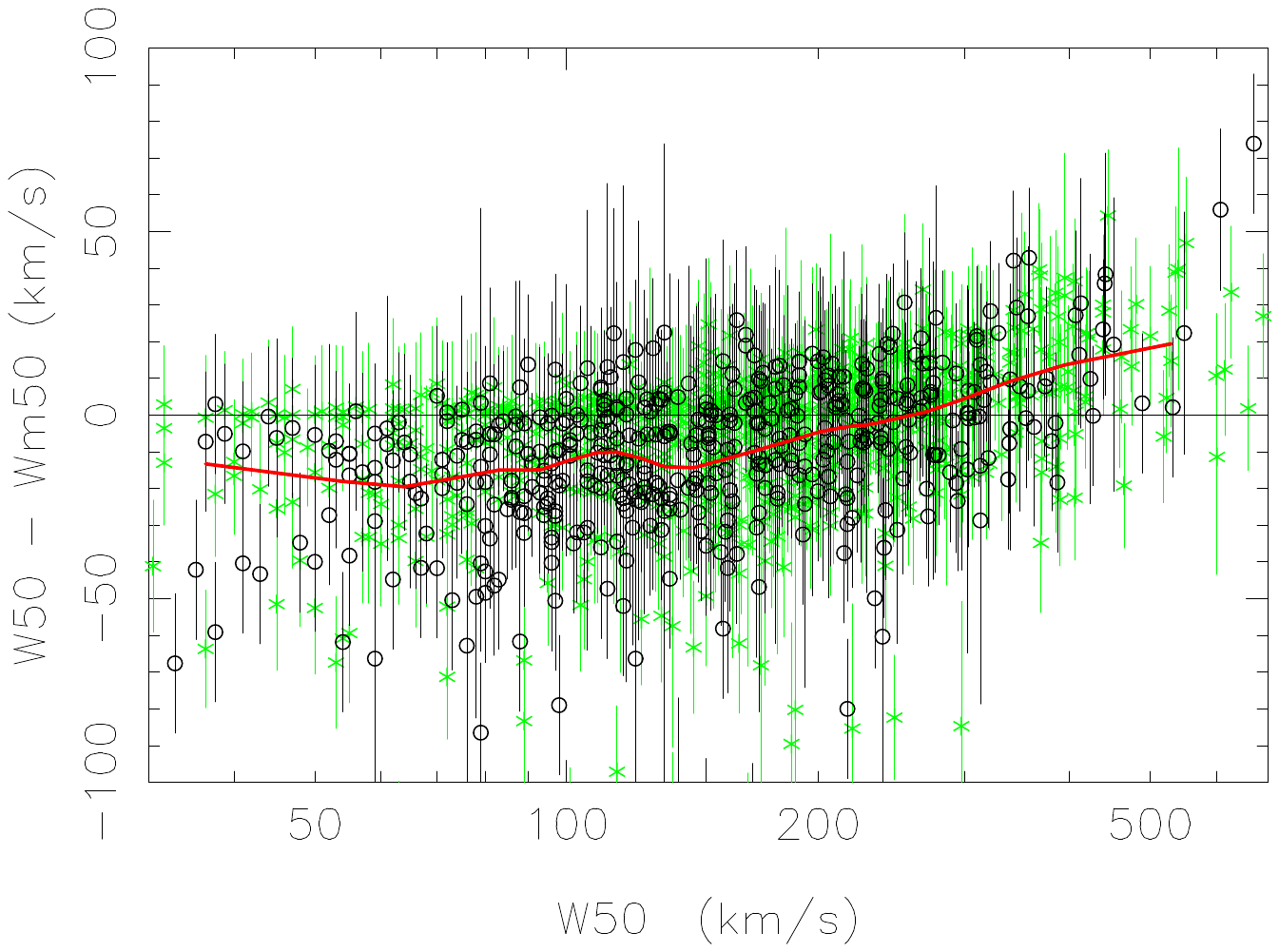} %FIG A1b
\end{figure}
\subsection{Alternative axial ratios}
In \S3.2 we compared IRAF measurements of axial ratios on $g$ band DECaLS images with kinematic inclinations. The comparison showed considerable scatter.
Other sources of inclinations are worthy of consideration and we show them in Figure A2. The measurement of WISE total magnitudes also produces axial ratios and these are shown in red in the figure. They are biased towards higher inclinations than the kinematic inclinations. We also wrote an ellipse fitting program on a different principle from that of the IRAF task aiming at the 25th $g$ magnitude isophote. Minimization of the deviation from an ellipse with (a, b, \& PA) as
the parameters of the isophotal pixels yields the solid black symbols with error bars in Figure A2. The horizontal axis errors derive from the covariance matrix of the fit. They seem unrealistically small, but, while some inclined TFR galaxies are well fitted by ellipses in the outer disk, others are more irregular in their stellar light distribution, a problem acknowledged in the original TFR paper (Tully \& Fisher 1977). These inclinations are also biased towards higher inclinations, which the IRAF inclinations (Figure 13) are not.

\begin{figure}
	\vspace{-4mm}
\includegraphics[width=1.3\columnwidth]{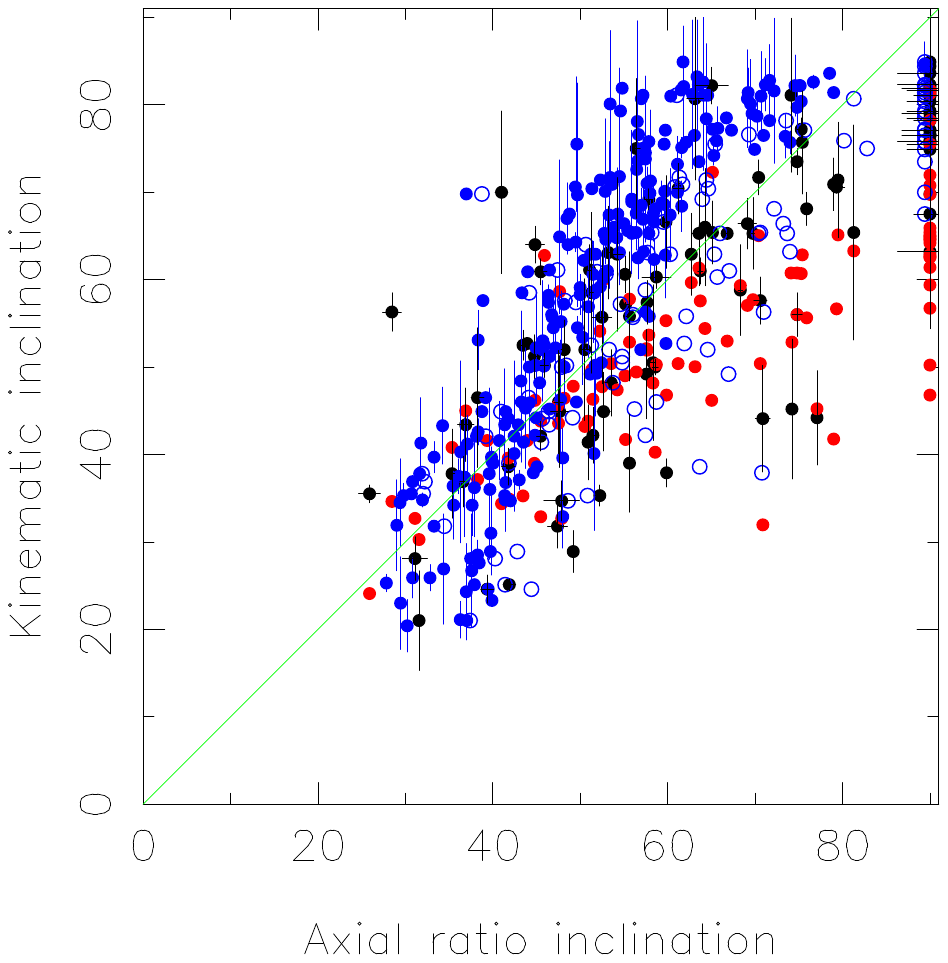}%FIG A2
\caption{Alternative sources of disk inclination. Solid black symbols are from ellipse fitting to $g$ band with experimental software; the red symbols are from WISE W1 axial ratios; the blue symbols are from resolved WALLABY hydrogen maps (open, with beam size correction), (solid, without). All these are plotted against kinematic inclinations by Deg \etal (2022). Pilot survey phase 1 data have been included in the solid blue category to more fully populate the graph, and these have error bars to distinguish them. } 
	\vspace{-4mm}
\end{figure}
The SoFiA source finding algorithm also provides estimates of a and b for the neutral hydrogen. Inclinations from this axial ratio are plotted in blue, the solid symbols for the raw b/a, the open symbols for the same galaxies after subtracting 5 pixels%$^{``}$
 in quadrature to allow for the WALLABY beam size. The raw values are biased, but the open ones, which are not biased, have more scatter. The sample with kinematic models is still small and will remain only a subset of WALLABY TFR galaxies. However, it is possible that bias in the raw hydrogen axial ratios can be calibrated out, when more data are available. This might be the path to dispensing with troublesome optical inclinations.

Finally, the stack of galaxies in Figure A2 at 90$^\circ$ axial ratio inclinations is due to galaxies being measured with b/a $<$ 0.2. This choice of minimum axial ratio should be revised when more data are obtained. For the time being we note that --8.7~$\log~\sin~80^\circ$ = 0.06 mag, a small correction.

\subsection{CosmicFlows4 distances}
The positions of WALLABY galaxies within the Cosmic Web
can be studied using the full matter (dark and luminous) density
contrast field (usually denoted $\delta$) reconstructed from the CosmicFlows-
4 Catalogue of peculiar velocities (Tully \etal 2023, Courtois \etal 2023).
This catalogue now uses the Baryonic TFR (Kourkchi \etal 2022) to calculate distances, and stellar masses are obtained from multiwavelength photometry.
Our WALLABY galaxies have W1 photometry, although within the decade multiwavelength galaxy photometry may be supplied by the Rubin Telescope.
For the 10,000 CF4 galaxies we have compared baryonic masses calculated with stellar masses from W1 as in $\S$2.4 with those published in CF4.
The result is in Figure A3, after a correction of 0.525 mag to the W1 magnitudes of Kourkchi \etal, which are from a different source from the W1 total magnitudes
used here and in Paper 1.
This allows us to use a baryonic TFR calibration relation from the adjusted CF4 galaxies, shown in Figure A4.
To be incorporated into CosmicFlows4, the distance moduli obtained with this calibration also needed a 0.15 mag change for the Hubble constant 
used in this paper and that calculated from the CF4 database. 
%\begin{figure*}
%	\setcounter{figure}{2}
%\includegraphics[width=.36\textwidth]{figa3.pdf}%FIG A3 & 3 more
%\includegraphics[width=.33\textwidth]{J131805a.pdf}
%\includegraphics[width=.33\textwidth]{J132022.pdf}
%\includegraphics[width=.36\textwidth]{figa3a.pdf}
%\includegraphics[width=.36\textwidth]{figa3b.pdf}
%\includegraphics[width=.36\textwidth]{figa3c.pdf}
%\caption{Optical images in $g$ band from the DECaLS survey of the dark galaxies in Table A1. The WALLABY
%position is marked by the very small red cross at the centre of each frame.}
%\end{figure*}
\begin{figure}
	\vspace{-4mm}
\includegraphics[width=1.2\columnwidth]{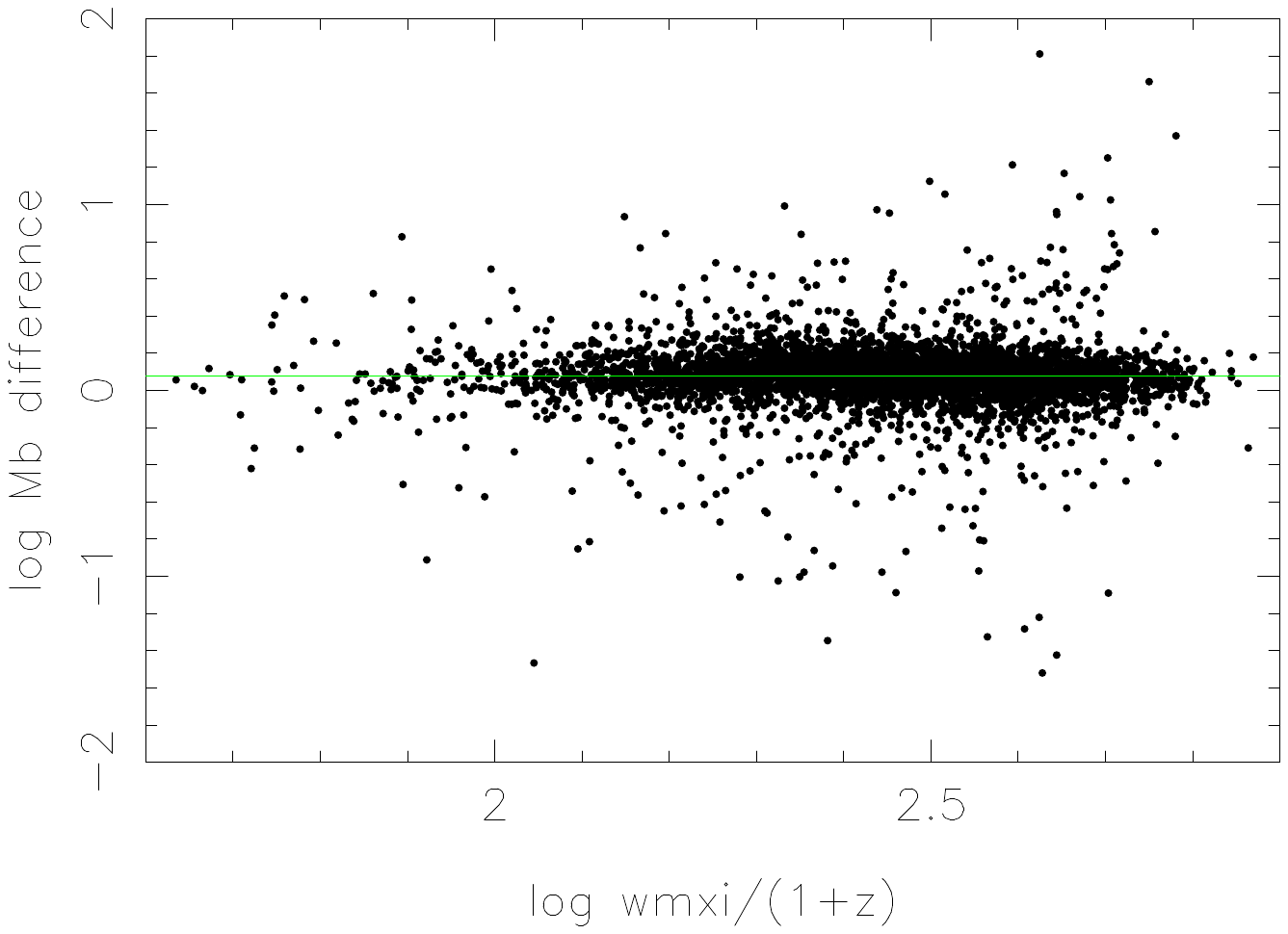}%FIG A3
\caption{Difference in baryonic mass calculated with W1 magnitudes only and with the multiwavelength photometry of Kourkchi \etal (2022).}
\end{figure}
\begin{figure}
	\vspace{-1cm}
\includegraphics[width=1.2\columnwidth]{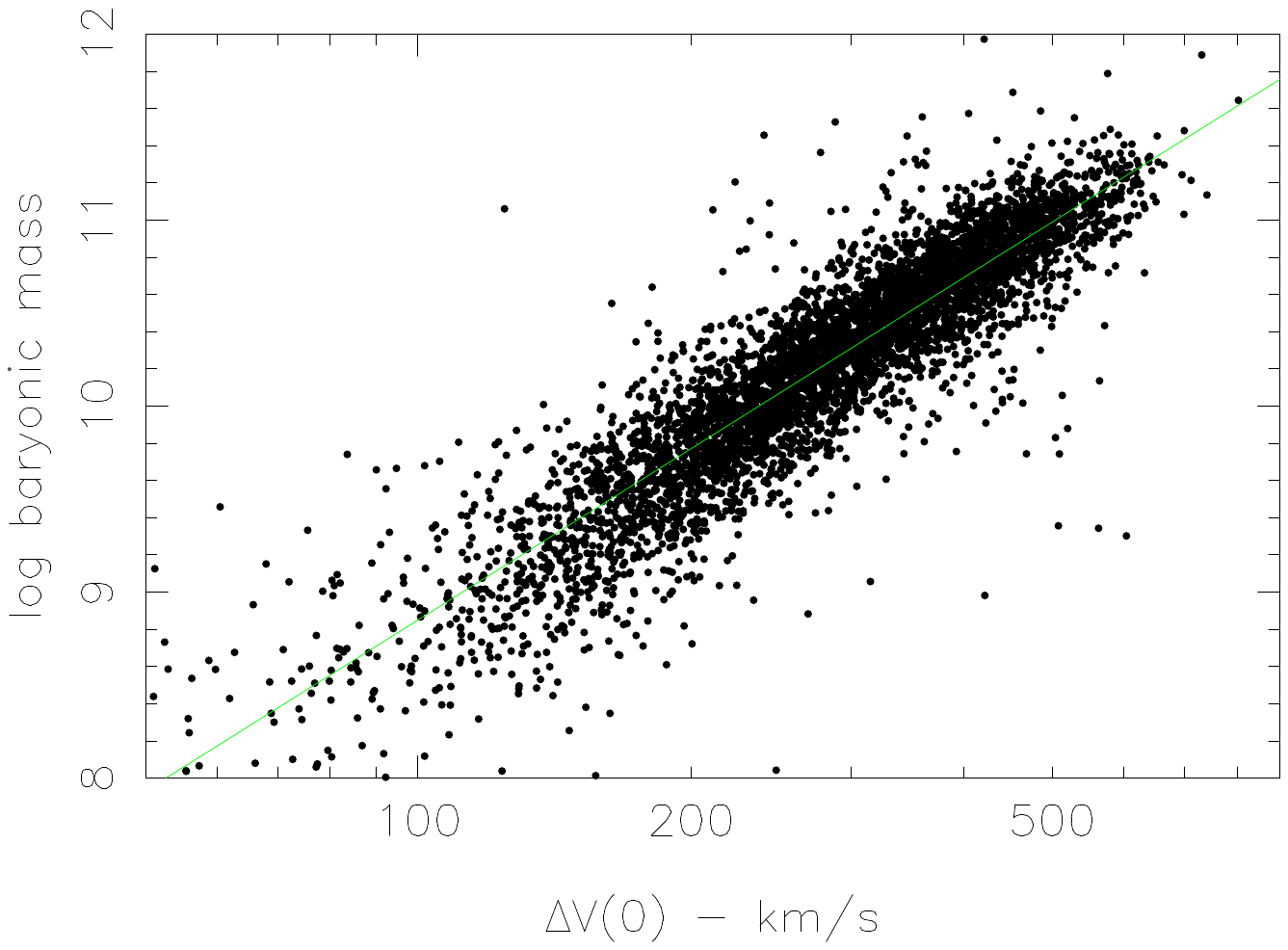}%FIG A4
\caption{The calibration used in $\S$5.3 for the baryonic TFR,\\  
 based on the CF4 galaxies of Kourkchi \etal (2022).}
\end{figure}
\subsection{Siena catalogue TFR for the NGC4808 field}
The Siena catalogue (Moustakas \etal 2023) has photometry in the $grz$ bandpasses for bright galaxies over half the sky and aspires to complete sky coverage outside the Galactic plane. These total magnitudes are measured by exponential disk fitting which also produces axial ratios. We show the TFR in the NGC 4808 field for $z$ magnitudes and Siena axial ratios in Figure A5. The normal cut is applied to inclinations, but no S/N cut is made. This is a promising new resource for the WALLABY project.

\begin{figure}
\includegraphics[width=\columnwidth]{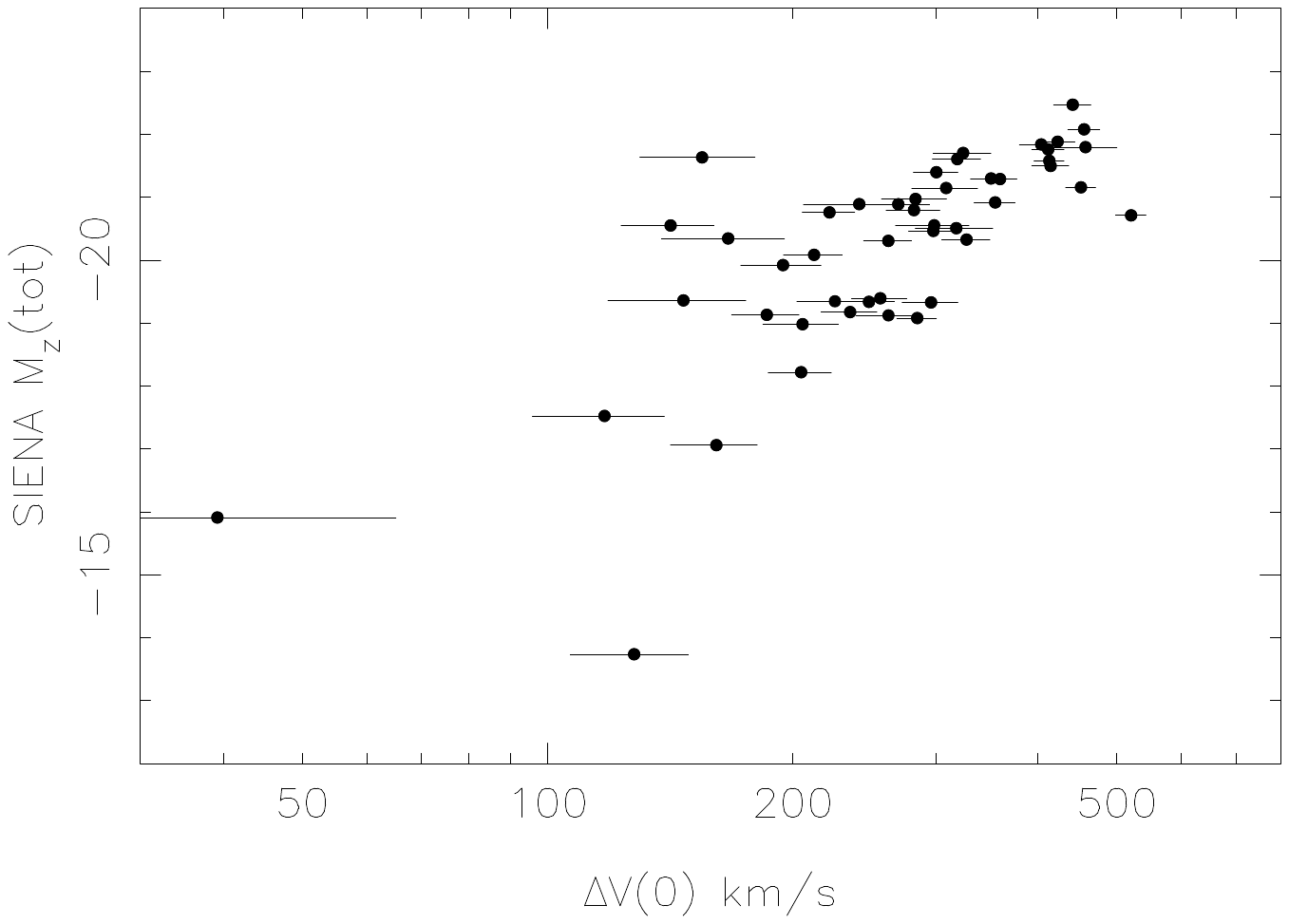}%FIG A5
\caption{The TFR for the NGC4808 field from Siena catalogue $z$ band  
 photometry. These magnitudes are on the AB system.}
\end{figure}
%%%%%%%%%%%%%%%%%%%%%%%%%%%%%%%%%%%%%%%%%%%%%%%%%%

% Don't change these lines

% End of mnras_template.tex
\bsp	% typesetting comment
\label{lastpage}
\end{document}